\documentclass[usenatbib]{mnras}%{aa} 
\usepackage{amsmath}
\usepackage{times}
\usepackage{graphicx}
\usepackage[usenames]{color}%\usepackage{txfonts}
\usepackage{hyperref}
\usepackage[]{amssymb}
\definecolor{pink}{rgb}{1,0.60,0.60}
\definecolor{green}{rgb}{0.,1.,0.}
\definecolor{lightblue}{rgb}{0.,0.6,.60}
\definecolor{orange}{rgb}{1,0.5,0}
\definecolor{brown}{rgb}{.7,0.3,.1}

\newcommand{\cross}[1][1pt]{\ooalign{%
  \rule[0.7ex]{0.7ex}{0.15ex}\cr% Horizontal bar
  \hss\rule{0.15ex}{.5em}\hss\cr}}% Vertical bar

\newcommand{\Msun}{M$_\odot$}
\title[interferometry and hydrodynamics in $\gamma^2$ Vel]{Numerical simulations and infrared spectro-interferometry reveal the wind collision region in $\gamma^2$ Velorum}

\author[Lamberts et al.]{A.~Lamberts$^{1}$ \thanks{lamberts@caltech.edu} \thanks{Based on observations made with ESO Telescopes at the La Silla Paranal Observatory under programme ID 60.A-9054, 074.A-9025, 078.D-0656, 082.D-0146, 082.D-0452, and 084.D-0400.},
F. Millour$^{2}$ \thanks{fmillour@oca.eu},
A. Liermann$^3$,
L. Dessart$^{2}$,
T. Driebe$^4$,
G. Duvert$^{5}$,
\newauthor W. Finsterle$^{6}$,
V. Girault$^{2}$,
F. Massi$^7$,
R. G. Petrov$^2$,
W. Schmutz$^6$,
G. Weigelt$^8$,
\newauthor and O. Chesneau $^{2}$\thanks{ O. Chesneau started the collaboration for this article. He unfortunately passed away before publishing it.}\\
$^{1}$ Theoretical Astrophysics, California Institute of Technology, Pasadena, CA 91125, USA\\
$^2$ Universit\'e C\^ote d'Azur, Observatoire de la C\^ote d'Azur, CNRS, Laboratoire Lagrange, Nice, France\\
$^3$ Leibniz-Institut f\"ur Astrophysik Potsdam (AIP), An der Sternwarte 16, D-14482 Potsdam, Germany\\
$^4$ German Aerospace Center (DLR), Space Administration,  Königswinterer Str. 522-524, 53227 Bonn, Germany\\
$^5$ UJF-Grenoble 1, CNRS-INSU, Institut de Plan\'etologie et d'Astrophysique de Grenoble (IPAG), UMR 5274, Grenoble, France\\
$^6$  Physikalisch-Meteorologisches Observatorium Davos and World Radiation Center (PMOD/WRC), Dorfstrasse 33, CH-7260 Davos Dorf, Switzerland\\
$^7$ INAF - Osservatorio Astrofisico di Arcetri, Largo E. Fermi, 5, I-50125 Firenze, Italy\\
$^8$ Max-Planck Institut f\"ur Radioastronomie, Auf dem H\"ugel 69, D-53121, Bonn, Germany
 }

\begin{document}
\date{\today}% Accepted **. Received **; in original form **}

\pagerange{\pageref{firstpage}--\pageref{lastpage}} \pubyear{2011}

\maketitle

\label{firstpage}

\begin{abstract}
Colliding stellar winds in massive binary systems have been studied through their radio, optical lines and strong X-ray emission for decades. More recently, near-infrared spectro-interferometric observations have become available in a few systems, but isolating the contribution from the individual stars and the wind collision region still remains a challenge. In this paper, we study the colliding wind binary $\gamma^2$ Velorum and aim at identifying the wind collision zone from infrared interferometric data, which provide unique spatial information to determine the wind properties. Our analysis is based on multi-epoch VLTI/AMBER data that allows us to separate the spectral components of both stars.  First, we determine the astrometric solution of the binary and confirm previous distance measurements. We then analyse the spectra of the individual stars, showing that the O star spectrum is peculiar within its class. Then, we perform three-dimensional hydrodynamic simulations of the system from which we extract model images, visibility curves and closure phases which can be directly compared with the observed data. The hydrodynamic simulations reveal the 3D spiral structure of the wind collision region, which results in phase-dependent emission maps. Our model visibility curves and closure phases provide a good match when the wind collision region accounts for 3-10 percent $\gamma^2$ Vel's total flux in the near infrared. The dialogue between hydrodynamic simulations, radiative transfer models and observations allows us to fully exploit the observations. Similar efforts will be crucial to study circumstellar environments with the new generation of VLTI instruments like GRAVITY and MATISSE. 
\end{abstract}


\section{Introduction}

$\gamma^2$ Velorum~\footnote{other names denominations include WR 11, HD 68273} ($\gamma^2$ Vel hereafter) is the closest Wolf-Rayet (WR) star, located at 336\,pc \citep[see for instance][]{2007MNRAS.377..415N}. It is therefore a unique laboratory to scrutinize the physics of WR stars. They display strong, broad emission lines due to their powerful winds 
 \citep[see][for a review]{2007ARA&A..45..177C}. The resulting mass loss rate is crucial to the final evolution of these massive stars. In binaries with massive stars, like $\gamma^2$ Vel, they are thought to be the progenitor systems of the compact object mergers detected with gravitational wave observatories \citep{2016ApJ...818L..22A}. Due to their very short lifetime, WR stars trace recent massive star formation, which is the main source of stellar feedback and chemical enrichment in galaxies. 
  
 $\gamma^2$ Vel is composed of a WC\,8 component in a close binary system with an O star companion in a 78.5 day orbit. It has been extensively studied by means of photometry, spectroscopy, spectropolarimetry, and spectro-interferometry, making it likely the best known WR + O system in our galaxy. Masses, spectral types and flux ratios of the components, as well as  the distance of the system have been subject of numerous studies 
 \citep[e.g.][]{1997A&A...328..219S, 1997ApJ...484L.153S, 1997NewA....2..245V,2007MNRAS.377..415N,  2007A&A...464..107M}.  Detailed modelling of the optical and infrared spectra, based on radiative transfer models in expanding atmospheres determine the stellar and wind parameters for both stars \citep{1999A&A...345..163D,2000A&A...358..187D}. \citet{1999A&A...345..163D,2001NewAR..45..135V} establish the O star as an O7.5 III-I type, which is hotter than previous estimates, where blends with lines from the WR star were not accounted for \citep{1971A&A....11...83B,1972ApJ...172..623C,1997ApJ...484L.153S}. 
 
\citet[][hereafter Paper~I]{2007A&A...464..107M} refined the distance to the system ($368^{+38}_{-13}$ pc) using an accurate astrometric measurement  from the Astronomical Multi-BEam Recombiner  \citep[AMBER,][]{Petrov2007} operated at the Very Large Telescope Interferometer (VLTI). In that first paper, we also revised the spectral types of the WR and O star through the comparison with radiative transfer models for both components. We discussed the origin of strong residuals in the absolute visibility curves, speculating they would come from the free-free emission of the wind collision zone (WCZ) between the two stellar winds, but could not provide conclusive evidence. The detection of the WCZ is a major observational challenge as the semi-major axis of the binary is 3.5  thousandths of an arcsecond (milli-arcsecond or mas -- 1.2 AU at the distance of the binary) and, according to   \citet{1970MNRAS.148..103H,2007MNRAS.377..415N,2007A&A...464..107M}, the stellar radii are below a milli-arcsecond.

\begin{figure}
   \centering
 \includegraphics[width = .45\textwidth]{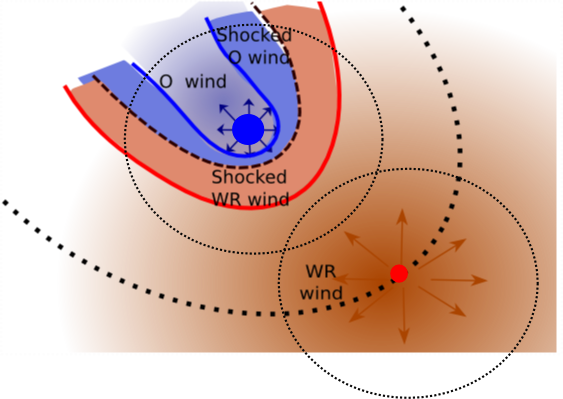}
  \caption{Schematic view of $\gamma^2$ Vel in the plane of the binary showing the WR wind (red) and O wind (blue) and their shocked counterparts. Both winds are separated by a contact discontinuity (dashed black line). The dotted arc shows the orbit of the WR star around the O star. The dotted circles show the regions referred to as the WR star and the O star.The stellar disks, binary separation and shocked regions are roughly to scale.} 
  \label{fig:schema}
 \end{figure}

The collision of the two hypersonic winds results in a dense, hot shocked region, as sketched in Fig.~\ref{fig:schema}. The shocked region manifests itself in various ways. First evidence comes from periodic variability in  emission lines \citep{1993ApJ...415..298S}, as the shocked region eclipses line forming regions during a fraction of the orbit.  Due to its $10^7-10^8$\,K temperature, the WCZ is directly detected in X-rays and analysis of the lightcurves and spectra constrain the opening angle of the shocked region and mass loss rates of the winds \citep{1995A&A...298..549W,1996MNRAS.283..589S,2000MNRAS.316..129R,2004A&A...422..177S}. Direct detection of the WCZ has been elusive in other wavebands.  Non-thermal radio emission, which is often detected in colliding wind binaries as a result of particle acceleration at the shocks remains undetected \citep{1990MNRAS.244..101W}. Dust emission often found in binaries with a WR component \citep[WC9 subtypes, see e.g. WR 104 in][]{2008ApJ...675..698T} is also notably absent \citep{2002ASPC..260..331M}. Direct observations of the WCZ in the optical and infrared remains a challenge, as it is strongly overshadowed by the wind and/or photospheric emission of the binary. It has only been possible for the very bright $\eta$ Carinae system, where the angular separation between the stars is much larger \citep{2016A&A...594A.106W}. While signatures of the WCZ are likely present in the data presented in Paper~I, a firmer answer requires more data and a detailed model of the structure and dynamics of the interaction region. 

The shocked winds are separated from each other by a contact discontinuity (see Fig.~\ref{fig:schema}), which may be subject to various instabilities \citep{Stevens:1992on}. Close to the binary, the hydrodynamic structure is essentially determined by the momentum flux ratio of the winds and radiative cooling.  In $\gamma^2$ Vel, we expect the shocked structure to be heavily bent towards the O star, which has a less dense wind. Further out, orbital motion turns the shocked structure into a spiral \citep{2012A&A...546A..60L}.  While analytic estimates provide the location of the shocked region between the stars and the asymptotic opening angle of the WCZ \citep{1990FlDy...25..629L}, hydrodynamic simulations are necessary to determine the complete 3D geometry of the interaction region, especially at distances where the spiral structure cannot be neglected. Only numerical simulations provide a complete model of the density, velocity and temperature conditions in the WCZ \citep{2009MNRAS.396.1743P}. These can then be used to model data  to be compared with observations. For instance, \citet{1999IAUS..193..298W,2005MNRAS.356.1308H} computed  X-ray data based on hydrodynamic simulations. Comparison with observations suggested that radiative braking is probably at work in $\gamma^2$ Vel. Radiative braking  \citep{1997ApJ...475..786G} may occur when the shocked region is very close to the secondary star and is pushed away by the strong radiation pressure of the latter. It results in a larger opening angle than would be expected from the wind parameters of the stars. 

Infrared interferometry offers a unique opportunity to  directly probe the WCZ and put constraints on its spatial extension and relative flux. With this paper, we report on the follow-up AMBER observations on this system (\S\ref{sec:obs}). First, using a geometrical model, we refine the orbital solution of the binary using the wide orbital coverage of the data. Our continuum and spectral analysis provides further information on the residuals discussed in Paper~I, that were tentatively attributed to the wind-wind collision zone between the two stars (\S\ref{sec:data_model}).  We then present the 3D hydrodynamic simulations that establish the properties of the WCZ (\S\ref{sec:model}). Based on the simulations, we compute model images and then compare the resulting visibility curves and closure phases with the data (\S\ref{sec:mock_data}). We then discuss how the direct combination of numerical simulations and observational data opens new possibility for the study of massive binary systems (\S\ref{sec:discussion}) and conclude (\S\ref{sec:conclusion}).

\begin{table*}
  \caption{Log of $\gamma^2$ Vel observations, sorted by date, including the data of Paper~I, which we have reprocessed for the purpose of the current paper. The spectral configuration distinguishes between low resolution (LR), medium resolution (MR) and high resolution (HR) configurations. \#Obs. is the number of observations of Gamma Vel during the considered night. "Orb." stands for "Orbital solution determination" and  "Spec." stands for "Spectral separation" in the "Used for?" column.  In the dataset from 07/02/2006, baselines are all aligned in a direction perpendicular to the binary direction, making it useless (see text and Fig.~\ref{fig:plan_UV} for details).}
  \label{tab:logObs}
  \centering
  \begin{tabular}{lcccccccccccc}
    \hline
    \noalign{\smallskip}
    Night & Calibrators HD & Spec. conf. & Band &  \#Obs. & Used for? & MJD & Orb. phase \\ 
    \noalign{\smallskip}
    \hline
    \noalign{\smallskip}
    25/12/2004 & 75063         & MR & K    & 4 & Orb., Spec. & 53365.18--20 & 0.315 \\
    07/02/2006 & 73155         & MR & H    & 1 & -           &53774.15 & 0.523 \\
    29/12/2006 & 34053, 53840, & LR & H, K & 1 & Orb.        & 54099.28 & 0.663 \\
     & 81720  \\
    07/03/2007 & 37984, 69596  & LR & H, K & 1 & Orb.        & 54166.13 & 0.514 \\
    31/03/2007 & 68763         & MR & K    & 2 & Orb., Spec. & 54190.06--08 & 0.819 \\
    20/12/2008 & 68512, 75063, & HR & K    & 6 & Orb.        & 54821.18--37 & 0.857 \\
     & 69596  \\
    21/12/2008 & 35765, 68512, & MR & K    & 8 & Orb., Spec. & 54822.15--37 & 0.870 \\
      & 69596   \\
    22/12/2008 & 68512, 69596  & MR & K    & 8 & Orb., Spec. & 54823.15--38 & 0.882 \\
    22/01/2010 & 68512, 69596  & MR & H    & 8 & Orb., Spec. & 55219.07--28 & 0.924 \\
    06/01/2012 & 23805, 36689, & LR & H, K & 1 & Orb.        & 55932.27 & 0.005 \\
     & 67582, 82188 \\
    \noalign{\smallskip}
    \hline
  \end{tabular}
\end{table*}

\section{Observations and data processing}\label{sec:obs}

\subsection{Observations}
AMBER is a 3-telescope spectro-interferometric combiner \citep{Petrov2007} operating at the VLTI. It allows one to observe in the infrared J, H \& K bands simultaneously at a low spectral resolution of $\lambda/\Delta\lambda=35$, or the H or K bands at a
medium spectral resolution of $\lambda/\Delta\lambda=1500$, and small chunks of the K band at high spectral resolution of
$\lambda/\Delta\lambda=12000$,  with an angular resolution of a few thousandth of an arcsecond.  Since Paper~I, $\gamma^2$ Vel was observed with AMBER several times using all the available spectral configurations. We cover a wide range of orbital phases shown in Fig.~\ref{fig:orbitGammaVel} over a little more than 7 years.\footnote{The interferometric data will be publicly available at the Optical Interferometry Database from the Jean-Marie Mariotti Center http://oidb.jmmc.fr/index.html.}

Table~\ref{tab:logObs} lists the different epochs available together with the corresponding orbital phases. Of interest, are extensive observing campaigns in 2008 and 2010 where we used full nights to stack up high-SNR data of the system. By far, these nights provide the most accurate spectro-interferometric datasets of $\gamma^2$ Vel at one single orbital phase each. The medium spectral resolution data of  Dec, 25 2004 was already used in Paper~I to determine the distance of the system. 

The other datasets, obtained with low(er) spectral resolution, give less insights to the details of the binary system, but provide us with astrometric measurements along its orbit. The (u,v) plane coverage of each observation is shown in Fig.~\ref{fig:plan_UV}.

The interferometric field of view is approximately equal to the Airy pattern of one telescope, i.e. typically 50\,mas for the K-band UT (Unit telescope) observations performed before 2008, or 250\, mas for the K-band AT (Auxiliary Telescope) observations in 2008 and after. The interferometer is sensitive to a fraction of this field-of-view without any bias, as the use of fibres to perform spatial filtering has an impact on the effective unbiased field of view \citep{2004A&A...418.1179T}. In our case, the whole system is much smaller than the Airy disk of the field of view, therefore we do not expect to have a significant bias on the $\gamma^2$ Vel measurements.

The formal angular resolution of our observations is set to be the resolution at the longest baseline length, here $\approx$130\,m,  i.e. $\approx3.5$\,mas at $2\mu$m. However, the interferometer is sensitive to much smaller objects thanks to the precision of the measurements. For example, with a 3 per cent accuracy on one visibility measurement with a  130\,m baseline, one can measure a 91 per cent fringe contrast at the $3\sigma$ level. At 2$\mu$m, this corresponds to  a uniform disk of 0.95\,mas diameter. This "super-resolution" effect can be further improved by stacking measurements.

\subsection{Data reduction}

The AMBER datasets consist of spectro-interferometric measurements: namely one spectrum, three dispersed squared visibilities, three dispersed differential visibilities, three dispersed differential phases, and one dispersed closure phase per individual measurement (see figure \ref{fig:modelfit_lines} for illustrations of these observables). The data reduction scheme is now relatively settled for bright sources \citep[see details in][]{2007A&A...464...29T, Chelli2009c}. We use the standard \texttt{amdlib} data reduction library, version 3.0.4, plus our own additional scripts  for calibration \citep{Millour2008c}. Some datasets present specific issues, namely:
\begin{itemize}
\item 07/02/2006: there is only one science and one calibrator record, and this dataset presents baselines all aligned in the same direction. Unfortunately, the binary axis is perpendicular to the baseline direction, and the measurement is not used in our analysis.
\item 06/01/2012: the system is close to periastron, therefore the binary star is not formally resolved by the interferometer.
\end{itemize}

\section{Analysis of AMBER data with a geometrical model}\label{sec:data_model}

To model the interferometric data, we use the software \texttt{fitOmatic} \citep{2009A&A...507..317M}, which was developed in-house according to the lessons learned from Paper~I to account for chromatic interferometric datasets\footnote{fitOmatic is available at \url{http://fitomatic.oca.eu}}. \texttt{fitOmatic} is written in the scientific data language \texttt{yorick}\footnote{developed by  D. H. Munro, Lawrence Livermore National Laboratory, \url{http://yorick.github.io}}. It reads oifits  files (Optical Interferometry FITS,  \citet{2004SPIE.5491.1231P}). The user defines a model of the source, according to a set of pre-defined models including point sources, uniform discs, or gaussian disks \citep[as used in][]{Chesneau2014,Chiavassa2010, Millour2009a}. These models can be combined together to build up more complicated models, like the 3D-pinwheel nebula model used by \citet{Millour2009c}. Another option is to compute brightness distributions from models, as  for the rotating and expanding disk model used in \citet{Millour2011}. Once the parameters are set, the Fourier Transform of the modelled brightness distribution is computed and synthetic observables (squared visibilities, differential visibilities, differential phases, and closure phases) are computed at the wavelengths and projected baselines of the observations. The specific aspect of \texttt{fitOmatic} is its ability to define vector-parameters, i.e. parameters which may depend on wavelength and/or time.

These synthetic observables are then compared to the observed ones by computing square differences, which are minimized by varying the model parameters. A simulated annealing algorithm is used to scan the parameter space and optimize the fit globally (finding the least squares absolute minimum). In this section we present an analysis of the squared visibilities and closure phases based on geometrical models of the system. The complex visibility is the Fourier Transform of the brightness  distribution of a source. Its amplitude is related at first order to the size of the source, and its phase to its position and structure. The squared visibility is simply the squared modulus of the complex visibility, and therefore is related to the source's size. The closure phase is an "instrument free" combination of the phases for each baseline in a triplet of apertures  and relates to the asymmetry of the source. The differential phase is another way of measuring phases. At first order, it is related to the source position (photocenter) at a given wavelength, relative to its position at another wavelength. More details can be found in \citet{2003A&A...400..795L}. The analysis based on model images stemming from the hydrodynamic simulation is presented in Section~\ref{sec:mock_data}. 

\subsection{Continuum analysis}\label{sec:data_continuum}

\begin{figure}
\centering
    \includegraphics[width=0.48\textwidth]{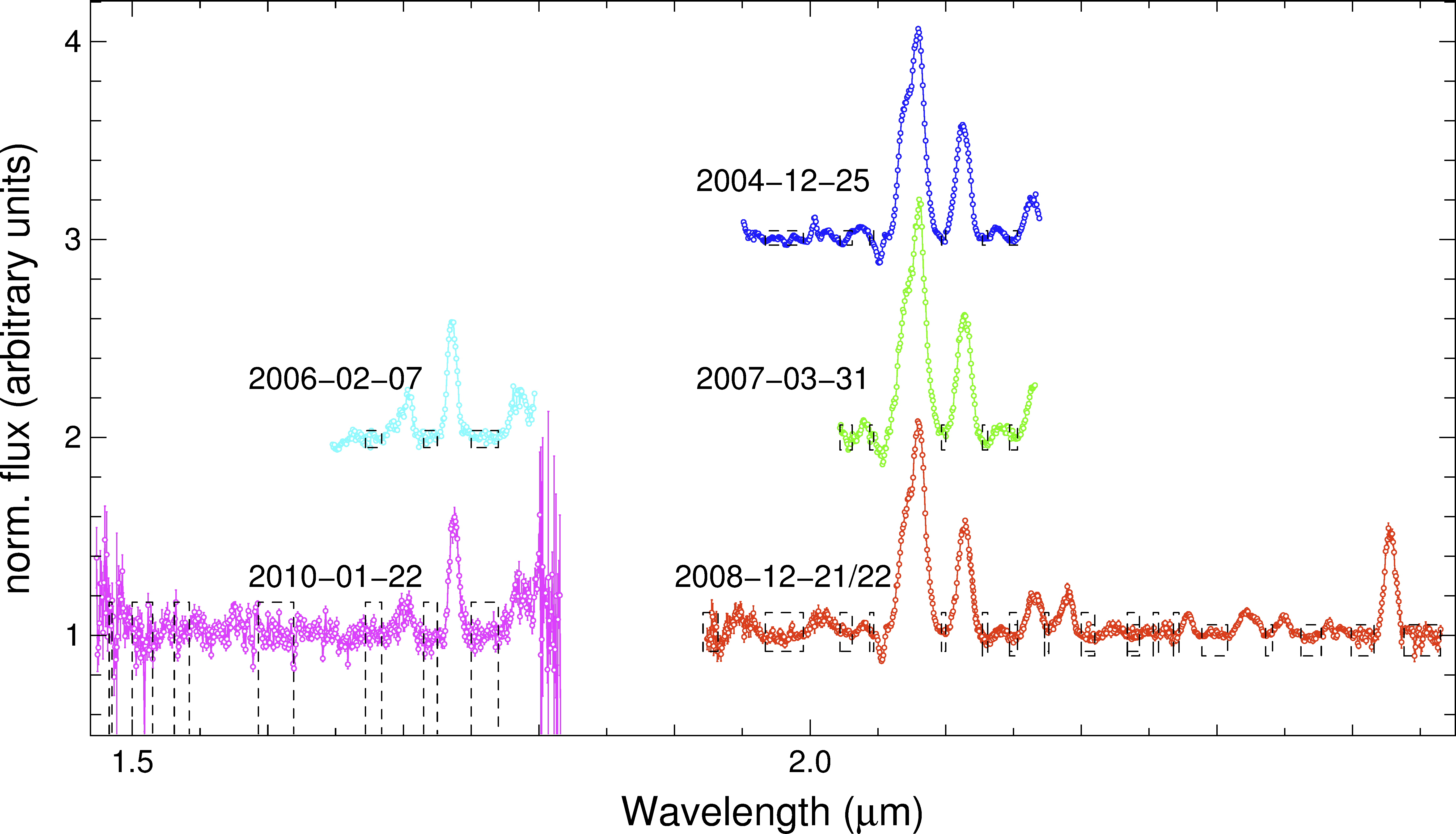}
  \caption{
Spectra of the $\gamma^2$ Vel system obtained with AMBER. The continuum normalization was determined within the dashed boxes. The colours correspond to different epochs: 2004 (blue, orbital phase $\phi=0.32$), 2006 (cyan, $\phi= 0.52$), 2007 (green, $\phi = 0.82$), 2008 (red, $\phi = 0.87$), and 2010 (magenta, $\phi= 0.92$). Spectra of the different epochs are offset for clarity.}
\label{Fig:spectraGammaVel}
\end{figure}

We  first focus on the continuum regions in our spectra (see Fig.~\ref{Fig:spectraGammaVel} in section~\ref{sec:data_spectrum}). We perform this first analysis on the two nights of December 2008 with medium spectral resolution ($R=1500$) since these two nights offer the best stability of the instrument  and atmosphere transfer function (better than 3 per cent for both nights). We checked that the results obtained here are consistent throughout the whole dataset.
 
\begin{figure*}
   \centering
  \includegraphics[width = .48\textwidth]{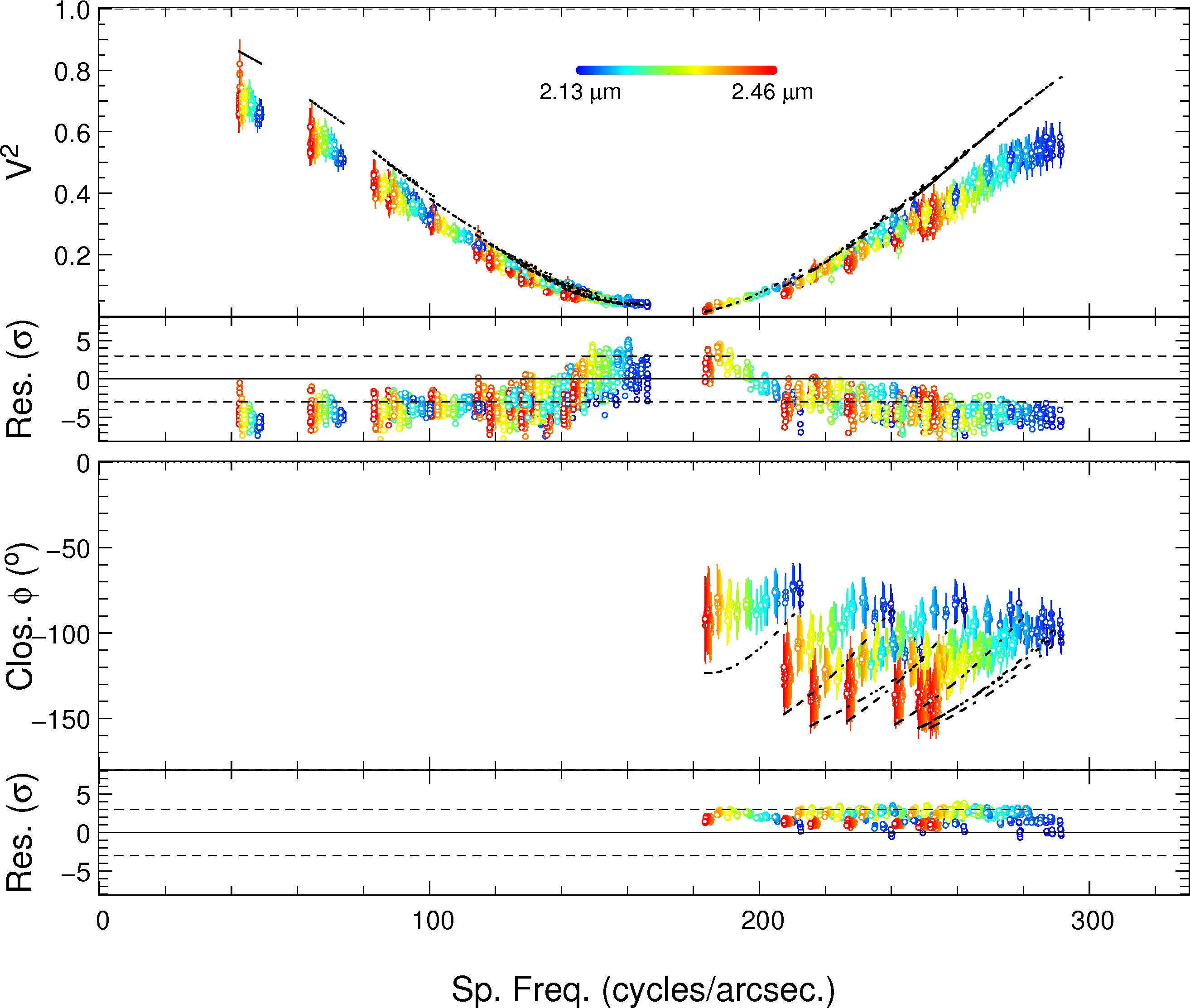}
  \includegraphics[width = .48\textwidth]{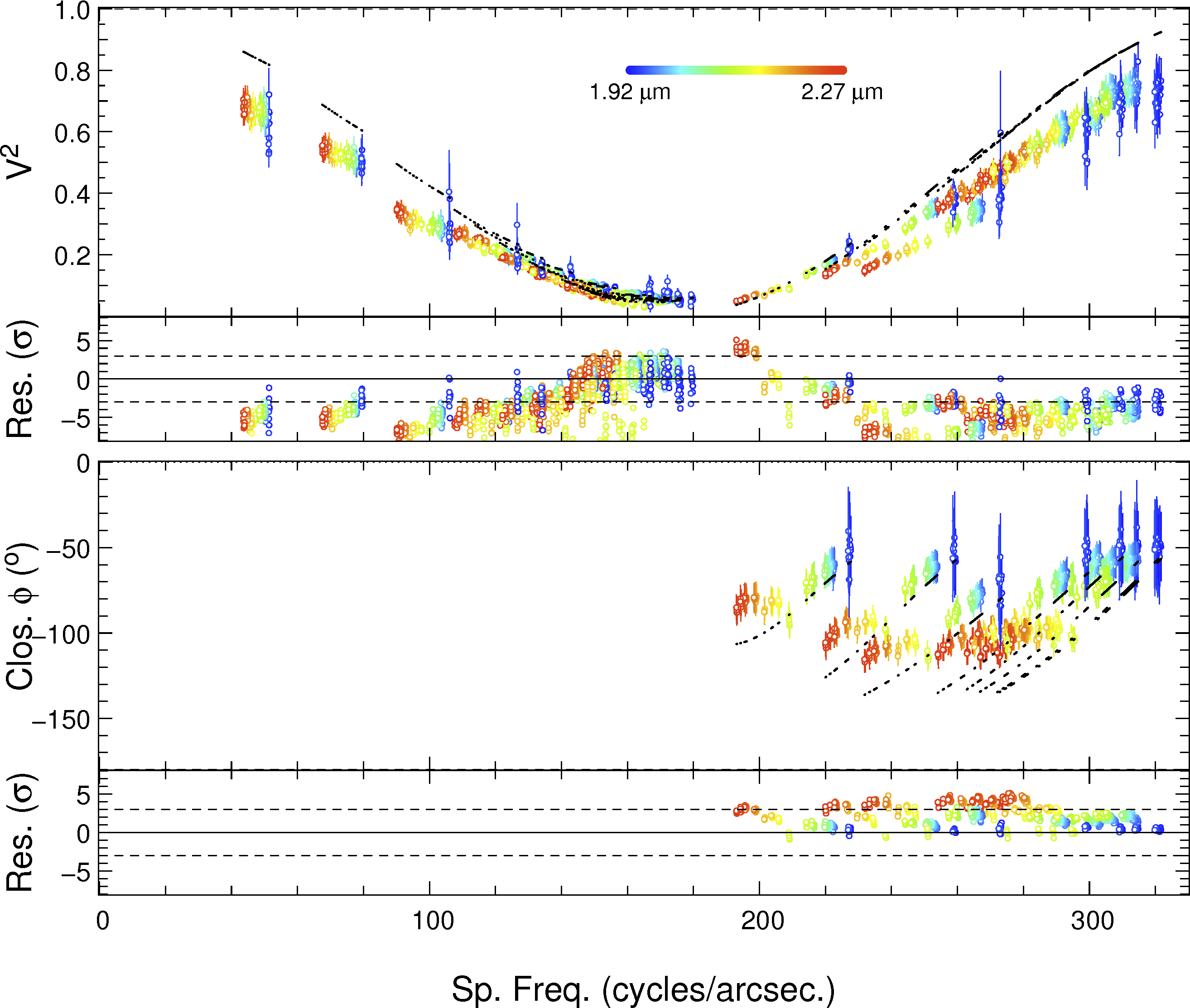}
  \caption{Continuum visibilities (top plots) and closure phases (bottom plots), together with residuals (smaller plots), from our high-fidelity observing campaign from 2008, projected along the binary star direction (to ease the reading of the binary star modulation). Left: 21/12/2008,  right: 22/12/2008. The dashed lines representing the 2 point sources geometrical model are too high in visibility and too far from zero in closure phase when compared with the measurements, showing that a "two point sources" model cannot perfectly fit the data. The colors show the wavelength dependence.}
  \label{fig:visibility1}
 \end{figure*}

As in Paper~I, we first compare  the AMBER data with a two point sources model, representing the two stars. Fig.~\ref{fig:visibility1} shows the best-fit models for the squared visibilities and closure phases for 21/12/2008 and 22/12/2008. The reduced $\chi^2$ is not good: 10.7 for the 21$^{\rm st}$ and 12.5 for the 22$^{\rm nd}$ and there are several discrepancies which are clearly above the noise limit, as in Paper~I. It is interesting to note that the higher visibilities in the model (in the 40 to 100 and the 280 to 320 cycles/arcsec intervals) are systematically larger than the measurements, and the measured closure phase is systematically closer to zero than the binary model values.

In Paper~I, we added a fully resolved uniform component to explain this effect. This additional flux greatly improves the agreement with the data (reduced $\chi^2$ of 1.1 and 2.3 for the 21$^{\rm st}$ and 22$^{\rm nd}$, respectively). However, such a uniform component around the binary has little physical justification. Therefore, we now focus on more physically relevant alternatives including resolved stellar components discussed here, and the WCZ model presented in Section~\ref{sec:mock_data}. 

We first use a Gaussian model for the WR and its wind and a uniform disk model for the O star. Our approach is similar to  \citet{2007MNRAS.377..415N}: the Gaussian model for the WR star is imperfect but good enough for this study, as the star and its wind are marginally resolved by the interferometer. Its full width at half maximum (FWHM) is related but not necessarily equal to the continuum-emitting region of the WR star. The FWHM definition of the WR size differs from the definition given in Paper~I (uniform disk (UD) diameter) and the relation between both is detailed in  Table 3 of Paper~I. However, since the stars are barely resolved, we can use a fixed factor of $\approx$0.65 between Gaussian FWHM and uniform disk diameter to compare both \citep[see e.g. values in Table 1 of][]{1995MNRAS.276..640H}. The resolved stellar components model (reduced $\chi^2$ of 1.3 and 3.1 for the 21$^{\rm st}$ and 22$^{\rm nd}$, respectively) match the data equally well as the model with a resolved uniform component, and makes more physical sense. We also combined the nights of 21$^{\rm st}$ and 22$^{\rm nd}$ to refine the values of the resolved diameters.  The reduced $\chi^2$ is less good in this case because binary motion is not included in this simple model.

Table~\ref{tab:models_chi2} provides a summary of the different models tested here. We note that the best match to the data results in rather large stellar diameters. We find a typical size of $\approx$1.1\,mas for the diameter of O star, and a typical full width half maximum of $\approx$0.9\,mas for the WR star. These values are about a factor two to three higher than the values computed in Paper~I (diameter of 0.48\,mas for the O star and 0.28 for the WR star) and \citet{2007MNRAS.377..415N}. Given the distance of the system, this corresponds to a radius of 51$R_{\odot}$ and 36 $R_{\odot}$ for the O and WR star respectively.

To strengthen this unexpected result, we performed the same analysis using the  LITpro software\footnote{available at \url{http://www.jmmc.fr/litpro}} \citep{2008SPIE.7013E..44T}. LITpro corroborates our findings and also provides a correlation matrix which shows a strong anticorrelation (-0.9) between the two stellar sizes. This means that the sizes may be interchanged, meaning a stellar diameter of 0.9\,mas for the O star and 1.1\,mas for the WR star. The latter situation would be a bit more in line with spectral analysis for the O star \citep{1999A&A...345..163D}, but would be worse for the WR star.

\begin{table*}%[htbp]
  \caption[]{Comparison between the AMBER data and several geometrical models in the continuum. The best-fit reduced $\chi^2$ is given for each case. The last line combines data from both nights. }
  \label{tab:models_chi2}
  \centering
    \begin{tabular}{lcccccccc}
    \hline
    \noalign{\smallskip}
    Night& $\chi^2$ 2 pt$^1$ & $\chi^2$ 2 pt + cont.$^2$  & $\chi^2$ G$^3$ + UD$^4$ & $\theta_{\rm G}$ (WR star) & $\theta_{\rm UD} $(O star) \\
    \noalign{\smallskip}
  \hline
    \noalign{\smallskip}
    21/12/2008 & 10.7 & 1.1 & 1.3 & $\theta_{\rm G} \leq 1$\,mas &  $1.5 \leq \theta_{\rm UD} \leq 1.9$\,mas \\
    22/12/2008 & 12.5 & 2.3 & 3.1 & $0.5 \leq \theta_{\rm G} \leq 1.1$\,mas & $0.5 \leq\theta_{\rm UD} \leq 1.3$\,mas\\
    21 + 22    & 13.8 & 3.2 & 3.8 & $0.39 \leq \theta_{\rm G} \leq 1.32$\,mas & $0.75 \leq  \theta_{\rm UD} \leq 1.63$\,mas\\
   \noalign{\smallskip}
    \hline
  \end{tabular}\\
  $^1$Point source, $^2$Fully resolved component, $^3$Gaussian disk, $^4$Uniform disk
\end{table*}

\subsection{Astrometry}

One product of the model fitting process is the astrometry of the binary star (separation, position angle) at all the observed epochs, listed in Table~\ref{tab:astrometry}. The three different models presented in Table~\ref{tab:models_chi2} provide the same astrometric measurements within 0.02\,mas, or 20 micro-arcsecond.

We adjust these data points with a classical Newtonian orbit with 7 parameters (period $P$, semi-major axis $a$, eccentricity $e$, periastron passage date $T_0$, longitude of periastron $\omega_{\rm WR}$, position angle of node $\Omega$, and inclination angle of the orbital plane relative to the line of sight $i$). To do so, we use a Markov Chain Monte Carlo algorithm with a simulated annealing algorithm for the hundreds of chain links we used, out of which we kept a few orbital parameters sets within 1 sigma error.

This allows us to provide an independent orbital solution of the system compared to \citet{2007MNRAS.377..415N}, obtained with optical interferometry. 
We first computed a solution with all the parameters free, including the eccentricity and period. This solution provides a final eccentricity ($0.29\pm0.01$) which is quite far from the previous best estimate based on radial velocities \citep{1997A&A...328..219S}. It also results in an orbital period in tension with previous results.  Similarly to \citet{2007MNRAS.377..415N},  we find  that the period $P$ and the time of periastron passage $T_0$ are strongly mutually dependent. 

\begin{table*}%[htbp]
 \caption{
 Parameters of the system from the literature and from the current work when the eccentricity $e$ is left as a free parameter or fixed to the \citet{2007MNRAS.377..415N} value or the \citet{1997A&A...328..219S} value. }
 \label{tableParameters}
\centering
 \begin{tabular}{lcccccccc}
 \hline
Parameter                           & Literature                                 & This paper ($e=0.334$) & This paper ($e=0.326$)\\
\hline \\
Distance (pc)                       & ${336^{+8}_{-7}}^1$, ${368^{+38}_{-13}}^2$ & $345\pm7$       & $346\pm7$ \\
\vspace{-0.3cm}\\
\hline \\
Period  $P$ (d)                     & $78.53 \pm 0.01^{3}$                     & \emph{fixed ($78.53$)} & \emph{fixed ($78.53$)}\\
Semi-major\protect\\ axis $a$ (mas) & $3.57\pm0.05^1$                            & $3.48\pm0.02$   & $3.47\pm0.02$\\
Eccentricity $e$                    & $0.334\pm0.003^1$, $0.326\pm0.01^3$        & \emph{fixed ($0.334$)} & \emph{fixed ($0.326$)}\\
Periastron $T_0$ (d)                & $50120.4\pm0.4^1$                          & $50120.7\pm0.2$ & $50120.6\pm0.2$ \\
Peri. long. $\omega_{WR}$ (\degr)   & $67.4\pm0.5^1$                             & $68.3\pm0.9$    & $67.8\pm0.9$\\
PA of node $\Omega$ (\degr)         & $247.7\pm0.4^1$                            & $247.6\pm0.4$   & $247.1\pm0.3$\\
inclination $i$ (\degr)             & $65.5\pm0.4^1$                             & $65.5\pm0.7$    & $65.4\pm0.8$\\
\hline \\
M$_{\rm O}$ (\Msun)                 & $28.5\pm 1.1^1$                            & $28.4\pm 1.5$   & $28.7\pm 1.7$ \\
M$_{\rm WR}$ (\Msun)                & $9\pm 0.6^1$                               & $8.9\pm 0.5$    & $9.0\pm 0.5$ \\
\hline \\
 \end{tabular}\\
$^1$\protect\citet{2007MNRAS.377..415N}, $^2$Paper~I, $^3$\protect\citet{1997A&A...328..219S}.
\end{table*}

We independently compute the eccentricity with our two measurements close to periastron (06/01/2012) and close to apastron (07/03/2007). Let $R$ be the ratio between the two astrometric separations, the eccentricity $e$ can be calculated as $e=\frac{1-R}{1+R}$. We find $e=0.34$ with this method. This further casts doubt on the result  with $e$ and/or $P$ as free parameters. The latter may stem from underestimated or overestimated error bars on some data points (especially in 2004), and/or not taking into account measurements correlations. As such, we discarded this first solution.

\begin{figure}
  \centering
    \includegraphics[width=0.9\hsize, angle=0]{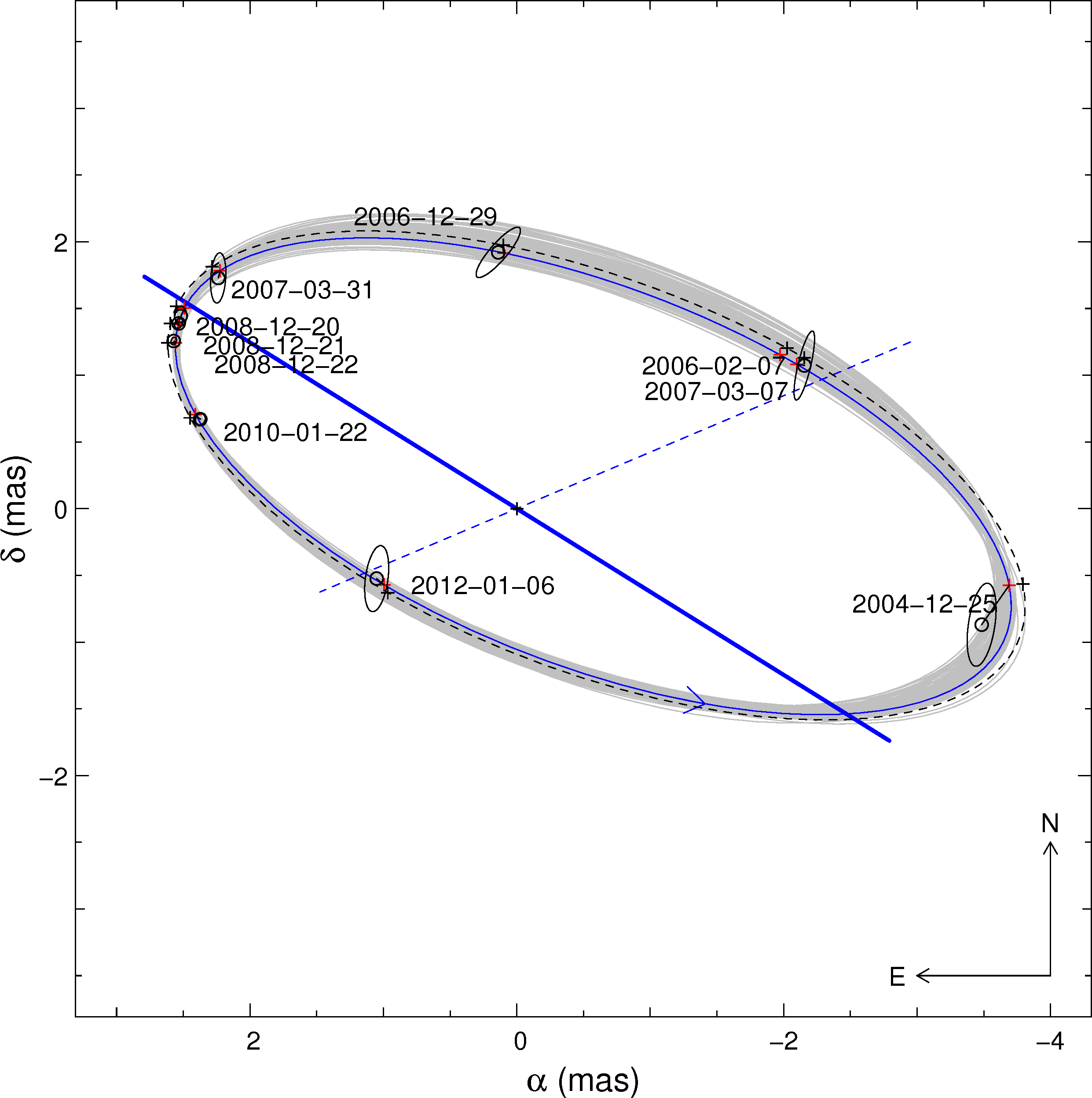}
  \caption{
    Orbital phases and positions of the WR star component relative to the O component for the different epochs related in this paper, with $e$ fixed to 0.334 and $P$ fixed to 78.53 days. The solid blue ellipse represents the new orbit, while the dotted ellipse represents the \protect\citet{2007MNRAS.377..415N} orbit. The small circles and ellipses represent the measured astrometric points, while the red crosses show the expected positions of the WR star with the new orbital solution. The blue thick line represents the line of nodes and the dashed line the semi-major axis of the system. The direction of motion is represented by an arrow. The gray ellipses show similarly significant results of the Markov chain computation.}
    \label{fig:orbitGammaVel}
\end{figure}

For the two presented solutions,  we adopted the value of $P$ to 78.53 days and $e$ to 0.326 or 0.334, as determined \citet{1997A&A...328..219S, 2007MNRAS.377..415N}, respectively. Overall, we find a solution which is very close to the one of \citet{2007MNRAS.377..415N} and which refines some error bars, especially the semi-major axis, which is more precisely defined and slightly smaller. As we use the same method as described in their paper, it is not surprising that we find a distance of $345-346\pm7$\,pc that is compatible with their distance. We also find stellar masses which are compatible with their values. These results are listed in Table ~\ref{tableParameters}. The astrometric data at the observed phases together with the best astrometric orbit of $\gamma^2$ Vel from \citet{2007MNRAS.377..415N} are plotted in 
Fig.~\ref{fig:orbitGammaVel}. 

Our distance estimate is consistent with \citet{2007MNRAS.377..415N} and  the revised  \textit{Hipparcos} measurement \citep{2007A&A...474..653V}. The current uncertainties on the distance of $\gamma^2$ Vel ($\approx 2$ per cent) are dominated by the error on the maximal radial velocities of the stars ($K_1$ and $K_2$ parameters in \citealt{1997A&A...328..219S}). A factor 10 improvement on these parameters would allow to reach sub-percent accuracy for the distance  of the system with the current astrometric dataset. Unfortunately, a 10 times better estimate of the semi-major axis would reduce the error on the distance by only 0.5 per cent. This calls for  a better understanding of the spectral variability of the emission lines of the  $\gamma^2$ Vel spectra in order to improve the radial velocities accuracy and $K_1$ and $K_2$ values. $\gamma^2$ Vel is too bright for direct parallax measurements with \textit{GAIA}, a new distance measurement will only be available with the last data release.

\subsection{Spectral Analysis}\label{sec:data_spectrum}

In Paper~I, we isolated continuum regions to determine the astrometric solution and then separated the spectral components using a radiative transfer model. Here, we can determine both simultaneously without the need for a spectroscopic model. The modelled binary system can be set in \texttt{fitOmatic} to have a free spectrum for each source.  We do not expect this to lead to a major difference in the final result compared to Paper~I.

The medium spectral resolution spectra of $\gamma^2$ Vel in 2004, 2006, 2007, 2008 and 2010 are shown in Fig.~\ref{Fig:spectraGammaVel}. As the spectra from the point sources are free parameters in our model, we can separate the spectrum from each star, the result of which is shown in Figs.~\ref{Fig:spectraGammaVelWR} and \ref{Fig:spectraGammaVelO}.  The best-fit model compared with the spectrally dispersed AMBER data is shown in figure \ref{fig:modelfit_lines}. The differential phases, combined with the spectrum and the orbital elements allow an exact separation of the spectra of the components of a multiple system \citep{1996AstL...22..348P}. The quality of this fit indicates that the spectra are successfully separated. However the persistent shift observed at all wavelength between the observed and the model squared visibility show that a binary model is insufficient. That triggered the development of the complete model described in \S \ref{sec:mock_data}.  We can nevertheless analyse the separated spectra as the datasets of 2008 and 2010 provide the full combined $K$-band and $H$-band spectra, respectively. Therefore, we identify several more lines compared to Paper~I (see Table\,\ref{tab:linelist}). In the following we describe the spectral classification and spectral variability observed in the different epochs.

\subsubsection{Spectrum of the WR star}

\begin{figure*}
\centering
\includegraphics[angle=-90,width=.92\textwidth]{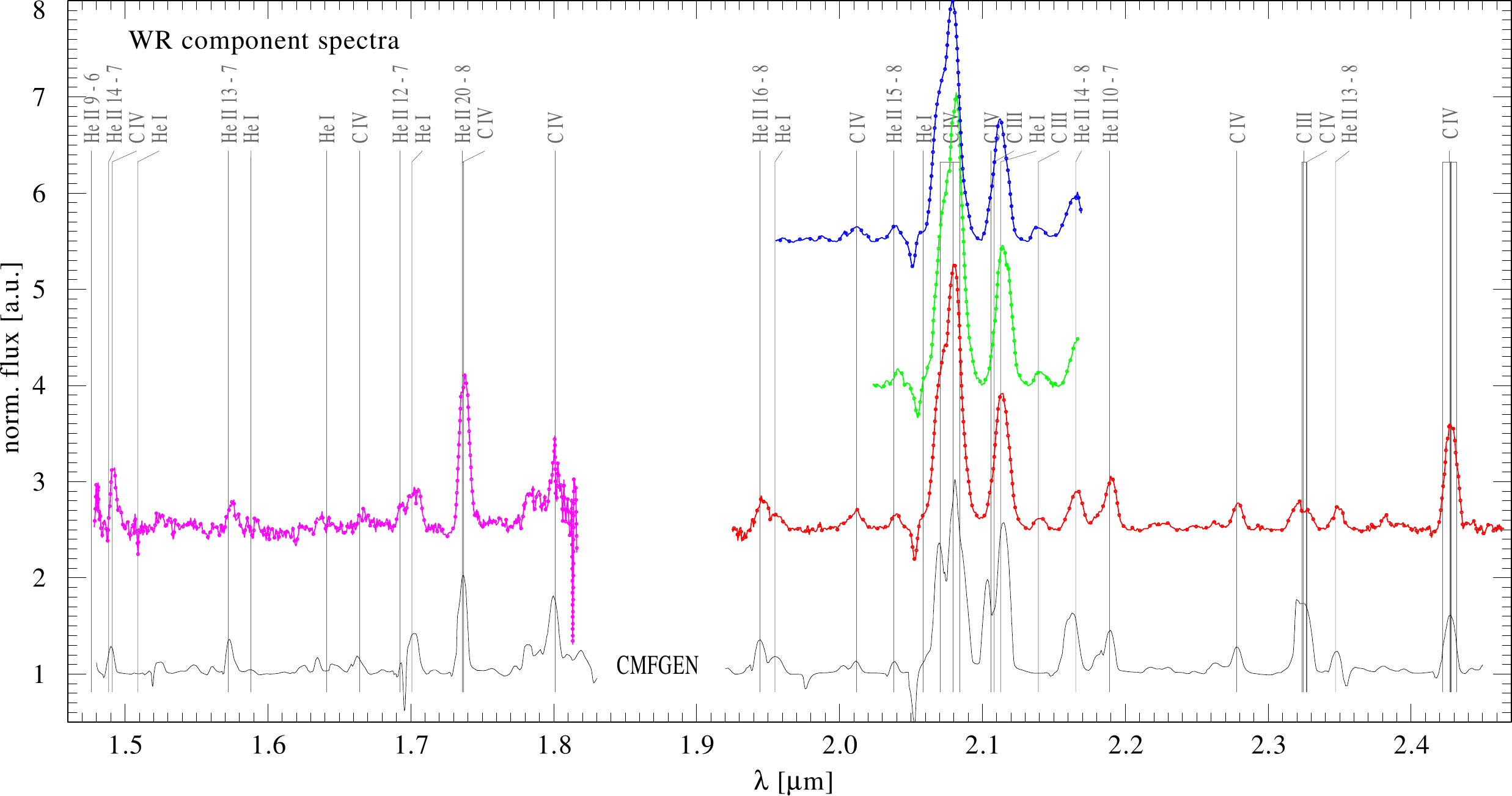}
  \caption{Normalized separated spectra of the WR component of $\gamma^2$ Vel with line identifications (see Fig.\,\ref{Fig:spectraGammaVel} for color-code of the different epoch data), shifted in flux by a constant for easier viewing.  At the bottom, we show the good match to the CMFGEN model spectrum from Paper\,I (black line).}
\label{Fig:spectraGammaVelWR}
\end{figure*}

Figure~\ref{Fig:spectraGammaVelWR} shows the full $H$- and $K$-band spectrum of the WR star alone, with line identifications also listed in Tab.~\ref{tab:linelist}. It exhibits similar features to Paper~I, i.e. very broadened emission lines of carbon and helium. The most prominent lines are the \ion{C}{iv} multiplet ($\lambda$\,2.0705-0842\,$\mu$m) and \ion{He}{ii} line series, as well as the \ion{He}{i} absorption ($\lambda$\,2.0587\,$\mu$m). Following \citet{Crowther+2006}, we use the $K$-band spectrum to measure the line ratio \ion{C}{iv 2.076}/\ion{C}{III 2.110} and confirm the spectral type of the WR star as WC\,8.

We  include the CMFGEN \citep{1998ApJ...496..407H,2000MNRAS.315..407D} model spectrum of the WR star presented in Paper~I, which was computed on a much larger wavelength range than presented at that time. It is shown for comparison in Fig.~\ref{Fig:spectraGammaVelWR}. CMFGEN solves the radiative transfer equation, in a non-local thermal equilibrium, spherically symmetric expanding atmosphere and determines the resulting spectra. The overall match with the spectrum extracted from the data is excellent at all epochs. 

\subsubsection{Spectrum of the O star}

\begin{figure*}
\centering
\includegraphics[angle=-90, width=.95\textwidth]{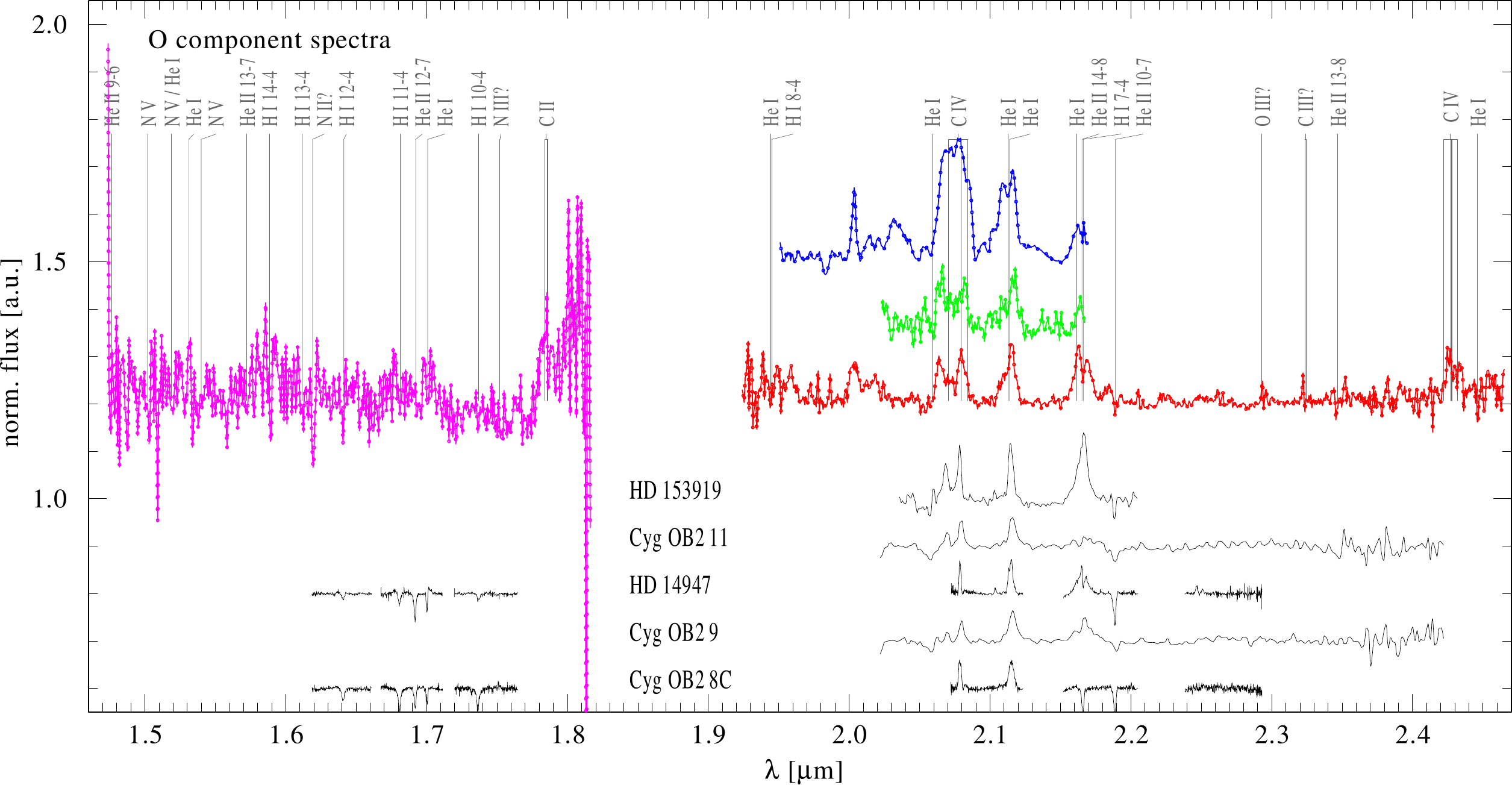} 
  \caption{
  Normalized separated spectra of the O star component of $\gamma^2$ Vel (the color code is the same as in Fig.\,\ref{Fig:spectraGammaVel}), the spectra are offset by a constant for clarity. For comparison, we include template spectra of O supergiants by \citet{1996ApJS..107..281H,2005ApJS..161..154H}. Although the latter provide the best match to the $\gamma^2$ Vel spectrum, the spectrum from $\gamma^2$ Vel is clearly more complex. }
\label{Fig:spectraGammaVelO}
\end{figure*}

The O star spectrum  is shown in Fig.\,\ref{Fig:spectraGammaVelO}. It displays a series of emission lines, which we tentatively identify (see Table\,\ref{tab:linelist}). There are hydrogen lines of the Brackett series as well as helium, carbon, and nitrogen features. We also see emission at $2.01\mu$m but the spectrum might be contaminated due to the presence of the CO$_2$ ro-vibrational atmospheric bands in this spectral region and residual \ion{C}{iv} emission from the WR component. 

As the interferometer only marginally resolves the binary star, the spectrum we refer to as "the O star spectrum" actually corresponds to an area around the star (see dashed circle around the stars on Fig.~\ref{fig:schema}). As such, it potentially contains emission from the O star, as well as the colliding wind region and some contribution of the free WR wind. 

For classification of the spectral type, we retrieve template spectra from \citet{1996ApJS..107..281H, 2005ApJS..161..154H}. These catalogs contain observed infrared spectra for a variety of stars, and especially O-type stars. In Fig.~\ref{Fig:spectraGammaVelO} we show the five closest matches to the O star component of the $\gamma^2$ Vel spectrum. We compare these spectra by calculating least squares between the templates and the observed spectrum.

Comparison with single O star spectra finds good agreement, for example with members of the Cygnus OB2 association: No.\,8C (O5\,If), No.\,9 (O5\,If+), and No.\,11 (O5\,If+) as well as and HD\,14947 (O5\,If). \citet{2013NewA...25....7D} argues that for  supergiants star like HD\,14947, the soft thermal X-ray emission is rather under-luminous  compared to regular single O stars. This is  because of the higher wind density, which also give rise to emission lines in the optical and infrared regime. Following these templates, we find a preference for an O5\,If classification of the O component. 

However, the best-matching spectrum in the near-infrared seems to be the one of HD153919, a high-mass X-ray binary with a  O6.5Iaf+ star \citep{1973ApJ...181L..43J}. By definition, high-mas X-ray binaries are composed of a massive star and a compact object (neutron star or black hole). The X-ray emission is orders of magnitude higher than for colliding stellar winds and results from the heating of the stellar material as it is accreted onto the compact object \citep[see e.g.][and references therein]{2015MNRAS.448..620J}. This fact hampers  further comparison with $\gamma^2$ Vel.  Still, the resemblance between spectra highlights the difficulty to classify the O star and supports the evidence that our "O star spectrum" is in fact a composite of the O star and additional heated material, i.e. the WCZ.

\subsubsection{Spectral variability}

\begin{figure}
\centering
    \includegraphics[width=0.48\textwidth]{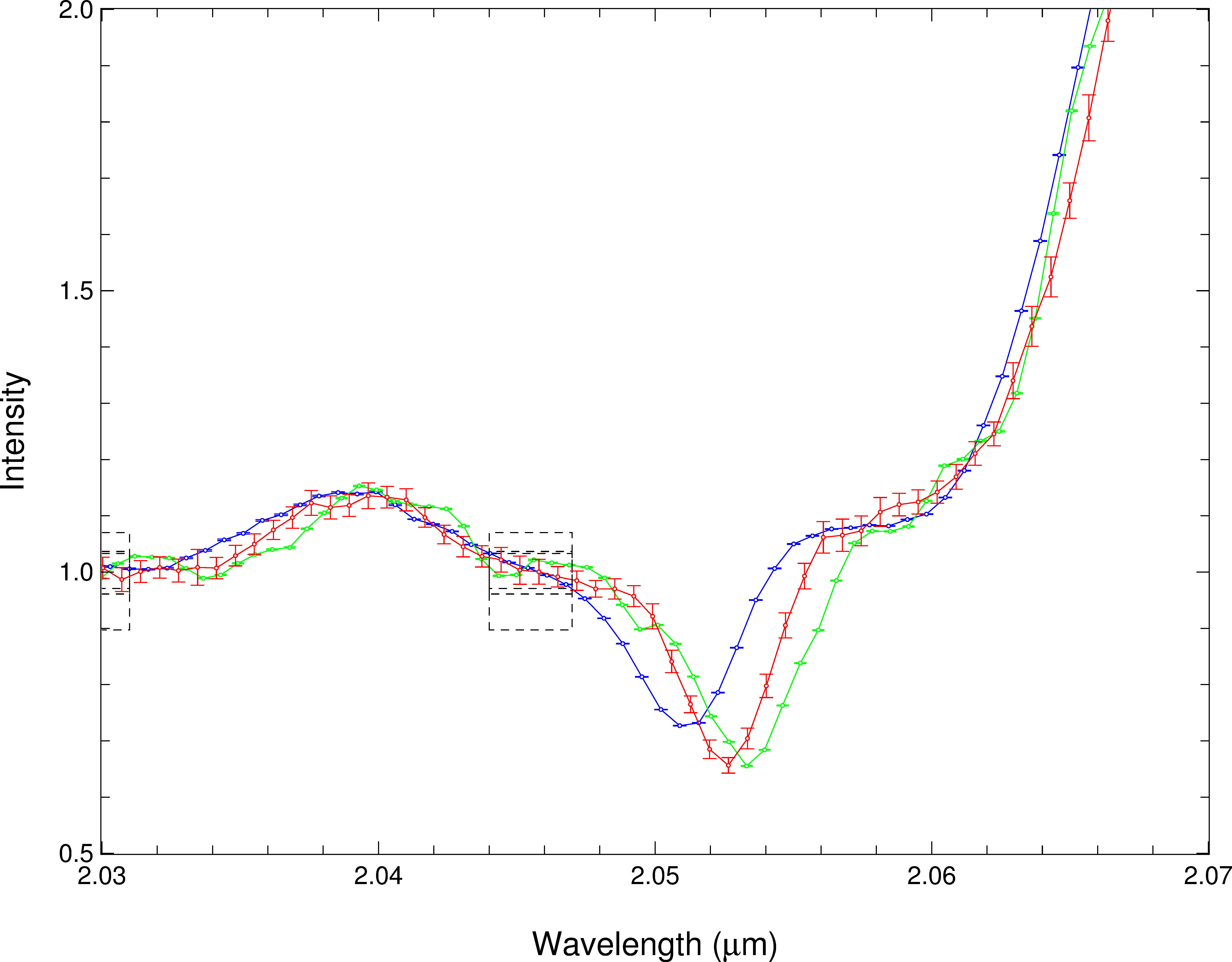}
  \caption{
    Wavelength shift and amplitude variation of the WR spectrum around the \ion{He}{ii} $\lambda$\,2.059$\mu$m absorption feature. The blue, green and red spectra correspond to 2004, 2007 and 2008, i.e. $\phi=0.36, 0.77$ and $0.83$, respectively.
    \label{Fig:zoomspectraGammaVel}
  }
\end{figure}

The spectral variability in the WR spectrum is shown in detail with the \ion{He}{i} absorption line at 2.0587\,$\mu$m in Fig.~\ref{Fig:zoomspectraGammaVel}. The variability is mainly made of a wavelength shift, which is likely due to the radial velocity of the star. We also note small amplitude variations, especially between 2004 and 2008. Variations are present between 2007 and 2008 (recorded at approximately the same orbital phase), though smaller, that may be due to the presence of clumps in the WR wind and/or instabilities in the shocked region, especially of the material along the line-of-sight to the inner part of the binary. Short term variability has already been reported in $\gamma^2$ Vel \citep{1990A&A...240..105B} and in other WR systems \citep[see e.g.][]{1997PASP..109..504L,1999MNRAS.302..549S}.

\begin{figure}
\centering
    \includegraphics[width=0.45\textwidth]{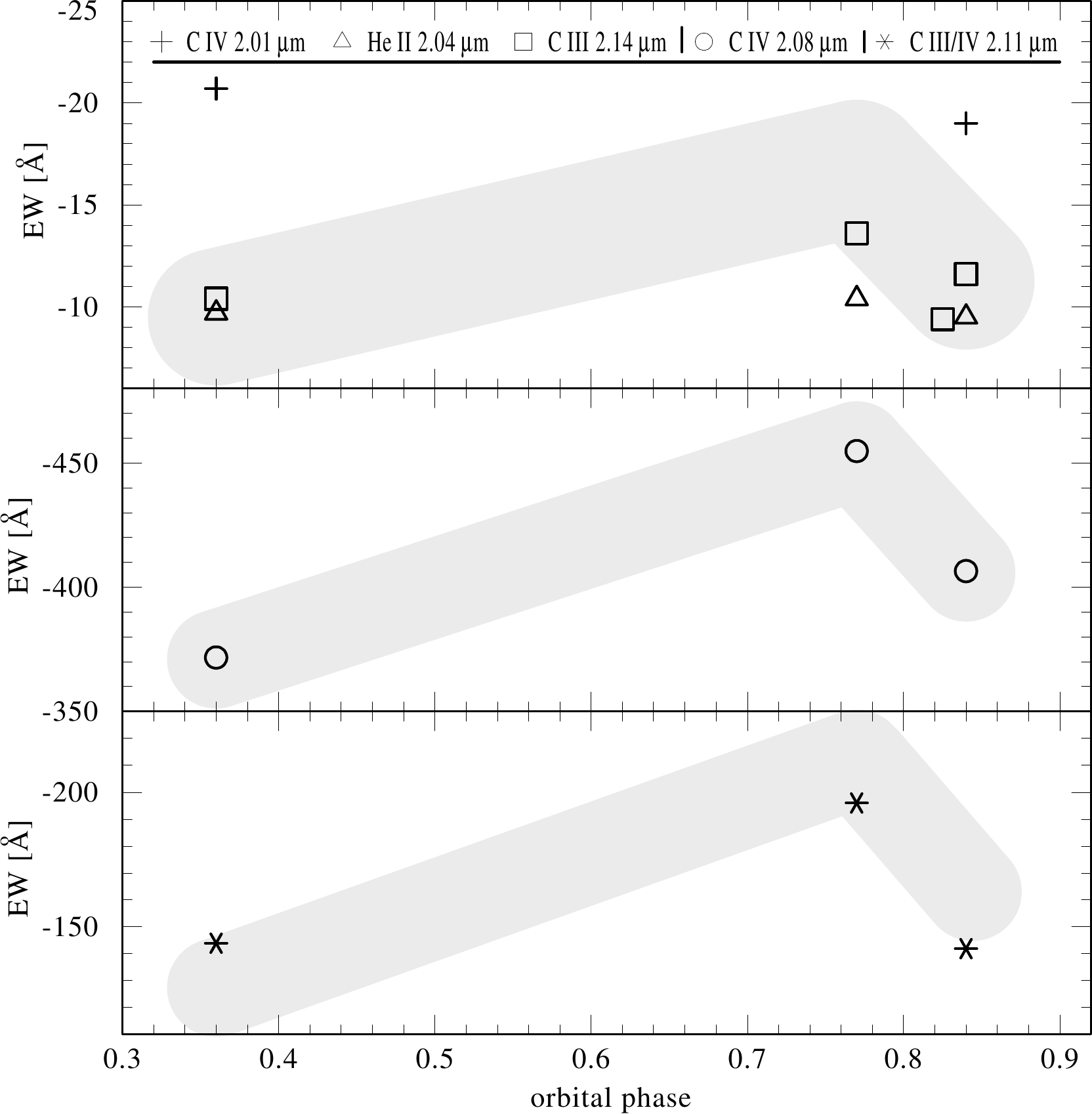}
  \caption{
  Measured equivalent widths of emission lines in the WR $K$-band spectra over orbital phase. A clear trend is visible with stronger emission approaching periastron passage.}
    \label{Fig:EW-orbitalphase}
\end{figure}

In addition, we measure equivalent widths (EWs) for the prominent emission lines in the WR $K$-band spectrum over the available data of all epochs. Fig.~\ref{Fig:EW-orbitalphase} shows the measured EWs, clearly changing with the orbital phase. \citet{Varricatt+2004} suggest that such trend can be explained by less dust emission diluting the spectra near periastron passage. However, we do not have a satisfactory explanation for this observation, as $\gamma^2$ Vel is not a dust-emitting WR star, thus eliminating the impact of dust on the spectral lines \citep{2002ASPC..260..331M}.

The O star spectrum also shows temporal variability, which seems to be strongly dependent on the orbital phase. Indeed, the 2007 and 2008 spectra (roughly at the same orbital phase) look almost identical, whereas the 2004 spectrum (taken at opposite phase) exhibits striking differences: for example, the C{\sc iv} lines exhibit a flat-top profile in 2004, while it looks asymmetric and double-peaked in the 2007 and 2008 spectra.

The line profiles and variations in relation to the orbital motion are discussed by \citet{2009MNRAS.395..962I} and \citet{2003MNRAS.346..773H}. They respectively focus on forbidden emission lines and X-ray lines forming in the WCZ, and clearly show line variations from flat-topped to double peak profiles.  Since the O star spectrum in our case most likely also contains contributions from the WCZ and maybe even the WR wind, we might expect to see spectral variability as described above along the orbit.  

In the next sections, we perform hydrodynamic simulations of the system and extract corresponding model data.  Thanks to the spatially resolved interferometric data, we focus on the detection of the shocked region around the O star. While we attribute the residuals in the O star spectrum as coming from the WCZ, detailed modeling of the line profiles in order to constrain wind parameters such as the opening an angle as suggested by \citet{2009MNRAS.395..962I} is beyond the scope of this paper. Focussing on the continuum emission from a large field of view, we determine the flux ratio of the WCZ.

\section{Hydrodynamic model}\label{sec:model}
\subsection{Numerics}

We use the hydrodynamics code RAMSES for our simulations \citep{2002A&A...385..337T}. This code uses a second-order Godunov method to solve the hydrodynamics equations
\begin{eqnarray}\label{eq:hydro}
    \frac{\partial\rho}{\partial t}+\nabla \cdot(\rho \mathbf{v})&=&0\\  \nonumber
	\frac{\partial(\rho \mathbf{v})}{\partial t}+\nabla (\rho \mathbf{v}\mathbf{v})+\nabla P 	&=& 0	\\
	  \frac{\partial E }{\partial t}+\nabla \cdot[\mathbf{v}(E+P)]	&=& n^2 \Lambda(T),	\nonumber
\end{eqnarray}

where $\rho$ is the density, $n$ the number density, $\mathbf{v}$ the velocity, and $P$ the pressure of the gas. The total energy density $E$ is given by

\begin{equation}
E= \frac{1}{2}\rho v^2+\frac{P}{(\gamma -1)},
\end{equation}

where $\gamma$ is the adiabatic index, set to 5/3 to model adiabatic flows. We include a passive scalar to distinguish both winds. $\Lambda$ is the radiative cooling rate of the gas, based on \citet{1993ApJS...88..253S}. This cooling curve assumes an optically thin gas in ionisation equilibrium. We assume solar abundances for both winds, which underestimates the cooling in the WR wind for temperatures below 10$^7$K \citep{Stevens:1992on} by a factor of a few.

We use the MinMod slope limiter together with the HLLC Riemann solver to avoid numerical quenching of instabilities. We perform 3D simulations on a Cartesian grid with outflow boundary conditions.   The size of the simulation domain is 16 times the binary semi-major axis, or about 60 mas, largely covering the emitting region of the system. The resolution is uniform set to  $N_x=1024$. This resolution is slightly lower than in other numerical work \citep{2009MNRAS.396.1743P,2012A&A...546A..60L} but is sufficient to capture the dynamics of the interaction region and allow comparison with observations.  Our physical resolution is about $2\times 10^{11}$ cm, which should yield limited numerical heat conduction at the interface between the winds \citep{2010MNRAS.406.2373P}. To check the impact of resolution, we perform a simulation of the system covering 8 times the binary separation with $N_x=1024$ (twice the effective resolution of our reference simulation) and find that differences in the emission maps are below 5 per cent, confirming that our resolution is sufficient.

To simulate the winds, we keep the same method as used in \citet{2011MNRAS.418.2618L}, which was largely inspired from \citet{Lemaster:2007sl}. Around each star, we create a wind by imposing a given density, pressure, and velocity profile in a spherical zone centred on the stars, of radius $l_O$ and $l_{WR}$ for each star. These zones are reset to their initial values at each time step to create steady winds. To allow for spherical symmetry, we set the $l_{WR}$ to 8 computational cells (in radius). The same large value for $l_O$ could impact the shocked structure, which is very close to the O star. Therefore, we set $l_O$ to 4 computational cells, which does not impact the formation of the WCZ. As a trade-off, the spherical symmetry of the wind is less good than for the WR star, but this does not strongly impact the large scale structure of the O star wind, which is set by the binary interactions (see Fig.~\ref{fig:hydro}). The orbital motion of the stars is determined using a leapfrog method.  As the orbital velocity is negligible with respect to the speed of the winds, each wind can be considered isotropic in the frame corotating with the corresponding star.   

\begin{table}
\caption[]{Wind parameters used for the hydrodynamic simulation. Data taken from a) \citet{1993ApJ...415..298S}, b) \citet{1999A&A...345..163D}, c) \citet{2007MNRAS.377..415N}.}
\label{tab:simu}
\begin{center}
\begin{tabular}{c c c}
\hline
\noalign{\smallskip}
 & WR & O \\
\noalign{\smallskip}
\hline
 \noalign{\smallskip}
 $v_{\infty}$ (km s$^{-1}$)  & 1550$\pm$ 150 (a) & 2500 $\pm$ 250 (b)\\
$\dot{M}$ ($M_{\odot}$ yr$^{-1}$) &  $8\pm 4 \times 10^{-6}$ (c)& $1.8\pm 0.36 \times10^{-7}$ (b) \\
\noalign{\smallskip}
\hline
\end{tabular}
\end{center}
\end{table}

Table\,\ref{tab:simu} shows the wind parameters for our simulation, found in the literature and compatible with our CFMGEN model for the WR star. We use the orbital parameters derived in section \S\ref{sec:data_model}. The temperature in the winds are set to $10^5$K, which is much higher than expected in stellar winds. However, given the speed and density in the winds, this value guarantees that the winds are highly supersonic, in which case the shocked region is unaffected by the value of the pre-shock temperature. For distances beyond $l_{WR}$ and $l_O$ the temperature evolves according to the equations of hydrodynamics and the cooling curve (Eq.~\ref{eq:hydro}). The momentum flux ratio of the winds, which is the main parameter determining the position and opening angle of the shocked region is given by \citep{1990FlDy...25..629L}
\begin{equation}\label{eq:eta}
\eta\equiv \frac{\dot{M}_{WR}\,v_{WR}}{\dot{M}_O\,v_O} \simeq 26.
\end{equation}	
This value is compatible with the $\eta=33$ from \citet{2000A&A...358..187D}, with the main difference coming from the mass-loss rate of the WR star, which depends on the clumping factor \citep{2004A&A...422..177S}.  Considering the uncertainties on the stellar wind parameters (see Tab.~\ref{tab:simu}), $\eta_{min}$ can be as low as 9 and  $\eta_{max}$ as high as 64. When instabilities and radiative effects are dynamically negligible, the intersection between the semi-major axis and the contact discontinuity between the winds is given by 

\begin{equation}
  \label{eq:standoff}
  r=a\frac{\sqrt{\eta}}{1+\sqrt{\eta}},
\end{equation}

where $a$ is the binary separation. Especially for the higher values of $\eta$, the wind collision is close to the O star, where the WR wind has reached its terminal velocity. At such close distance, the wind from the O star has not reached its terminal velocity yet.  Assuming a velocity law  $v_O=v_{O\infty}(1-R_O/r)^{\beta}$ with $\beta\simeq 0.8$ \citep{1996A&A...305..171P}, the O star wind can be as low as  $v_O\simeq 2070$ km s$^{-1}$ at the wind collision at periastron, which can yield an effective $\eta_{max} \lessapprox 70$.

However, this close to the O star, radiative breaking by radiation pressure of its photons on the WR wind could strongly slow the latter down. This locally  decreases the momentum flux ratio of the winds \citep{1997ApJ...475..786G}. According to the orbital and wind parameters of $\gamma^2$ Vel, radiative breaking is necessary  to allow for a wind collision region, otherwise the shocked region would crash onto the stellar surface. Based on models of X-ray spectroscopy,  \citet{2005MNRAS.356.1308H} suggest a shock opening half angle of $\simeq 85^{\circ}$ close to the apex of the shock.  While the value of the opening angle is still debated \citep[see e.g.][]{2000MNRAS.316..129R}, radiative breaking will largely increase $\eta_{min}$. As our simulations do not include radiative breaking, we set the winds of both stars to their terminal velocity. 

We have performed two simulations with $\eta=26$ and 36  and find that the difference in the hydrodynamic structure cannot be distinguished when comparing our model data with observations. As such, we use the $\eta=26$ results for the remainder of this paper.  While smaller or larger values of $\eta$ are plausible, the values we tested represented the most realistic range given the current data. Our simulation does not model clumps in the winds, which are likely destroyed in the WCZ \citep{2007ApJ...660L.141P}.

\subsection{Hydrodynamic structure}
\begin{figure*}
   \centering
 \includegraphics[width = .3\textwidth]{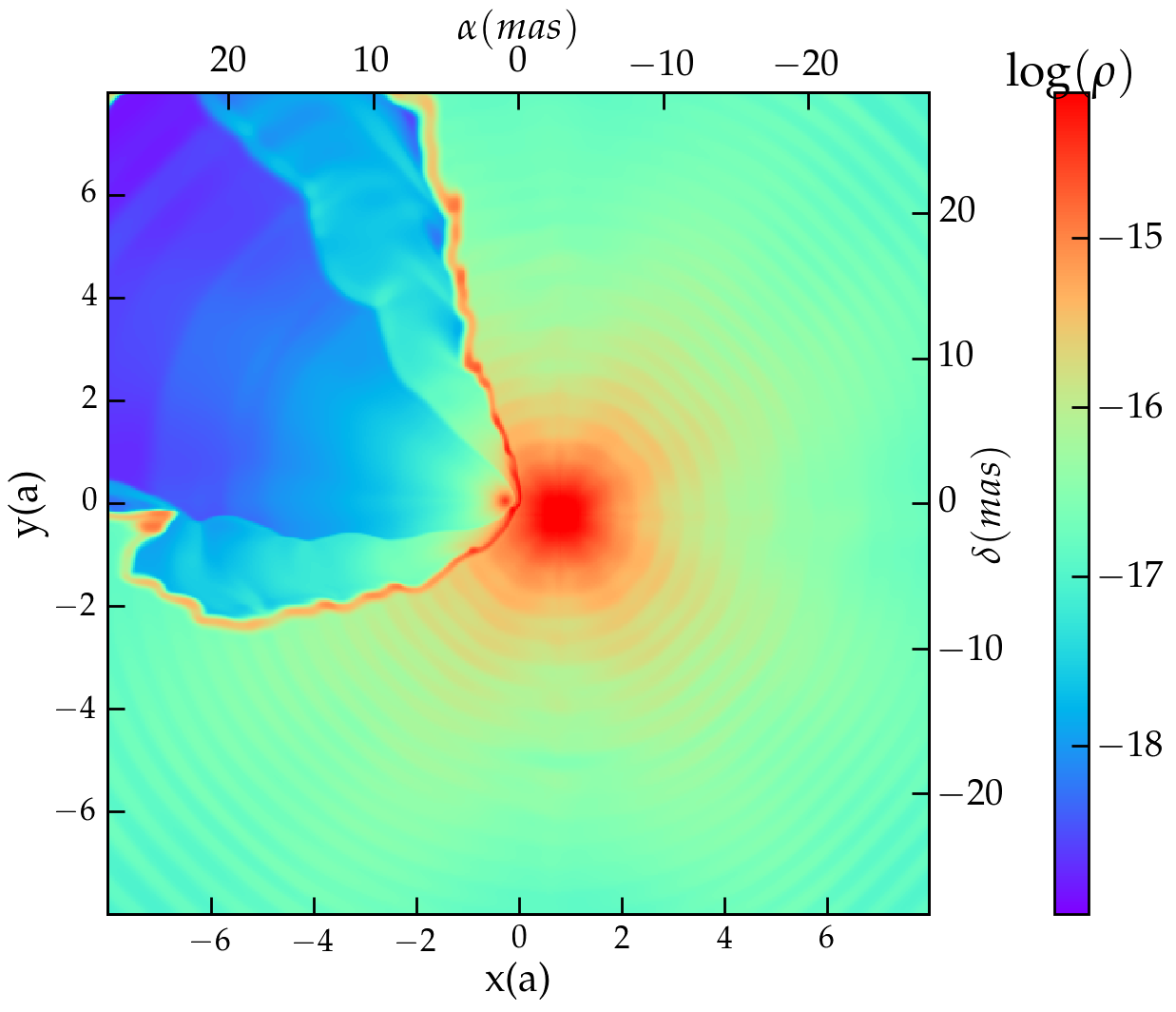}
 \includegraphics[width = .3\textwidth]{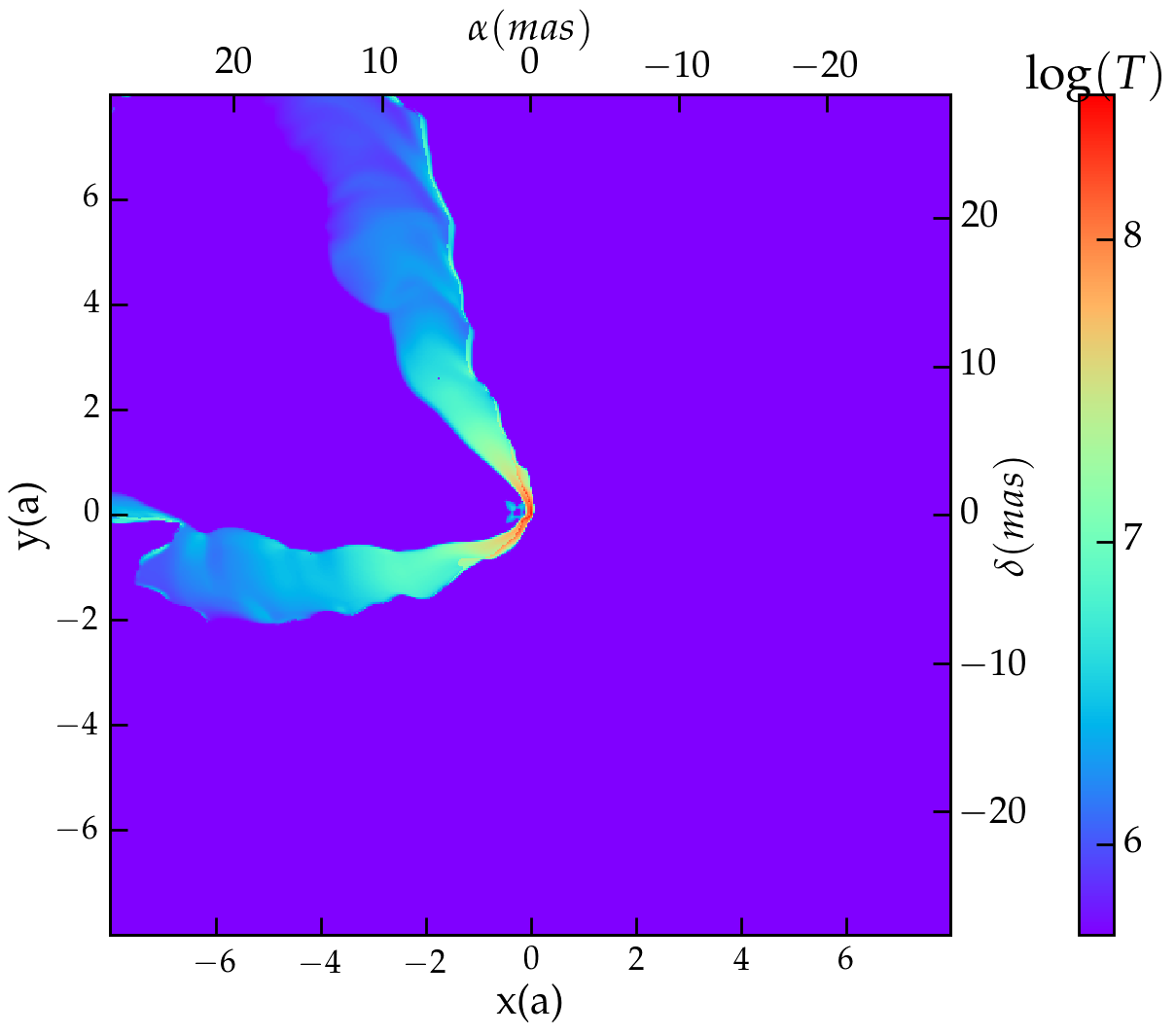}
 \includegraphics[width = .3\textwidth]{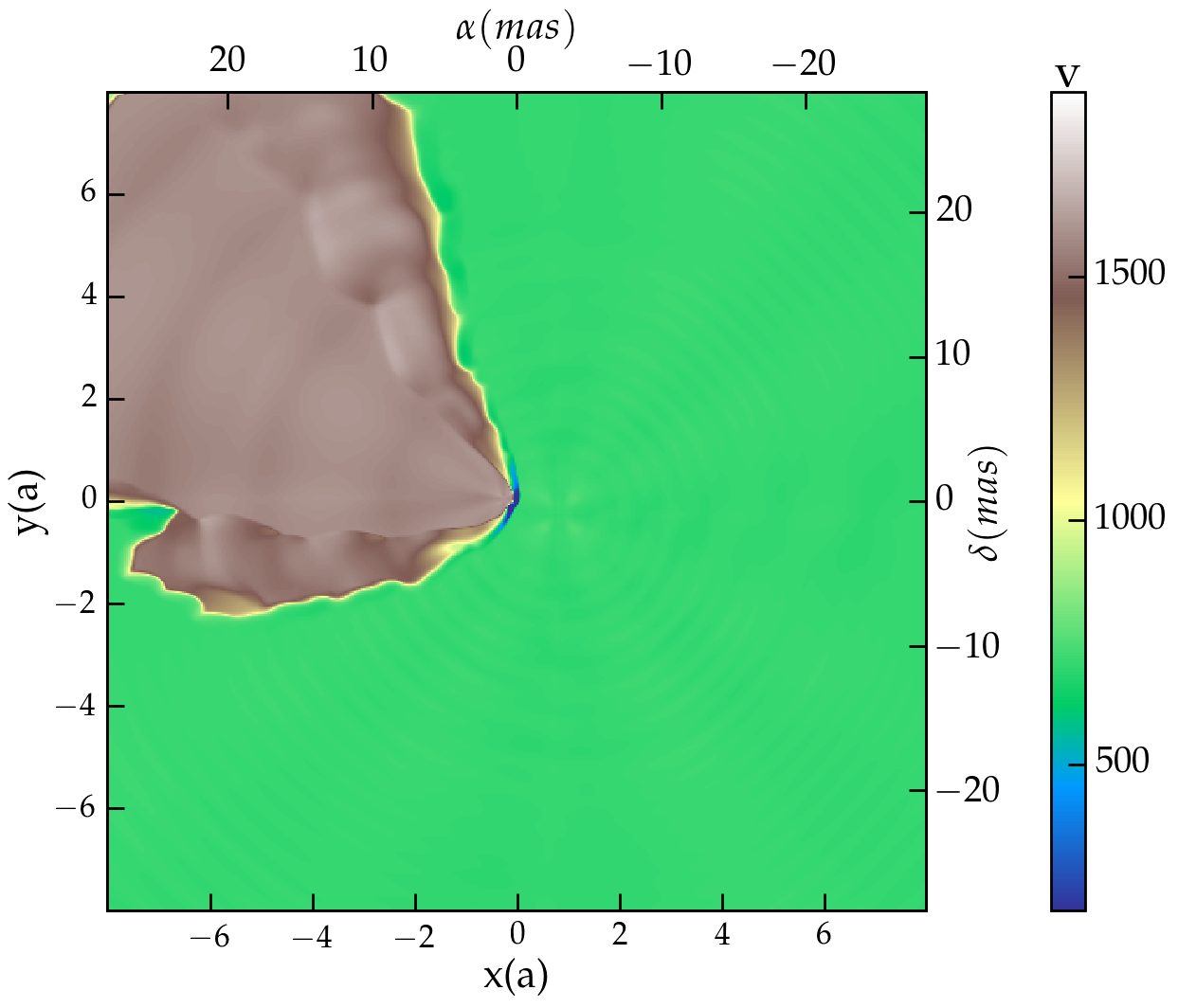}
  \caption{Density (g cm$^{-3}$), temperature (K) and  velocity (km s$^{-1}$) in the midplane of the binary for $\phi=0.3$. The WR star is located in the lower right corner and the O star in the upper left. The length scale is the semi major axis and the plot is centered on the center of mass of the binary. The colliding wind region is wrapped around the O star. }
  \label{fig:hydro}
 \end{figure*}

The density, velocity and temperature maps are shown in Fig.\,\ref{fig:hydro} for orbital phase $\phi=0.3$, i.\ e.\ when the binary is roughly in the plane of the sky, which is close to the 2004 observation. The results of these maps are compatible with the maps presented in \citet{2005MNRAS.356.1308H}, which focussed on the inner region of the binary. The density map shows both shocks (also seen on the velocity and temperature maps) and the contact discontinuity at the interface between both shocked winds.  Due to the lower density, the O star wind is less subject to radiative cooling and has a temperature of more than $10^8$ K at the apex of the shock. The WR wind on the other hand rapidly cools down to below $10^6$ K beyond the apex of the shocked region. The temperature in our model agrees well with \textit{Chandra} measurements of triplet-lines, associated with $4\times 10^6$ K gas at roughly twice the binary separation \citep{2001ApJ...558L.113S}. While both winds are highly supersonic before the shocks, their velocity is almost zero in the shocked region close to the apex. Further out in the shocked region, the winds progressively reaccelerate.

\begin{figure*}
   \centering
  \includegraphics[height=.23\textwidth]{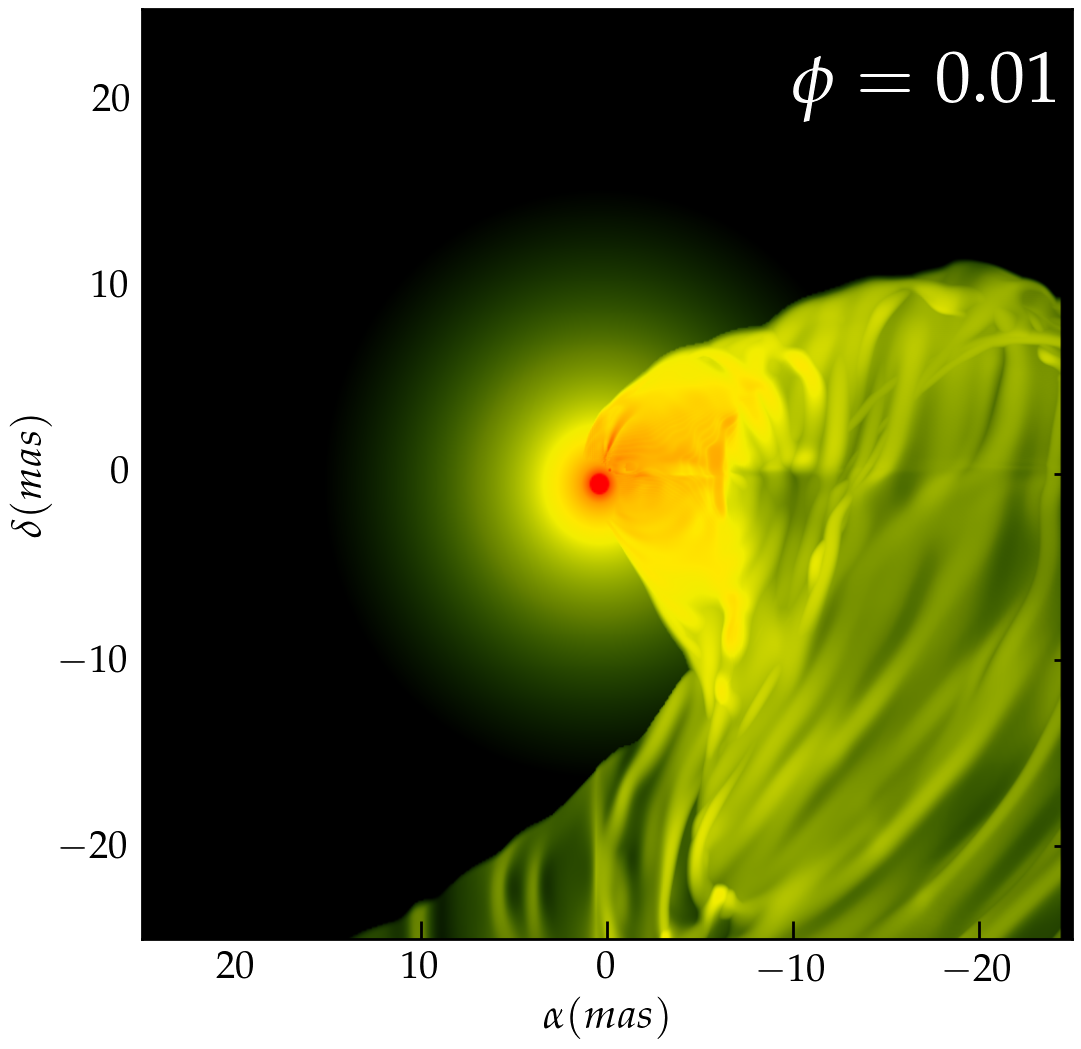}
   \includegraphics[height = .23\textwidth]{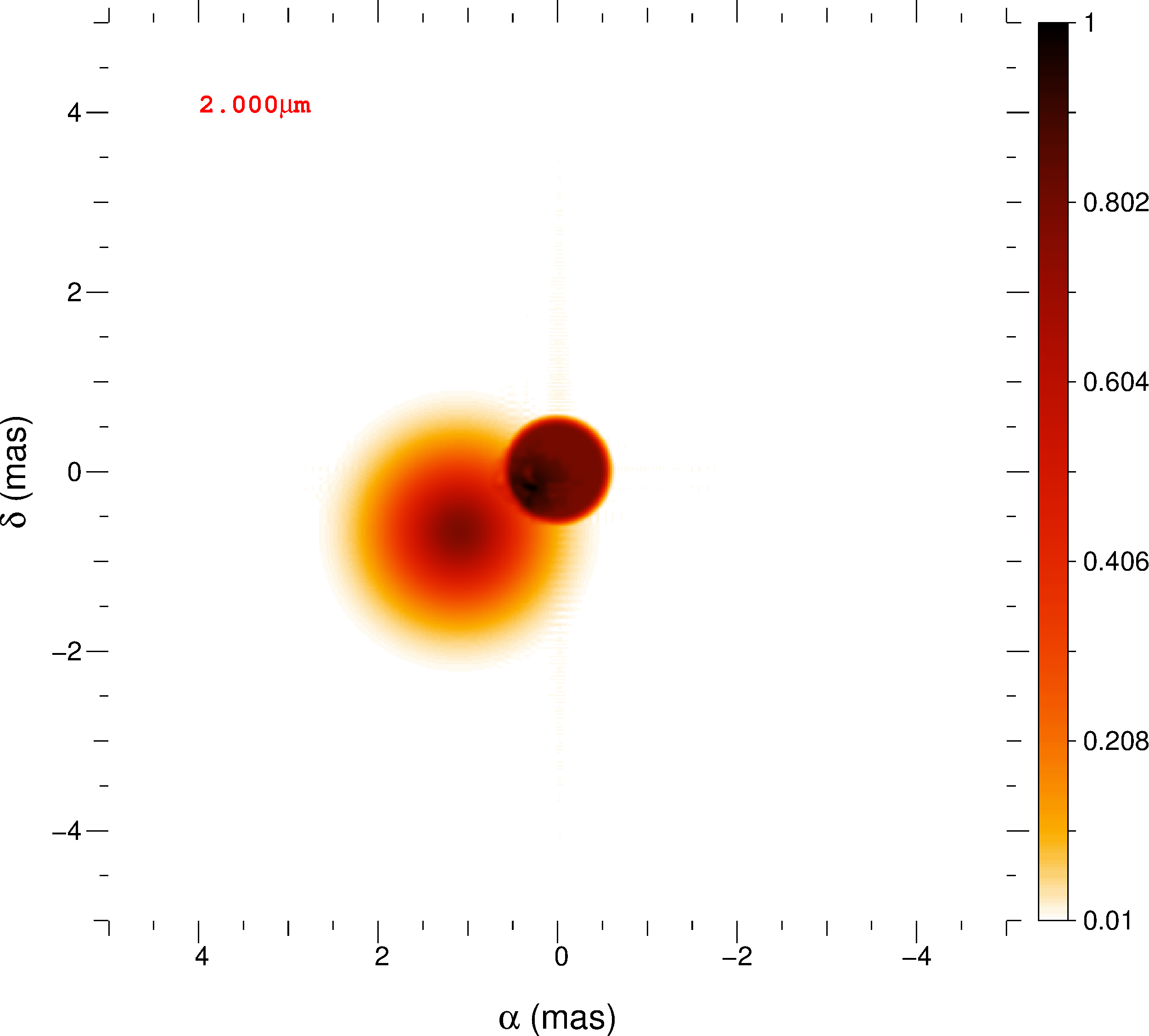}
     \includegraphics[height=.23\textwidth]{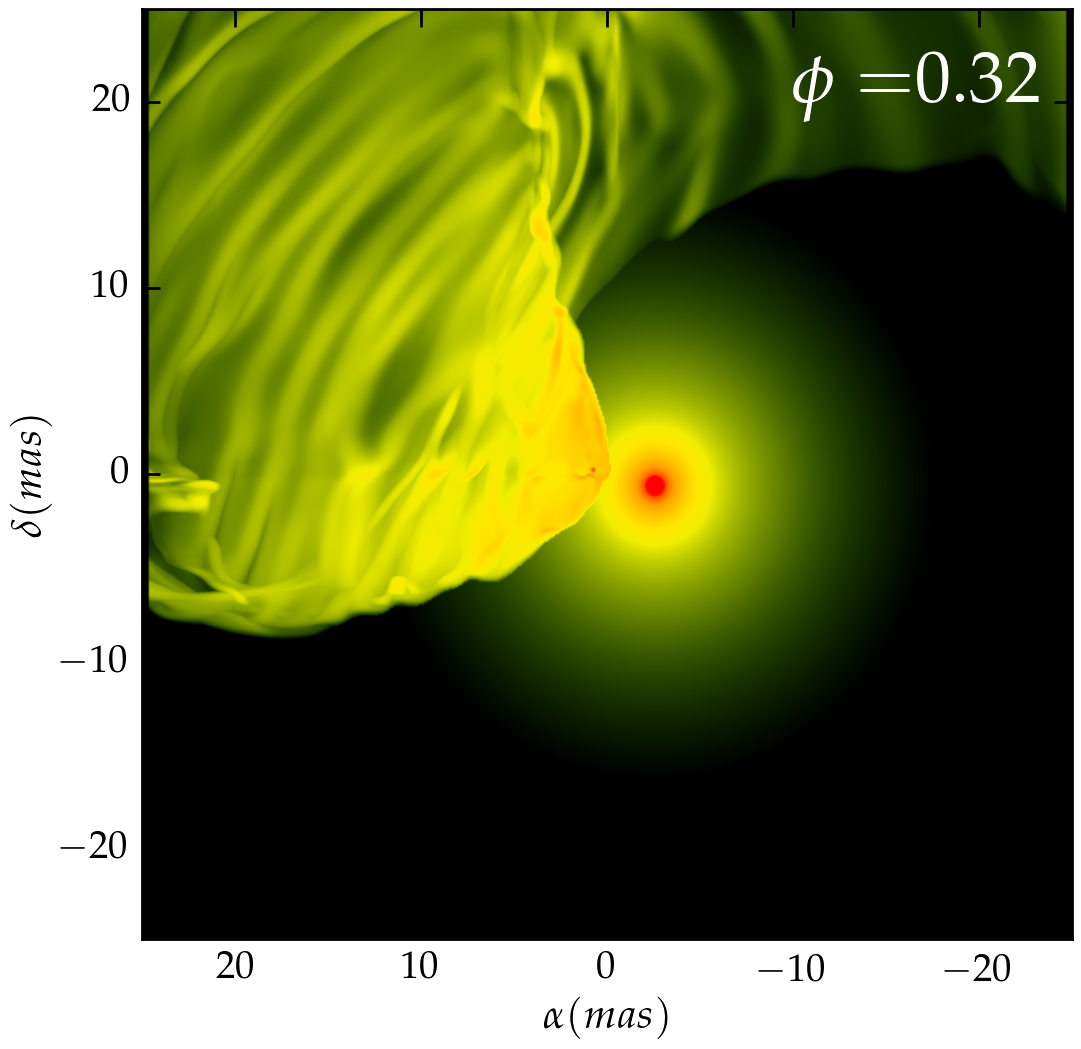}
 \includegraphics[height = .23\textwidth]{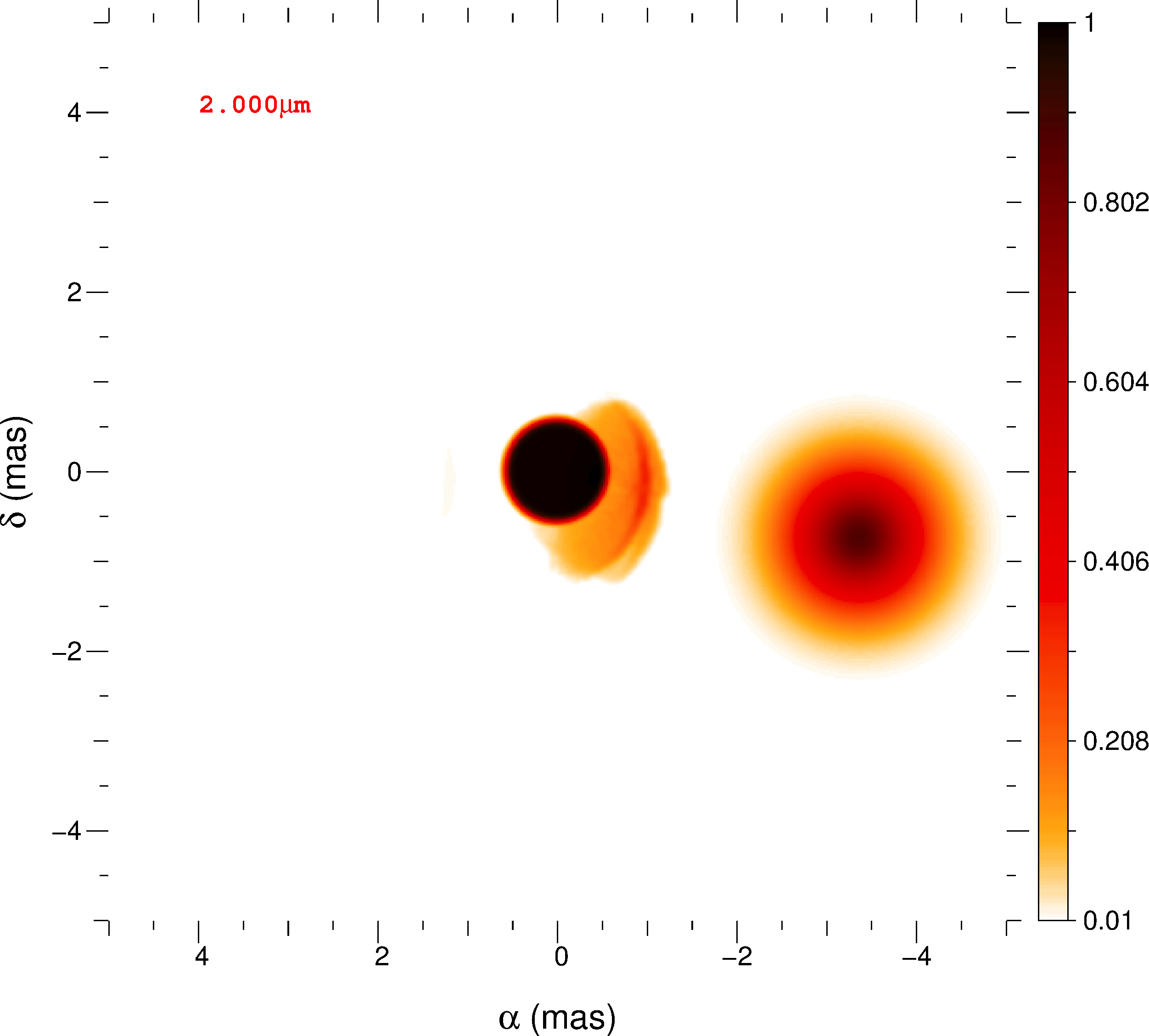}
 \includegraphics[height=.23\textwidth]{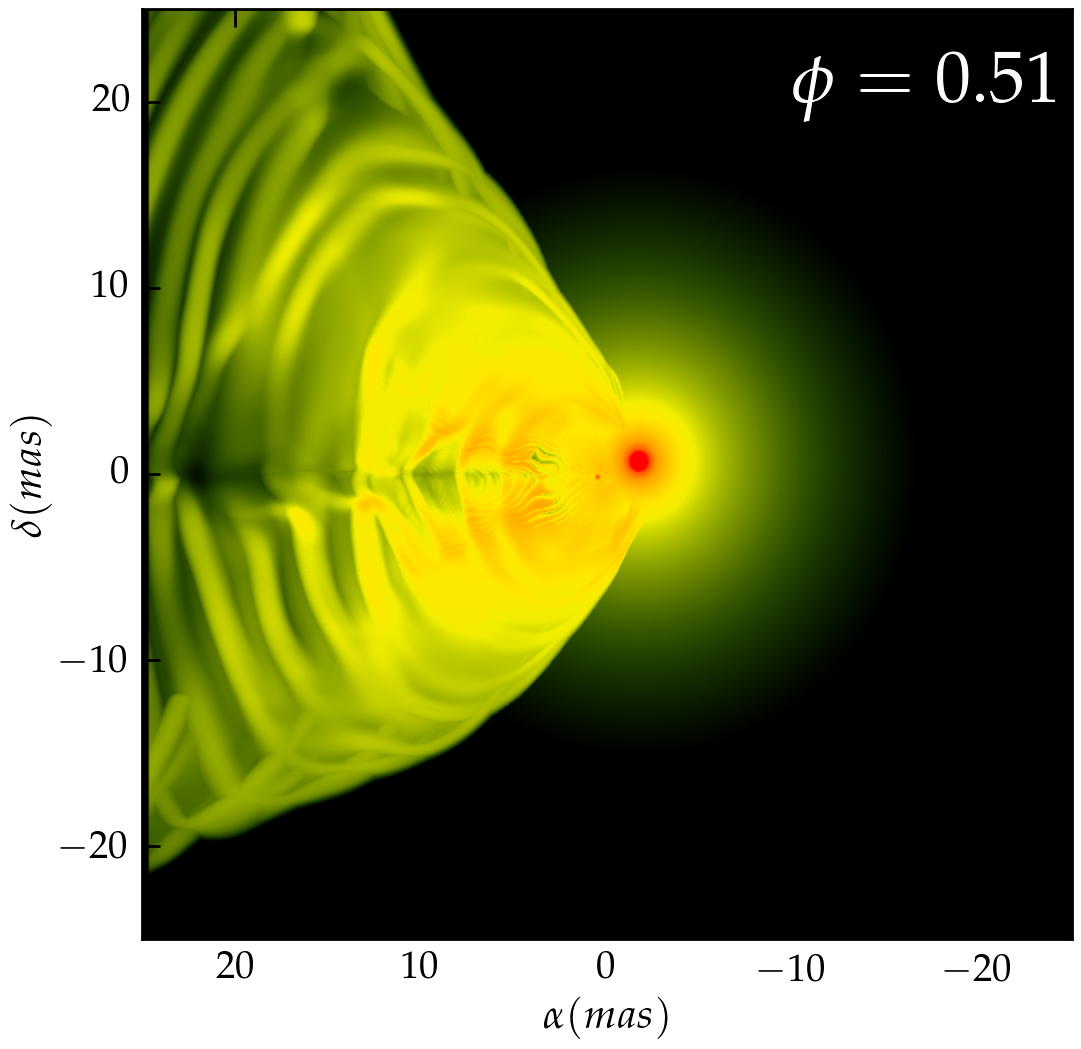}
 \includegraphics[height = .23\textwidth]{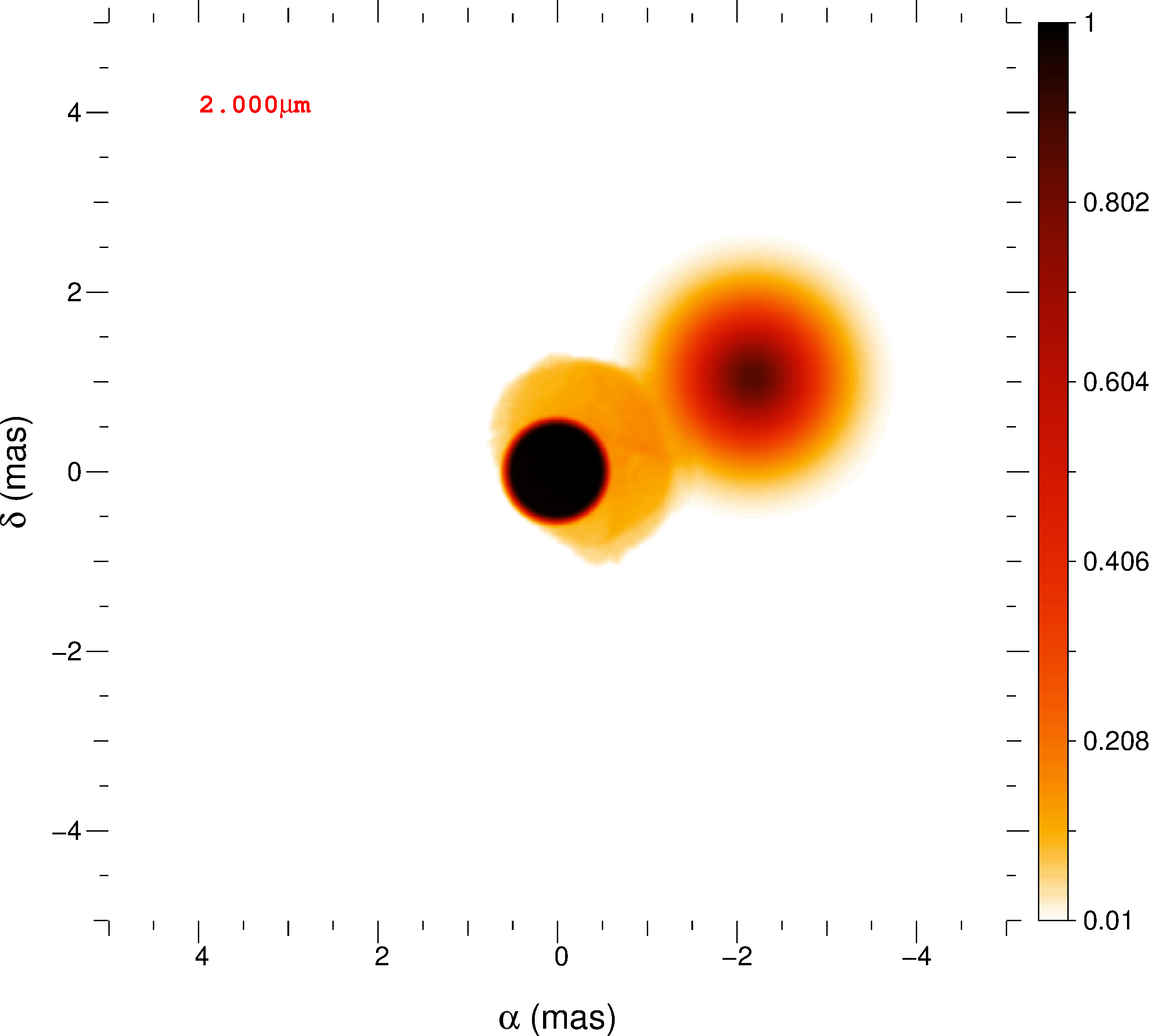}
   \includegraphics[height=.23\textwidth]{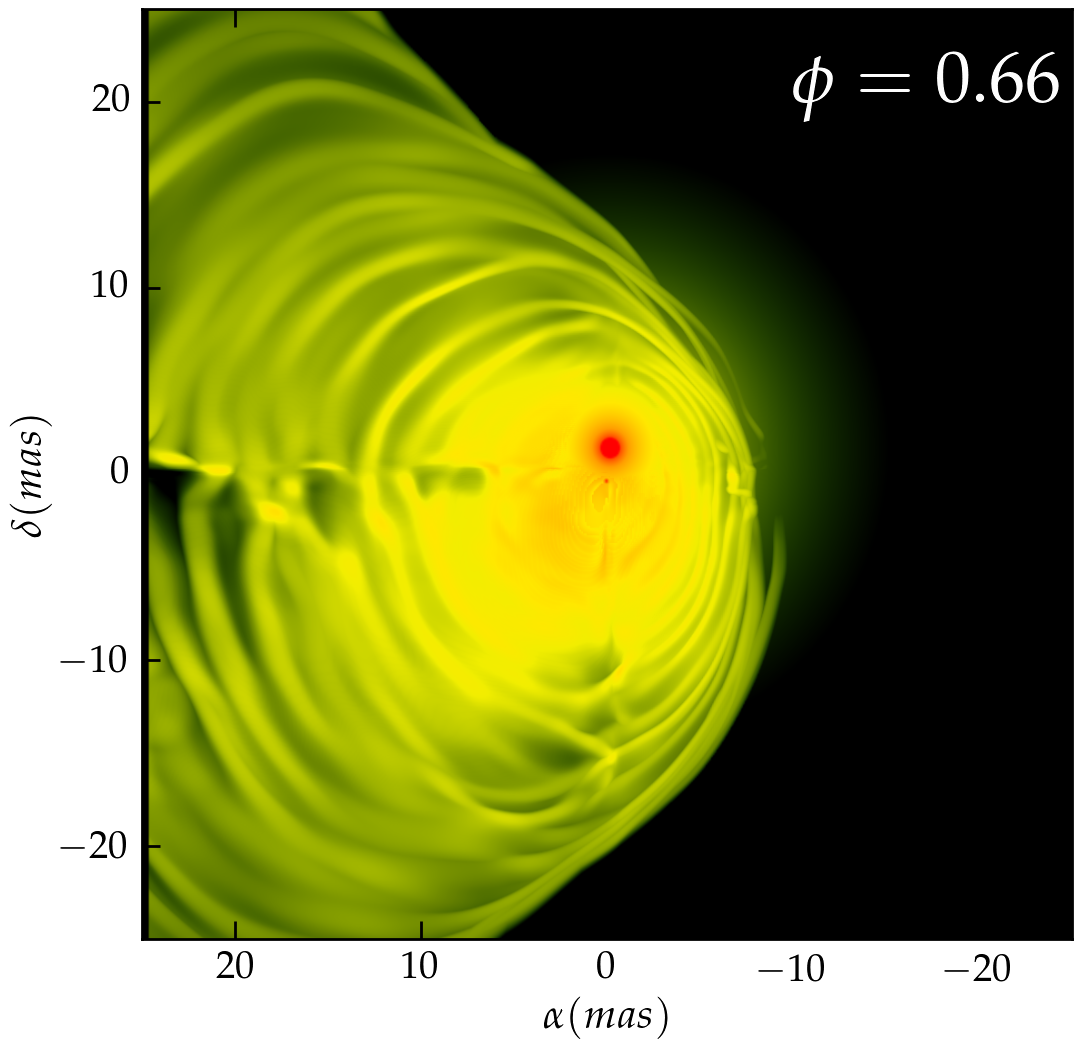}
 \includegraphics[height = .23\textwidth]{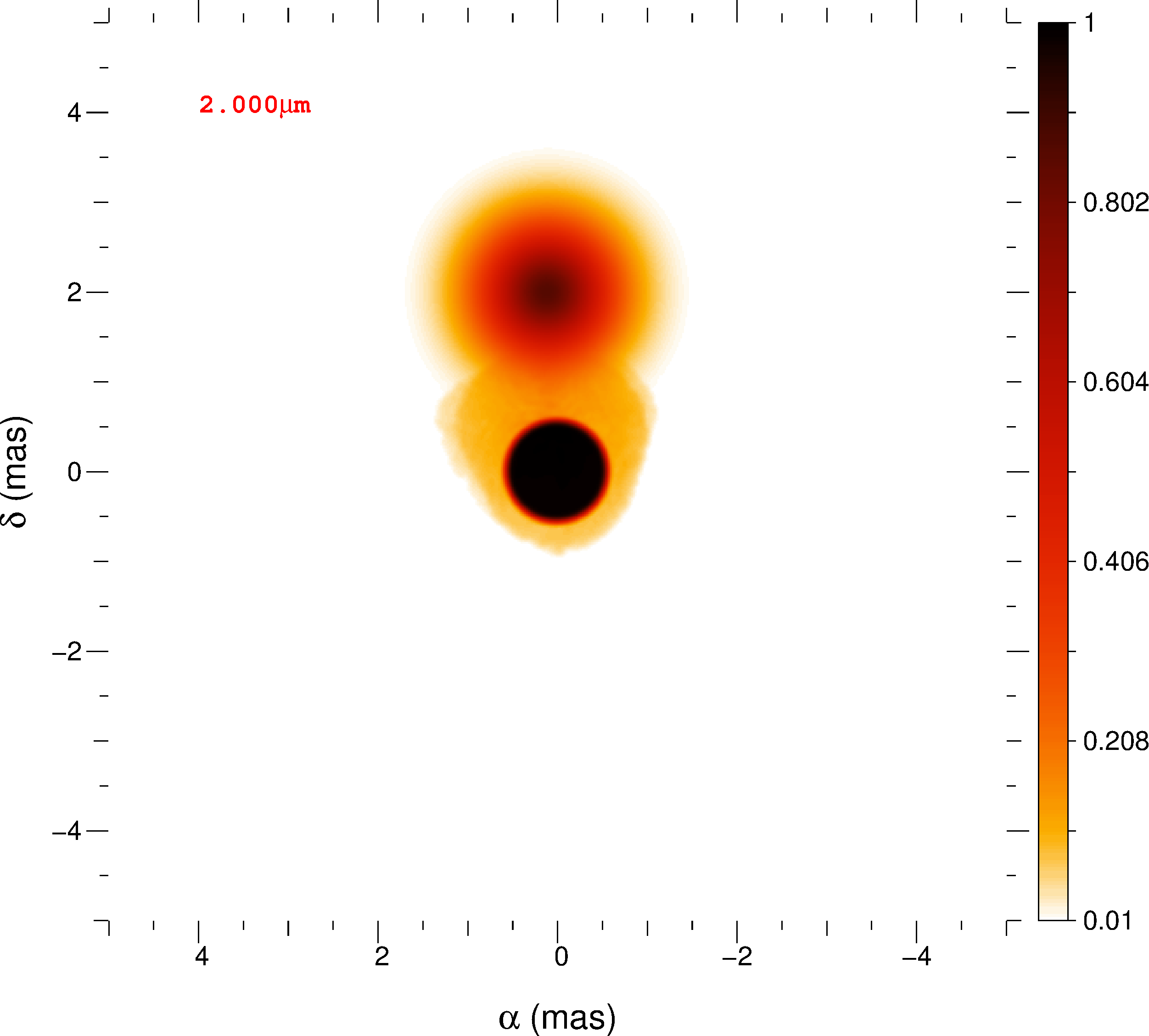}
   \includegraphics[height=.23\textwidth]{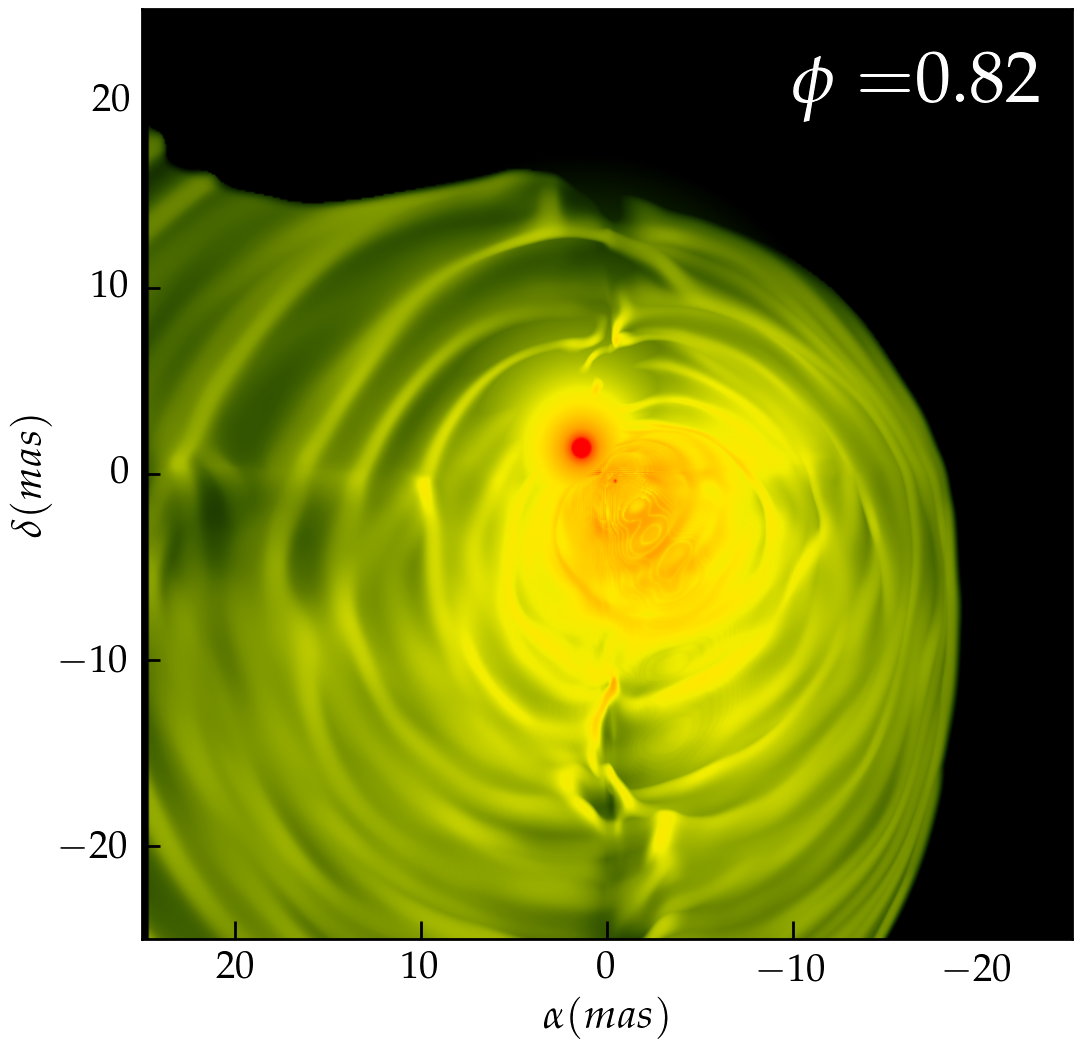}
 \includegraphics[height = .23\textwidth]{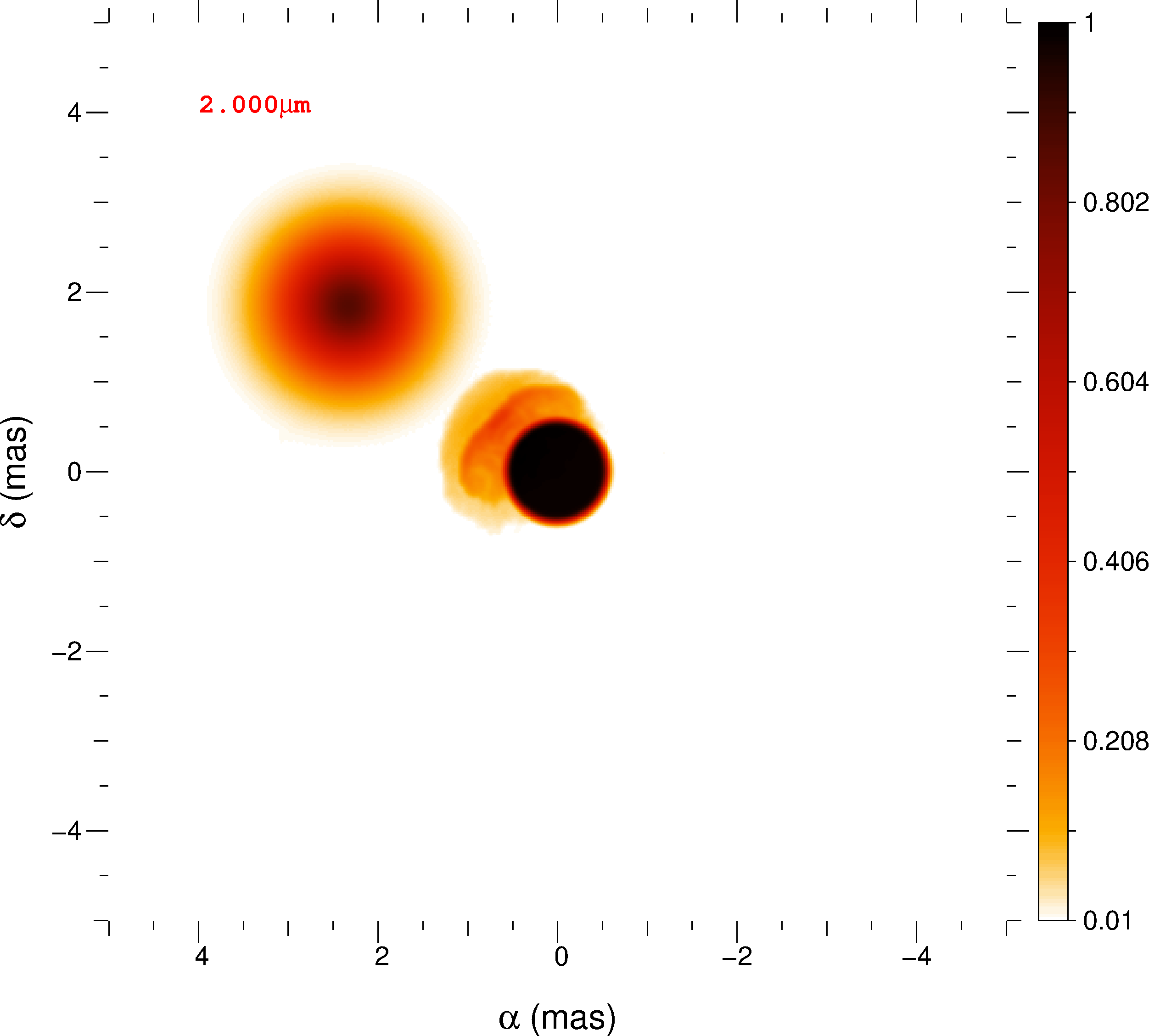}
 \includegraphics[height=.23\textwidth]{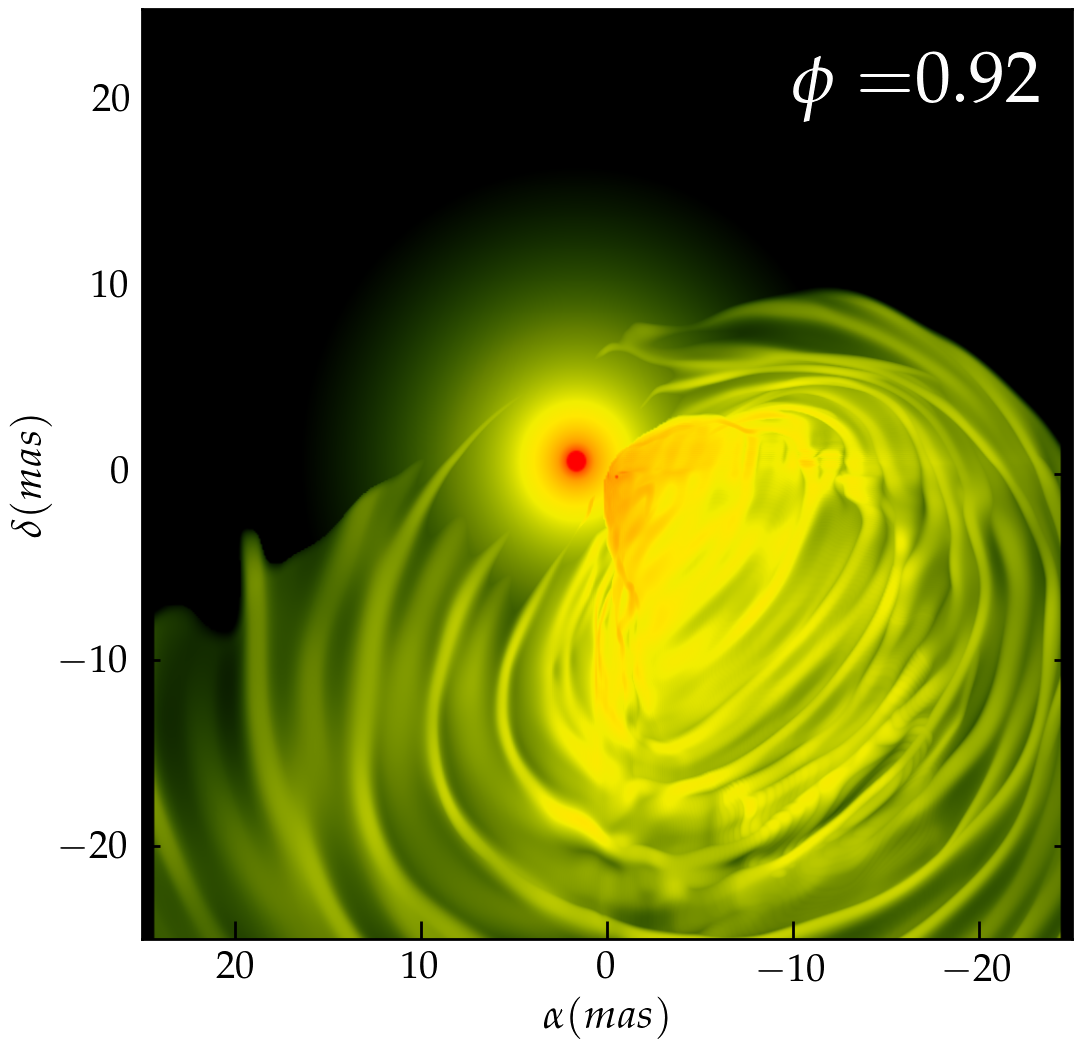}
 \includegraphics[height = .23\textwidth]{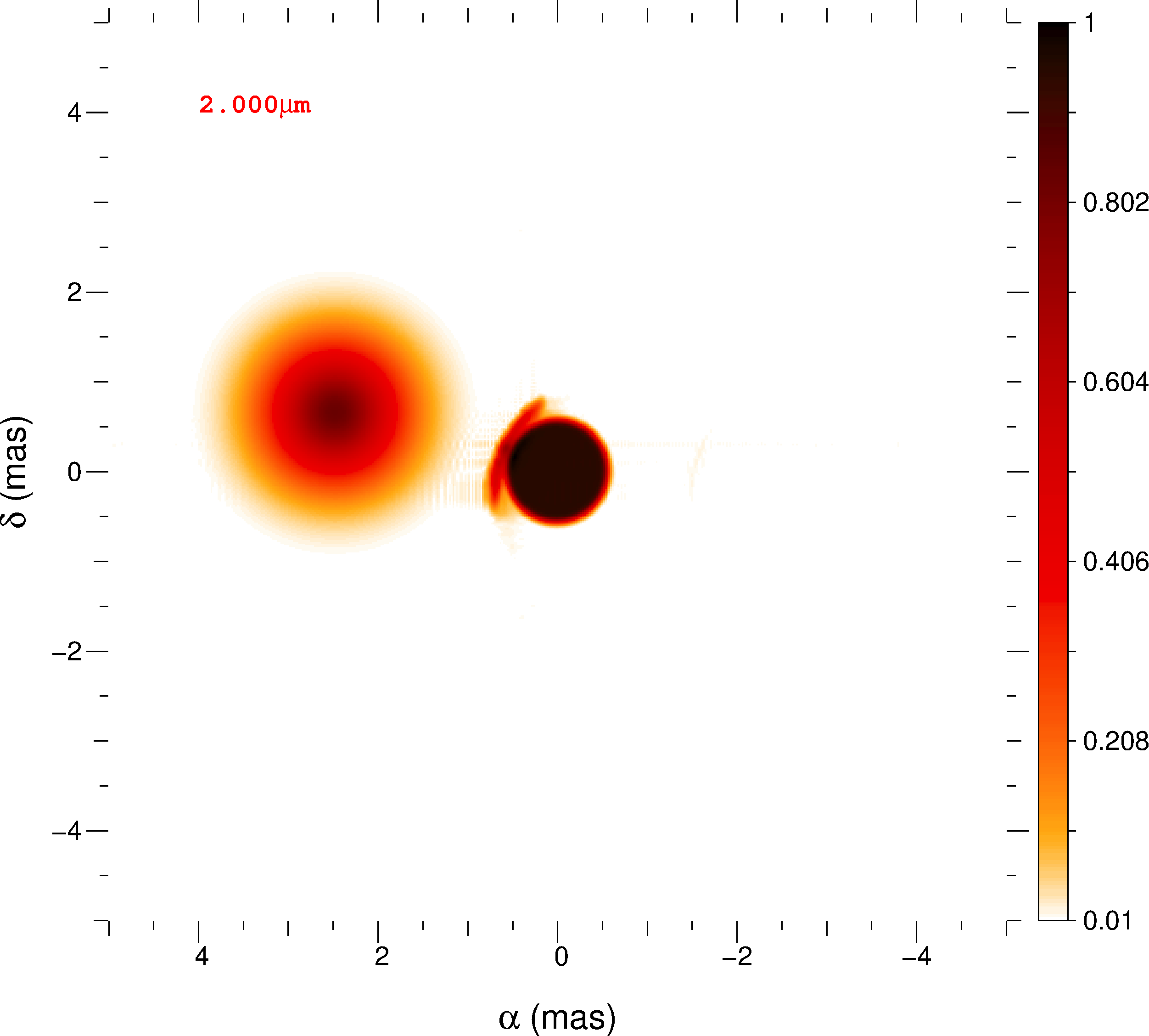}
  \caption{Volume rendering of the logarithm of the density in the shocked region (left) and the square root of the continuum  intensity maps at 2$\mu$m (right; different scale) at six of the observed phases ($\phi=0.01, 0.32, 0.51, 0.66, 0.82,0.92$ from upper left to lower right). The same fluxes of each component were used here: 54 per cent of the total flux for the O star (uniform disk), 41 per cent for the WR star (Gaussian disk) and 5 per cent for the wind collision zone. On the volume rendering, the WR star is a bright red spot while the O star is too embedded in the wind collision region to be distinguished.}
  \label{fig:emission}
 \end{figure*}

The first and third columns of Fig.~\ref{fig:emission} show a volume rendering of the 3D structure of the interaction region. It shows the logarithm of density at the outer edge of the shocked region, projected on the plane of the sky at the observed phases. Within $\simeq 5$ times the binary separation, the orbital motion of the binary has limited impact on the WCZ structure, which keeps its conical structure with symmetric arms, albeit off-centred with respect to the binary axis \citep{2008MNRAS.388.1047P}. Further out, the spiral structure becomes apparent. Its step depends on the velocity of both stellar winds as well as the orbital velocities \citep{2012A&A...546A..60L} and varies throughout the orbit. Around periastron (between $\phi=0.75$ and $\phi=0.1$), when the orbital velocity is highest,  more than a quarter of a step of the spiral is present within the simulated region and we see the shocked structure wrap around itself.

The shocked region can be subject to various hydrodynamic instabilities. They are responsible for the wobbly aspect of the shocked region in the volume rendering on Fig.~\ref{fig:emission}. The velocity difference between the winds leads to the development of the Kelvin-Helmholtz instability.  As the velocity gradient at the interface is small, the growth of the instability remains small and the instability is linear \citep{2011MNRAS.418.2618L}.  As energy is radiated away from the shocked region, the shocked shell gets narrower, and may be distorted by the thin-shell instability \citep{1994ApJ...428..186V}. Because cooling is limited in $\gamma^2$ Vel, the development of this instability is limited. Performing a simulation without orbital motion, we find that instabilities only change the total flux in the shocked region by at most 5 per cent  on timescales of a tenth of the orbital period.

\section{Model data}\label{sec:mock_data}

In this section we  compute model flux maps of $\gamma^2$ Vel at different observed wavelengths. We use the hydrodynamic simulation to determine the emission from the wind collision region and use the extended disk models described in \ref{sec:data_continuum} for the stars. We then use the resulting images to compute model visibility curves and closure phases with \texttt{fitOmatic}  and go beyond the geometrical point-source model from Paper~I (see \S\ref{sec:data_model}). We then compare the resulting continuum visibility curves with the observed ones  and estimate  the emission from the WCZ.

Based on our hydrodynamic model, we can estimate the IR continuum emission of the WCZ.  The bremsstrahlung (free-free) emissivity at frequency $\nu$ (in erg s$^{-1}$ cm$^{-3}$ Hz$^{-1}$) is given by
\begin{equation}
\label{emissivity}
\epsilon_{\nu}^b\simeq 6.81 \times 10^{-38} Z^2 n_e n_i g_b  T^{-1/2}e^{-h\nu/k_B T} ,
\end{equation}
where $Z$ is the average ionic charge, $n_e$ and $n_i$ the number density of electrons and ions respectively, $g_b$ the Gaunt factor set to unity and $h$ and $k_B$ the Planck and Boltzmann constants, respectively. 

We use the density and temperature provided by the hydrodynamic simulation to model the free-free  emission from the winds. Thanks to the passive scalar we can discriminate between the shocked O star and WR wind and take into account their distinct chemical compositions. The emitted flux is given by adding up the contributions of $\epsilon_{\nu}^b$ from both shocked region. We account for free-free absorption along the line of sight to the observer, using the absorption coefficient 
\begin{equation}
\kappa^b_{\nu} \simeq 3.7 \times 10^8 Z^2 n_e n_i g_b T^{-1/2}\nu^{-3}(1-e^{-h \nu/k_B T}) \quad \mathrm{(cm)}^{-1}.
\end{equation}

\begin{figure*}
\centering
\includegraphics[height=.2\textwidth]{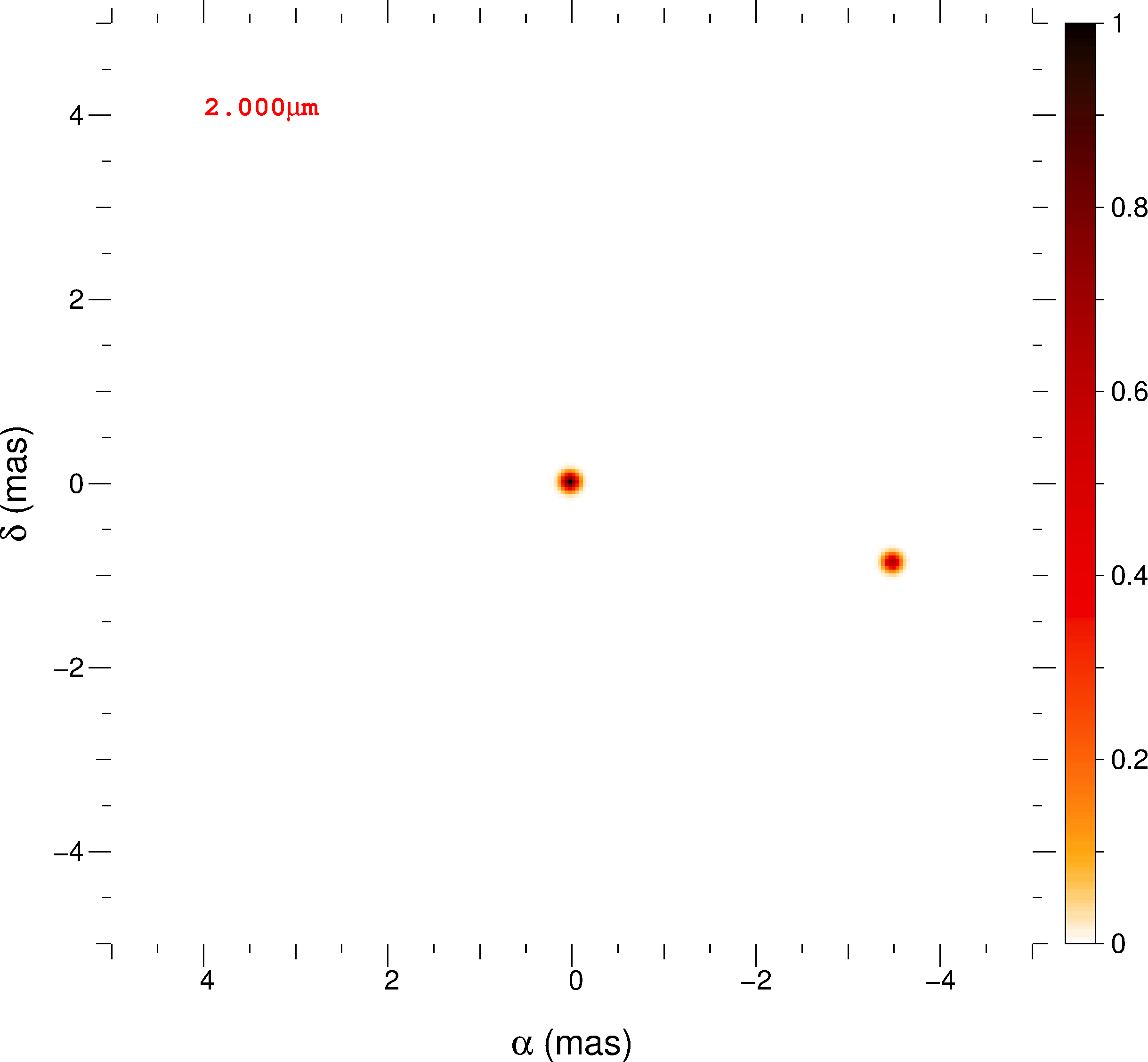}
\includegraphics[height=.2\textwidth]{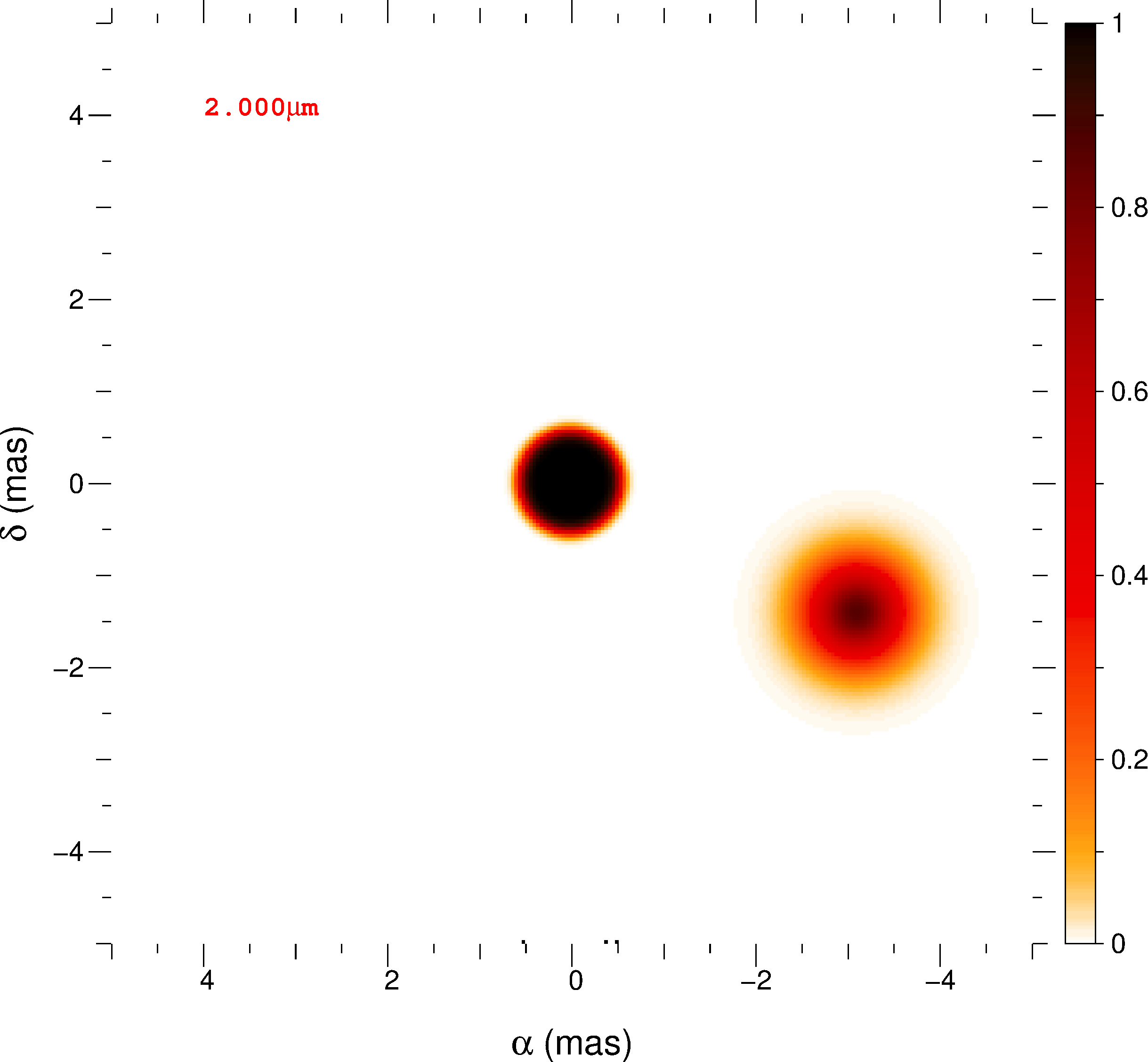}
\includegraphics[height=.22\textwidth]{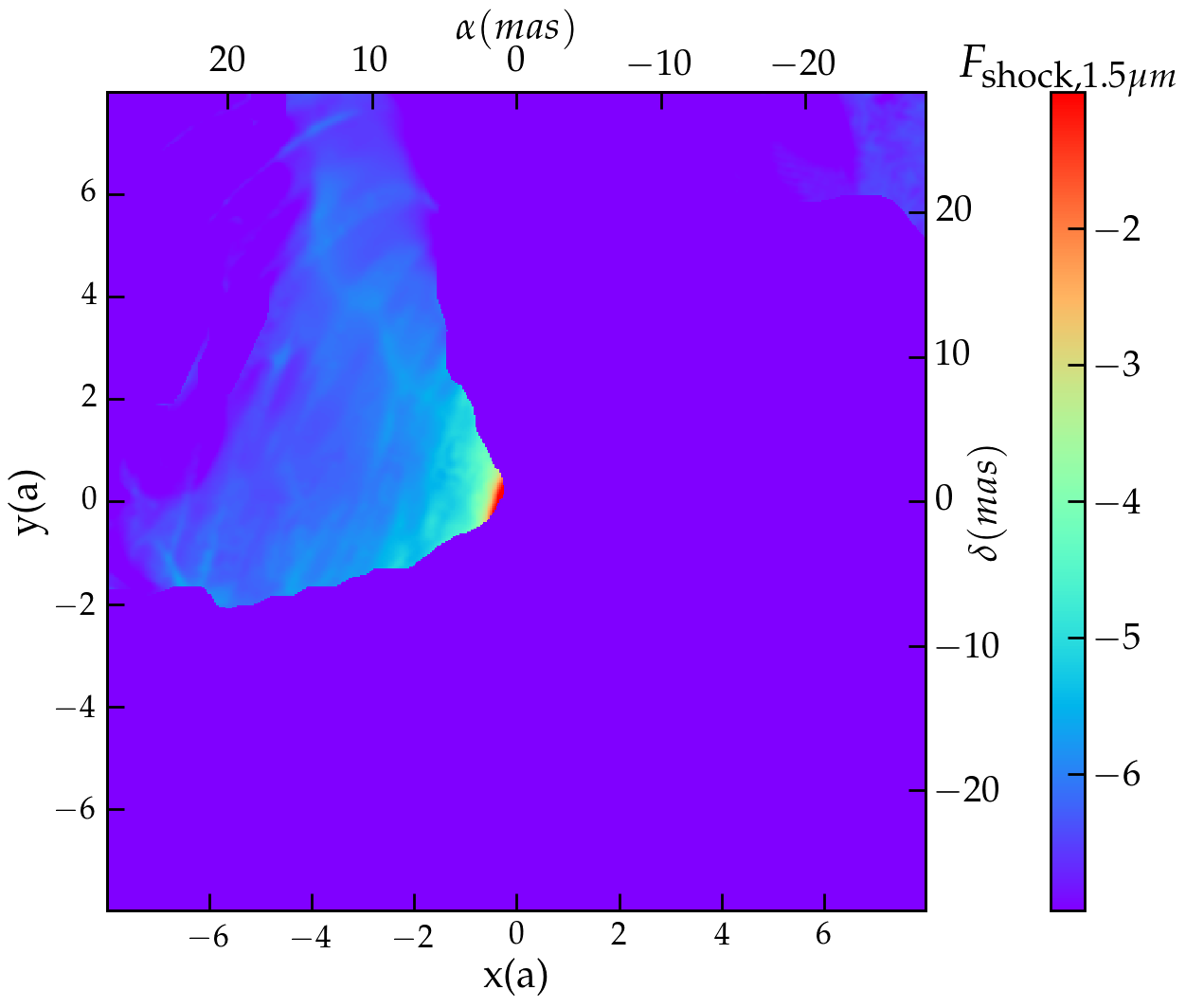}
\includegraphics[height = .2\textwidth]{SerieBig/2004-12-25_modelImage_ud_orb_gam_wcz} 
\caption{Model continuum emission at $\phi=0.3$ at 2.0$\mu$m. From right to left: model two point sources, geometrical model with  an extended Gaussian disk (WR star) and uniform disk (O star),  emission from the WCZ ,  total image with stellar disks and the WCZ. Images 1,2 and 4 represent the square root of the intensity, while the third one represents the log of the flux. The flux has been renormalized to the maximal flux in all images.}
\label{fig:mocks}
\end{figure*}

Fig.~\ref{fig:mocks} shows the build-up of the final model image when the binary is in the plane of the sky ($\phi=0.3$, close to the 2004 observation). The first and second image respectively show the images with the point sources and stellar disks described in \S\ref{sec:data_continuum}. The third image shows only the emission of the shocked regions, which we identify using density and pressure criteria.  The emission is very concentrated at the apex of the shock, due to the high temperature and density.  The final image is the complete composite image.

The second and last column of Fig. \ref{fig:emission} shows the total flux along the line of sight  at 2.0\,$\mu$m along the orbit using our complete model. On the left of each image is the volume rendering of the WCZ to guide the analysis of the model images.   At all phases, the emission of the WCZ is highly concentrated at the apex of the shock, folded over the O star. Its  flux rapidly drops  beyond a distance of roughly 1 mas.  The impact of the shocked region on the final images is modulated by the angular separation between the stars and the orientation of the shock cone with respect to the line of sight.

When the stars are at minimal projected separation ($\phi\simeq 0.01$), the highest emission of the WCZ is spatially coincident with, and outweighed by the emission of the stars. When the stars are further apart (large angular separation, e.g. around $\phi=0.32, 0.82$), the WCZ becomes more apparent, and distorts the spherically symmetric nature of the emission of the wind emission. In these cases, the total received flux is set by the orientation and shape  of the WCZ.  When the orbital plane is close to the plane of the sky, the shock cone is seen nearly edge on (e.g. $\phi=0.32, 0.82, 0.92$). As a result, the column density is very high near the apex of the shock, resulting in a small region of high flux.  When we are seeing the shock cone more face-on (e.g. $\phi=0.51, 0.66$), the column density of shocked material is lower, and we see a more extended region with lower flux.

Around periastron, our composite images show an \textit{apparent} crash of the wind collision region on the O star ($\phi=0.92$). As we will detail in the discussion (\S\ref{sec:discussion}), this is because the stellar radii turn out to be larger than expected. At such close distance to the stars, radiative braking from the O star is expected to push away the interaction region, preventing a crash.

\begin{figure*}
   \centering
   \includegraphics[width = .48\textwidth]{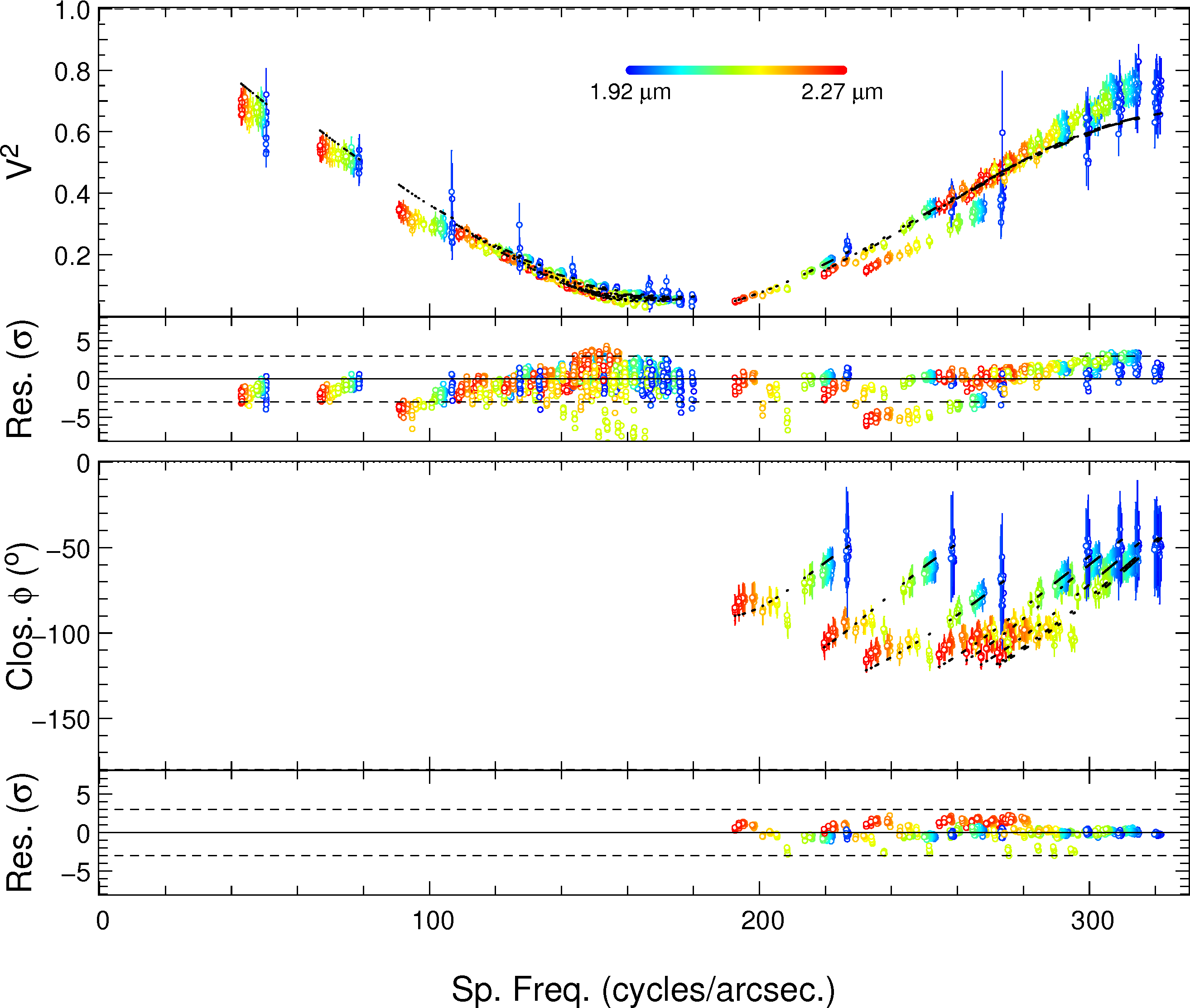}
   \includegraphics[width = .48\textwidth]{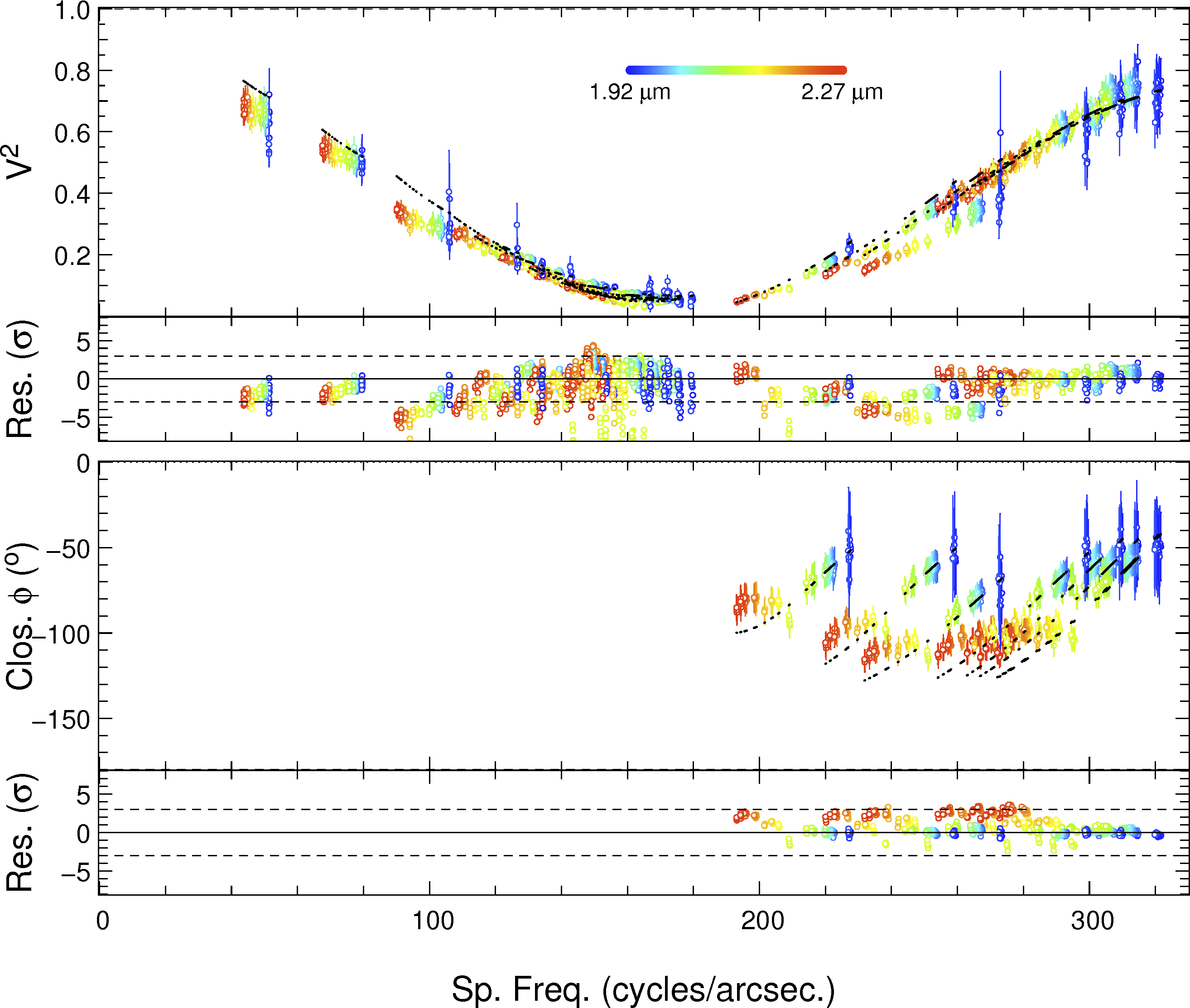}\\
  \caption{Continuum visibilities and closure phases from our high-fidelity observing campaign (night of 22/12/2008) with the goemetrical model (left) and the complete model (right) superposed in black. The colors show the wavelength dependence.} 
  \label{fig:visibility}
 \end{figure*}

Comparing a full hydrodynamic model coupled with simplified radiative transfer with interferometric data is a daunting task. At this point, our model does not include the line formation regions which would require to couple the hydrodynamics with a non-LTE radiative transfer model such as CFMGEN. Fig.~\ref{fig:visibility} shows visibility curves an closure phases of $\gamma^2$ Vel at $\phi=0.88$ (22/12/2008) based on our complete model (right). The left panel shows the model with resolved stellar disks (see \S\ref{sec:data_continuum}). The addition of the extended emission from the WCZ results in an improved squared visibility at the highest and lowest frequencies compared to the model with only the resolved stars with the side-effect of a slightly less good agreement for the closure phases. Especially at highest cycles/arcsec, the complete model shows better visual agreement with the data. In this model the WCZ accounts for a total flux fraction between 3 and 10 per cent of the total flux.

\section{Discussion}\label{sec:discussion}

In the former section, we analysed the spectro-interferometric AMBER data creating model observables to compare with, based on a complete hydrodynamic model of the WCZ. The high spatial resolution achieved with interferometry is necessary to resolve the binary separation.  Previous interferometric observations in the optical continuum \citep{1970MNRAS.148..103H,2007MNRAS.377..415N} have determined the orbital parameters of the binary. However, the stellar wind from the WR star and the WCZ benefit from observations at longer wavelengths due to their thermal free-free emission. Using AMBER data at one orbital phase, in Paper~I we showed that  an extended emission component of at most 5 per cent of the total flux was necessary to match the observed absolute visibilities. This additional component was tentatively attributed to the WCZ.

In this paper, the presence of the WCZ is more firmly established and its flux contribution to the continuum is estimated to be 3 per cent from the 2$\mu$m continuum observations. While this is a very small contribution, its strongly asymmetric and phase-dependent nature produce detectable variations in the interferometric signal. In order to go beyond the initial results from Paper~I, the hydrodynamic simulation was necessary to create model images. It reveals a complex density, velocity and  temperature structure. Due to the inclination of the system and its eccentricity, the aspect of the spiral changes drastically throughout the orbit and wraps around itself around periastron.   All of this contributes to the final observables and would not have been accounted for with analytic models. As the current dataset does not allow for complete reconstruction of the image (as in \citet{2016A&A...594A.106W} for $\eta$ Carinae), one has to construct a model for the system and compare its observables with the data. While the model we propose has a strong physical motivations, a unique solution for the system's emission cannot be formally identified with the continuum data.

Beside the continuum study, we have presented the first model independent spectral separation of the WR and O star component in the near-infrared. While the WR spectrum can be reproduced with well established models, we fail to explain the second component spectrum with any O star model alone. From the hydrodynamic simulation, we confirm that the latter is contaminated by the emission/absorption in WCZ, which also gives a likely explanation of its variability.  A full analysis of the line profiles, which are influenced by the geometry, density and temperature of the WCZ can ultimately determine the stellar wind parameters \citep{2005MNRAS.356.1308H}. According to the analysis of the "O star spectrum", we expect the WCZ line emission to account for as much as 20 per cent of the O star flux. As such, a similar analysis with the line emission would allow us to strongly constrain the emission properties (size and flux) of the WCZ.   

\begin{figure}
   \centering
  \includegraphics[width = .45\textwidth]{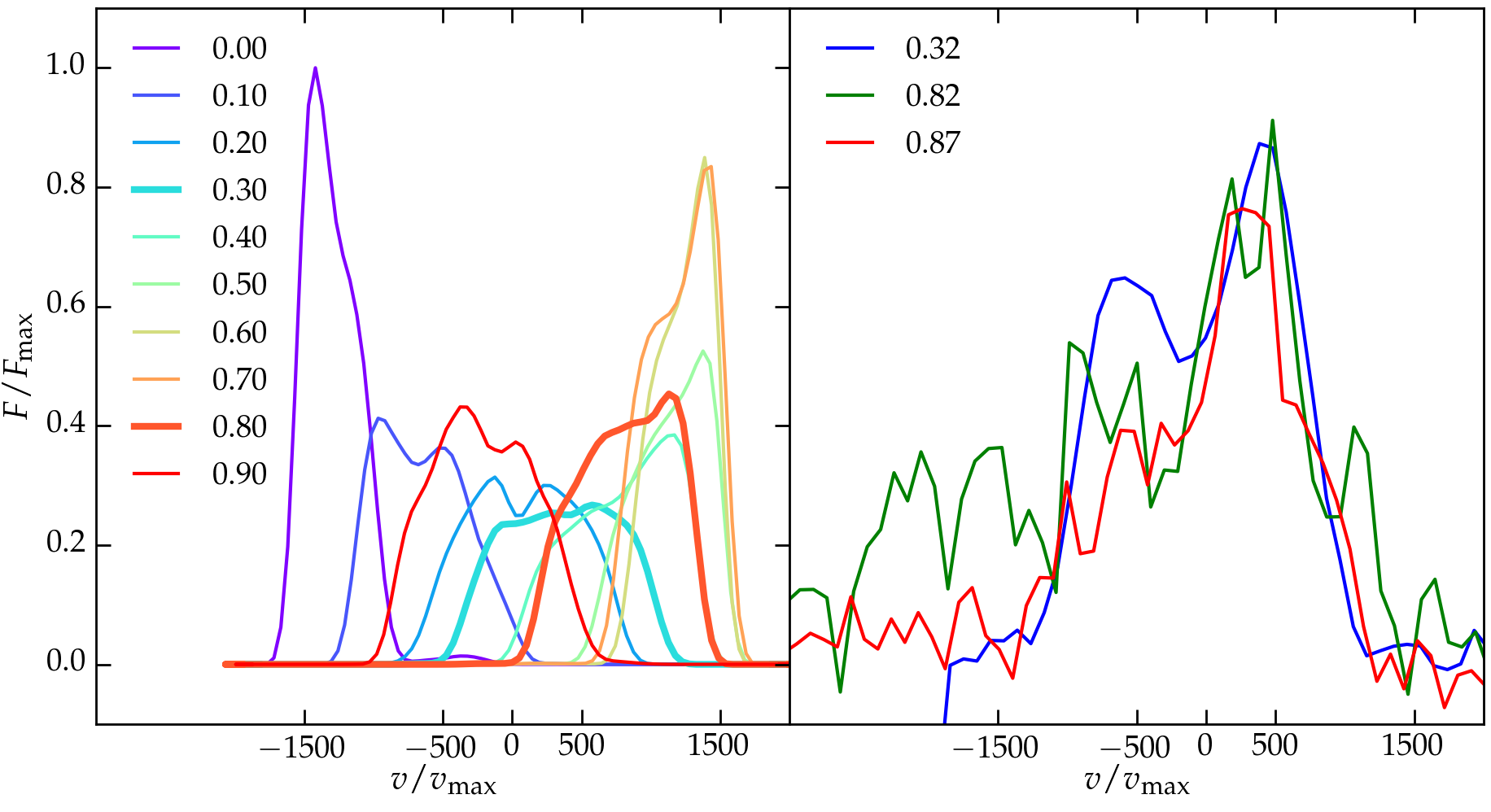}
    \caption{Left: Smoothed histogram of the flux-weighted line of sight velocity of the WCZ at different phases (color coding shown on the plot). The phases closest to the observed phases have thicker lines. Right: observed O-star spectrum for the  $\lambda 2.11$ $\mu$m\ion{He}{i} blended line.}
  \label{fig:histograms}
\end{figure}

A complete analysis of the line emission would requires a radiative transfer model for the whole system, which is clearly beyond the scope of this paper. We can have a foretaste of the result of such work by computing the flux-weighted line-of-sight velocity histograms of the WCZ from the hydrodynamic simulations, as is shown on Fig.~\ref{fig:histograms}. While this histogram only models the WCZ contribution to the line profiles, it shows that strong line variability is expected throughout the orbit, reminiscent of the observed variability in the O star spectrum.  According to the orientation of the WCZ along the orbit (see the 3D strucure in Fig.~\ref{fig:emission}), the line profile may vary from fully blueshifted to fully redshifted emission, including a flat-top phase. Similar trends were found by \citet{2009MNRAS.395..962I} in  forbidden line profiles based on a geometrical model of the WCZ. Unfortunately, our orbital coverage is too sparse to fully observe this phenomenon in the obtained spectra. Still, we find that our simulations predict higher velocities than observed in the emission lines.  On top of this, the WCZ, with its $10^6-10^8$ K temperature and enhanced density is also an additional photon source. As such, it may affect the ionization state and line emission in the free WR and O winds, possibly leading to the variability in the He II absorption profile in Fig.~\ref{Fig:zoomspectraGammaVel}. While these aspects are currently speculative, combining a fully hydrodynamic model with a complete radiative transfer model to compute the line profiles in spectro-interferometric data, while arduous, is clearly the way forward to fully exploit the next generation of instruments.

Our data analysis yields stellar radii in the K band which are about twice as large as estimated with optical interferometric data in \citet{2007MNRAS.377..415N} and predicted from spectral analysis and based on theoretical models for the stellar evolution and atmospheres \citet[e.g.][]{1999A&A...345..163D}. The latter still present important uncertainties, in particular for the  hottest, biggest and most massive stars. This is partially related to the rarity of these objects due to a combination of the initial mass function and short lifetime. The evolution of massive stars is highly sensitive to mass loss and rotation, the latter being very poorly constrained \citep{2000A&A...361..159M}. These uncertainties are enhanced by binarity, which is likely to have affected the evolution of $\gamma^2$ Vel \citep{2009MNRAS.400L..20E}.  Possible alternate explanations for the scale resolved emission is the presence of an additional emission component in the system such as fall-back shocked material. Similarly to the WCZ, the latter will  decrease the angular size of the stars needed to fit the model. While this explanation is highly speculative, it highlights the difficulty to understand these rare hot, massive and large stars.

A side-effect of the larger-than-expected stellar radii is that the WCZ appears to crash onto the O star photosphere around periastron. In our simulations, we have assumed instantaneous wind acceleration and did not directly model radiative effects such as the radiative braking of the WR wind expected at periastron. According to \citet{1997ApJ...475..786G,2000MNRAS.316..129R}, the latter is expected to be important in $\gamma^2$ Vel and will push the WCZ away from the O star and yield a stable shocked structure. In our simulation, we assumed immediate wind acceleration, which actually overestimates the opening angle of the WCZ at periastron, somewhat counterbalancing the lack of radiative braking. As such, we do not expect the resulting WCZ to be qualitatively different from the WCZ in our hydrodynamic simulations. Simulations of $\eta$ Carinae including radiative effects show a widened shock opening angle and also an increased development of instabilities \citep{2011ApJ...726..105P}. As the development of instabilities is limited in $\gamma^2$ Vel, the latter should not impact our comparison with the interferometric data.

\section{Conclusions}\label{sec:conclusion}

In this paper we have studied the colliding wind region in $\gamma^2$ Vel combining near infrared spectro-interferometric data from AMBER/VLTI with a 3D hydrodynamic model.

As our data provides good coverage of the orbital period, we confirm the orbital parameters from \citet{2007MNRAS.377..415N}. At phases where higher spectral resolution is available, we are able to separate the composite spectrum in a WR component and a component with and around the O star. Using CFMGEN, a model for radiative transfer in expanding stellar atmospheres, we confirm the WC 8 spectral type and wind parametres of the WR star compatible with \citet{2000A&A...358..187D}. We fail to satisfactory match the O-star component with the spectrum of an O-star alone, both with theoretical models and template spectra from other O-type stars. We interpret this as a signature of strong contamination by the wind collision zone, as was suggested in Paper~I. This is supported by the phase-locked line profile variability in the spectrum. Additional evidence comes from the hydrodynamic simulation showing the wind collision region shrouding  the O star. 

The 3D hydrodynamic simulation includes orbital motion of the binary and covers 16 times the binary separation, a physical extension never modeled before for this system. It reveals the spiral nature of the wind collision zone, particularly marked around periastron. Based on the hydrodynamic variables in the simulation, we determine the free-free emission in the wind collision region. We combine it with a model for the extended emission of the stars to produce model emission maps at different wavelengths. We find that most of the emission in the wind collision zone is concentrated at the apex of the shocked region, around the O star. Depending on the projected angular separation of the stars and the orientation of the shocked region, the wind collision zone has more or less impact on the final flux map. We use these model images to derive interferometric observables. Our model, where the WCZ contributes 5 per cent to the continuum emission, is in good agreement with the observed data. This constraint on the wind collision zone in $\gamma^2$ Vel with spectro-interferometric observations demonstrates how the hydrodynamic models enhance the scientific output of the data.

To improve our understanding of this emblematic system, there are clear paths forward, both on the observational and theoretical side:
\begin{itemize}
\item Updated, and especially more precise, radial velocity measurements would allow one to reach a sub-percent accuracy on the distance of Gamma Vel. Having such an accuracy on the distance of this system would be extremely interesting, as it could become a calibrator probe for GAIA distances to binaries.
\item Medium-spectral resolution interferometric data are needed at many more orbital phases (here the spectroscopic analysis focused on only two phases) to better characterize the system. Better spatial resolution is needed to resolve  the stars themselves, i.e. reach longer baselines and/or shorter wavelengths. This may be achieved with GRAVITY, or with a future visible interferometric instrument at the VLTI.  While the object might be too poorly resolved for full imaging on the VLTI, GRAVITY, thanks to its 6 baselines and its fringe tracker, will provide much more accurate data for modeling, and might open the possibility to evaluate to measure accurately the spectrum of different parts of the WCZ. 
\item A line analysis of the WCZ would then require to couple the hydrodynamic simulation presented in this paper together with a non-LTE radiative transfer code similar as CMFGEN.
\end{itemize}

\section*{}{Acknowledgements.}
We thank the CNRS for providing us with guaranteed time observations and the "Programme National de Physique Stellaire" (PNPS) of the CNRS for continuous support during the years of preparation for this article. This work made use of the JMMC and CDS tools. The MCMC orbital solution computations were performed on the ’Mesocentre SIGAMM’ machine, hosted by Observatoire de la C\^ote d’Azur. Hydrodynamic numerical calculations were run on UW-Milwaukee supercomputer Avi and supercomputer Pleiades from the NASA Supercomputing Division. Support for A. Lamberts was provided by an Alfred P. Sloan Research Fellowship, NASA ATP Grant NNX14AH35G, and NSF Collaborative Research Grant 1411920 and CAREER grant 1455342.

\bibliography{GVel2014}%>>>> bibliography data in report.bib
\bibliographystyle{mnras}%>>>> makes bibtex use spiebib.bst

\appendix\section{Data analysis}
\subsection{(u,v) coverage}
\label{sec:uvcov}
\begin{figure*}
  \centering
    \includegraphics[width=0.9\hsize, angle=0]{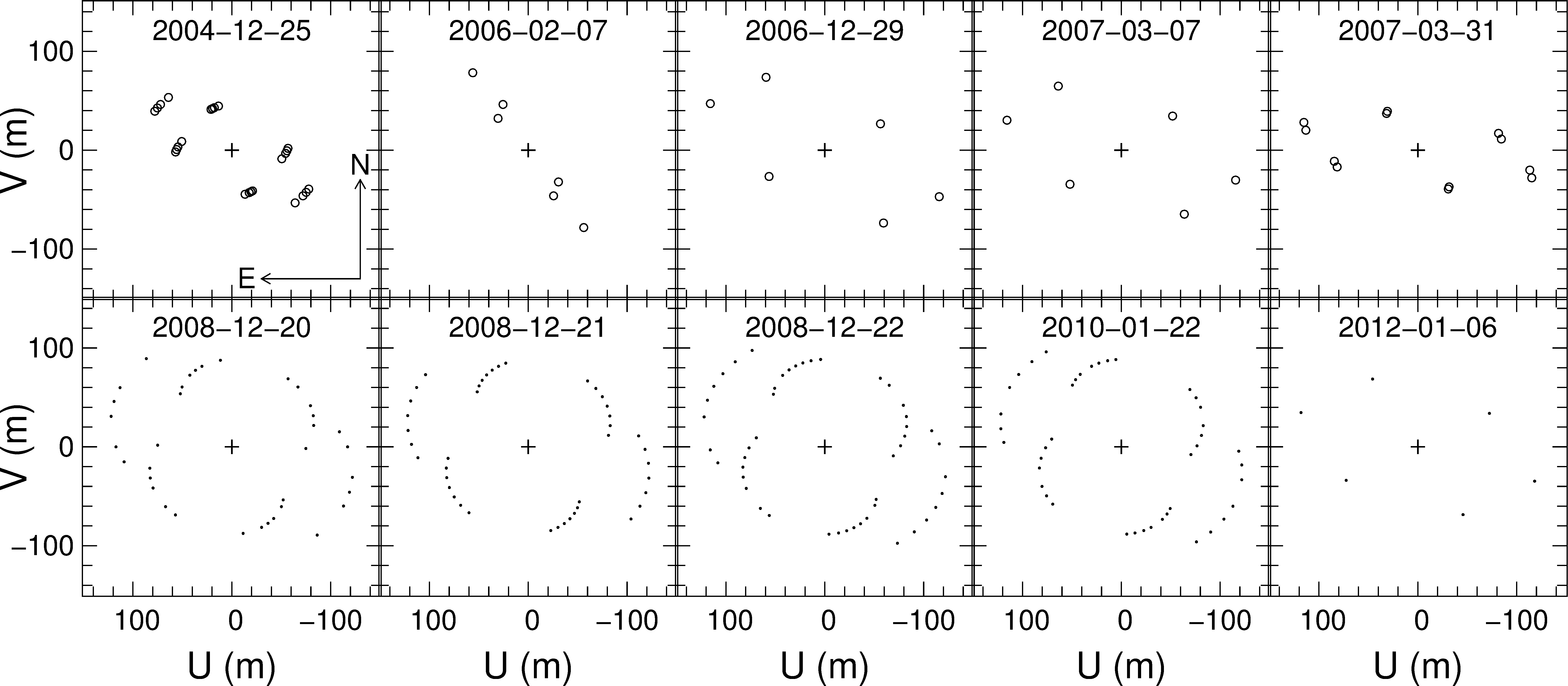}
  \caption{(u,v) plane of the observations for each epoch. Note that the North is up and the East is left, as if the (u,v) coordinates were seen on the plane of the sky. The circles represent the Unit Telescopes, the dots represent the Auxiliary Telescopes.}
    \label{fig:plan_UV}
\end{figure*}

Fig.~\ref{fig:plan_UV} shows the (u,v) plane coverage of our observations.

\subsection{$\chi^2$ maps and astrometry}
\label{sec:chi2maps}
\begin{figure*}
  \centering
    \includegraphics[width=0.9\hsize, angle=0]{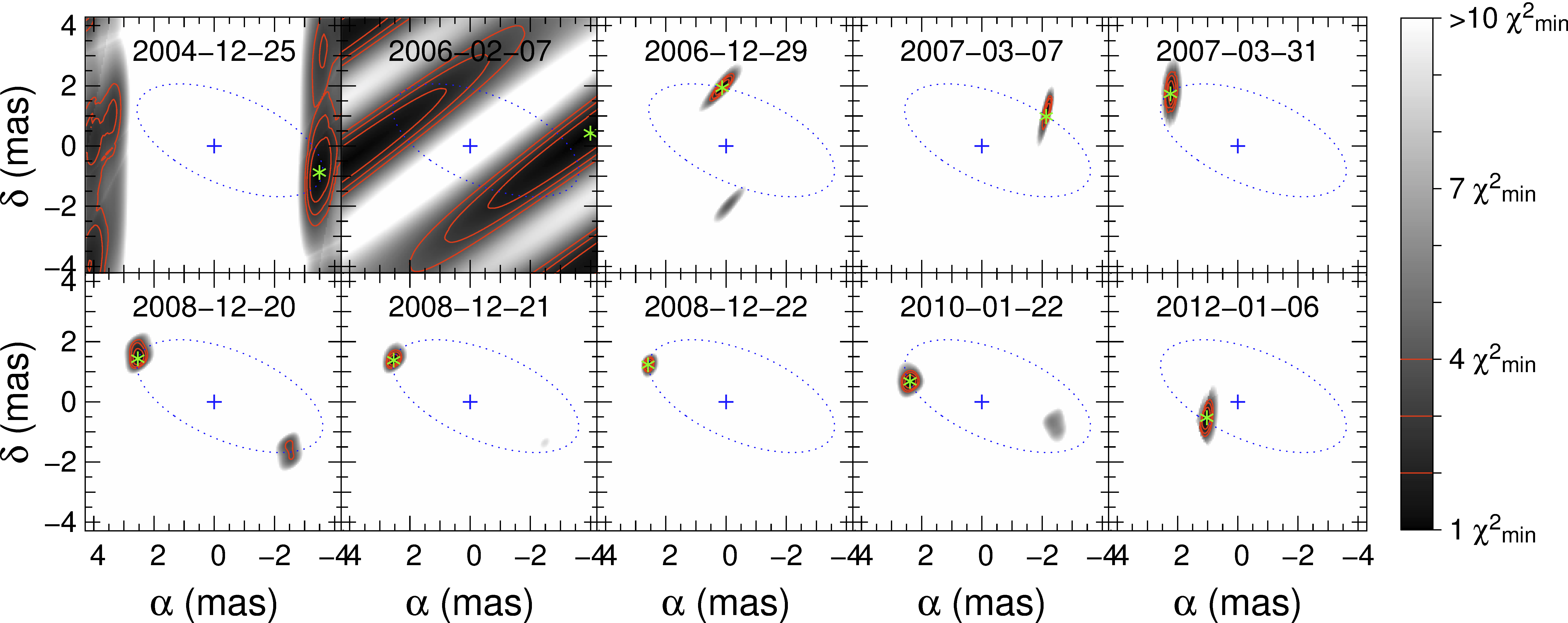}
  \caption{$\chi^2$ maps for our binary star model positions for each epoch of observation, showing the orbital motion of the system. The green asterisks show the best-match position to our data, while the red contours are the 1, 2 and 3$\sigma$ error contours.}
    \label{fig:astrometry}
\end{figure*}

Fig.~\ref{fig:astrometry} shows  the $\chi^2$ maps of our model along the orbit. There is a unique solution for most of our datasets except from 2006 for the reasons explained in the main body of the text. The resulting astrometric measurements are provided in Table~\ref{tab:astrometry}.

\subsection{Model fitting in the lines}
\label{sec:model_lines}

\begin{figure*}
   \centering
 \includegraphics[width = .33\textwidth]{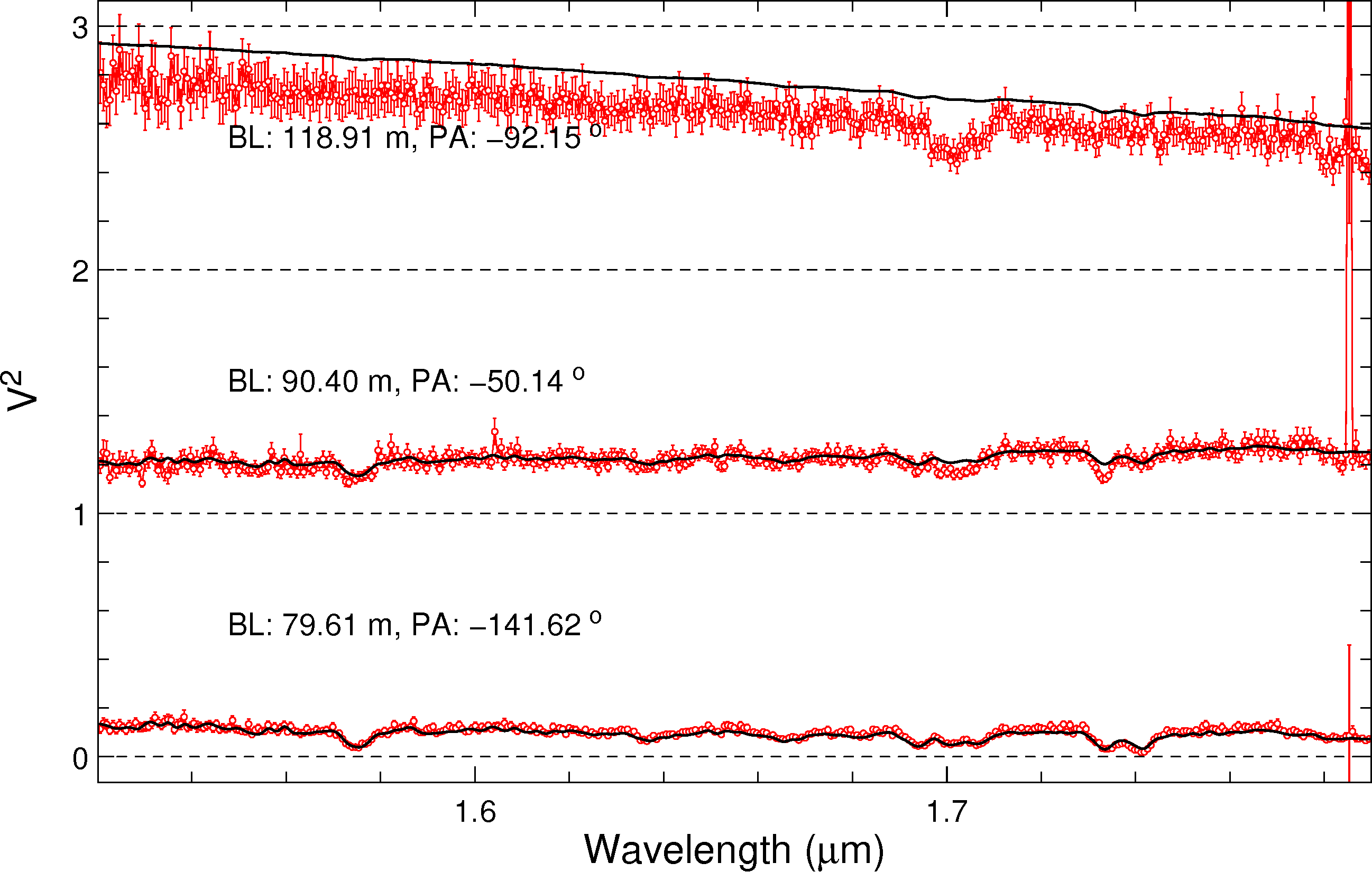}
 \includegraphics[width = .33\textwidth]{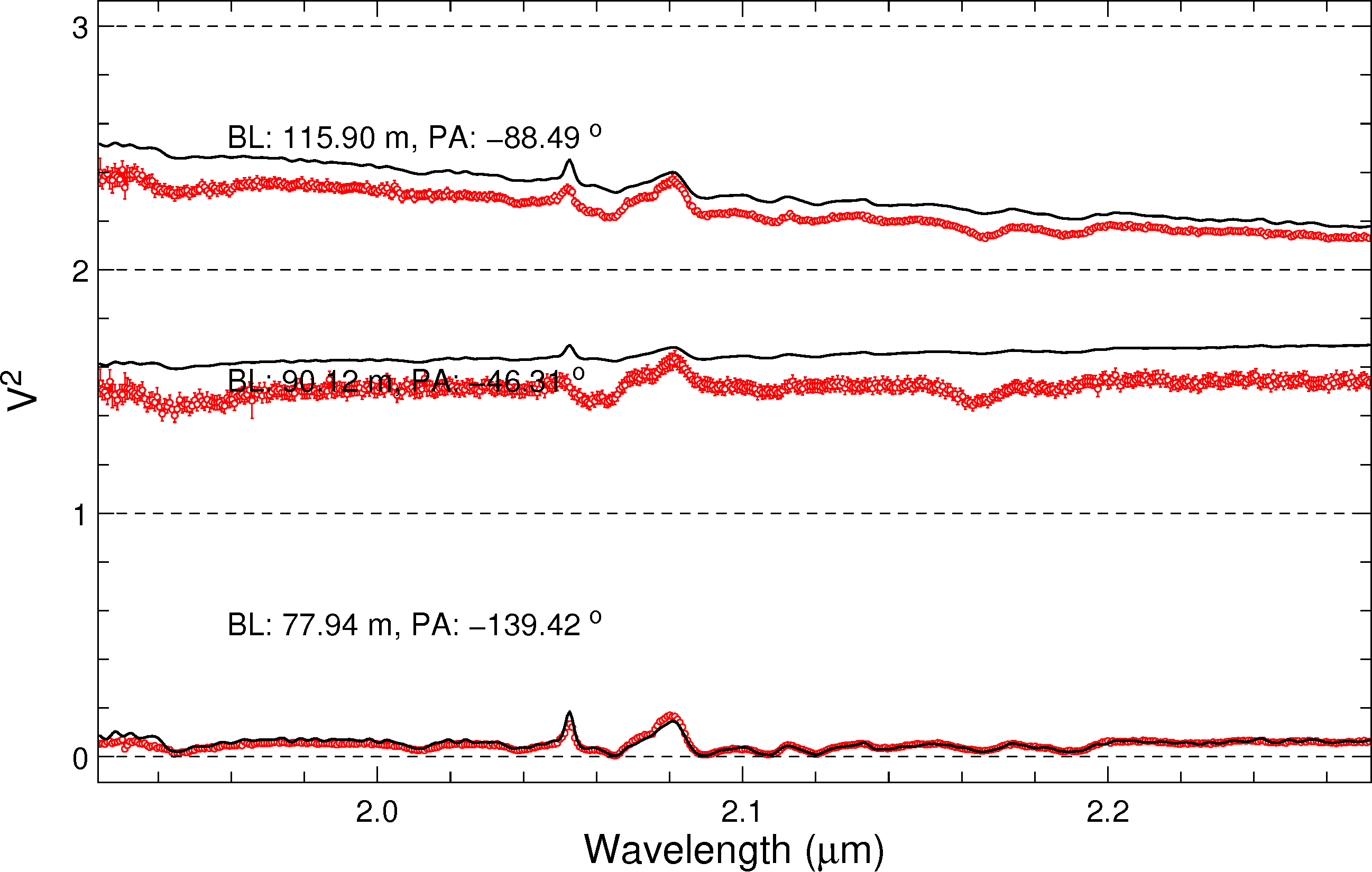}
 \includegraphics[width = .33\textwidth]{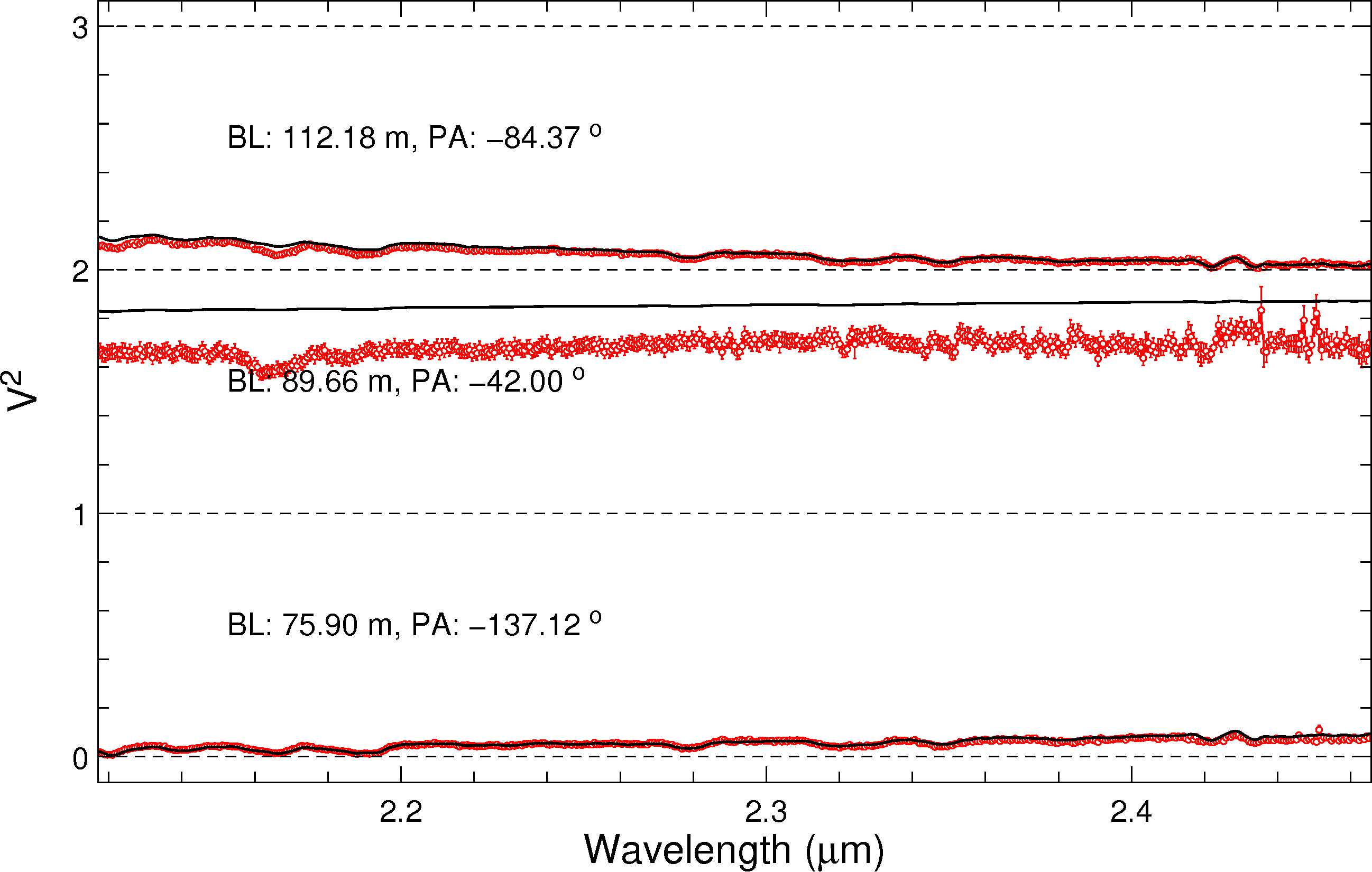}\\
 \vspace{0.2cm}
 \includegraphics[width = .33\textwidth]{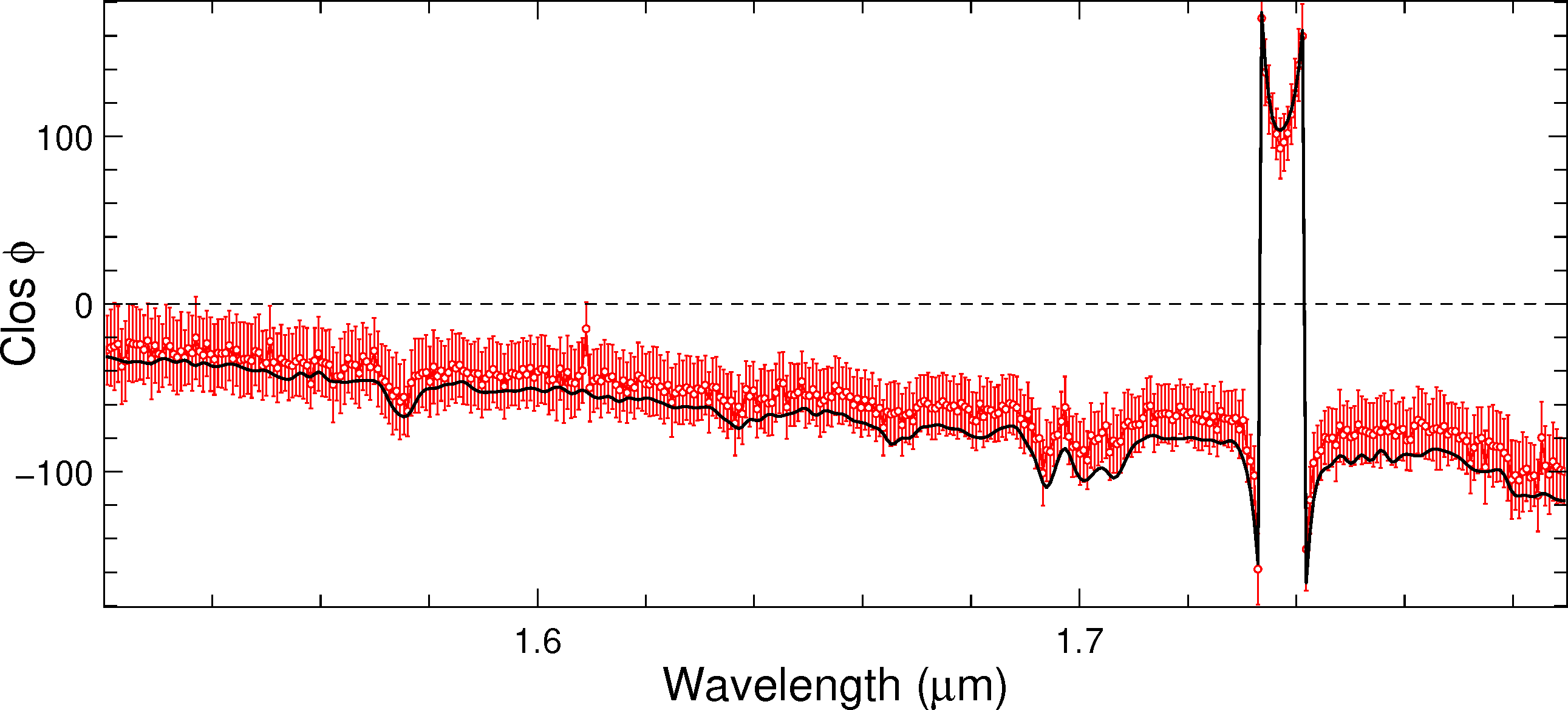}
 \includegraphics[width = .33\textwidth]{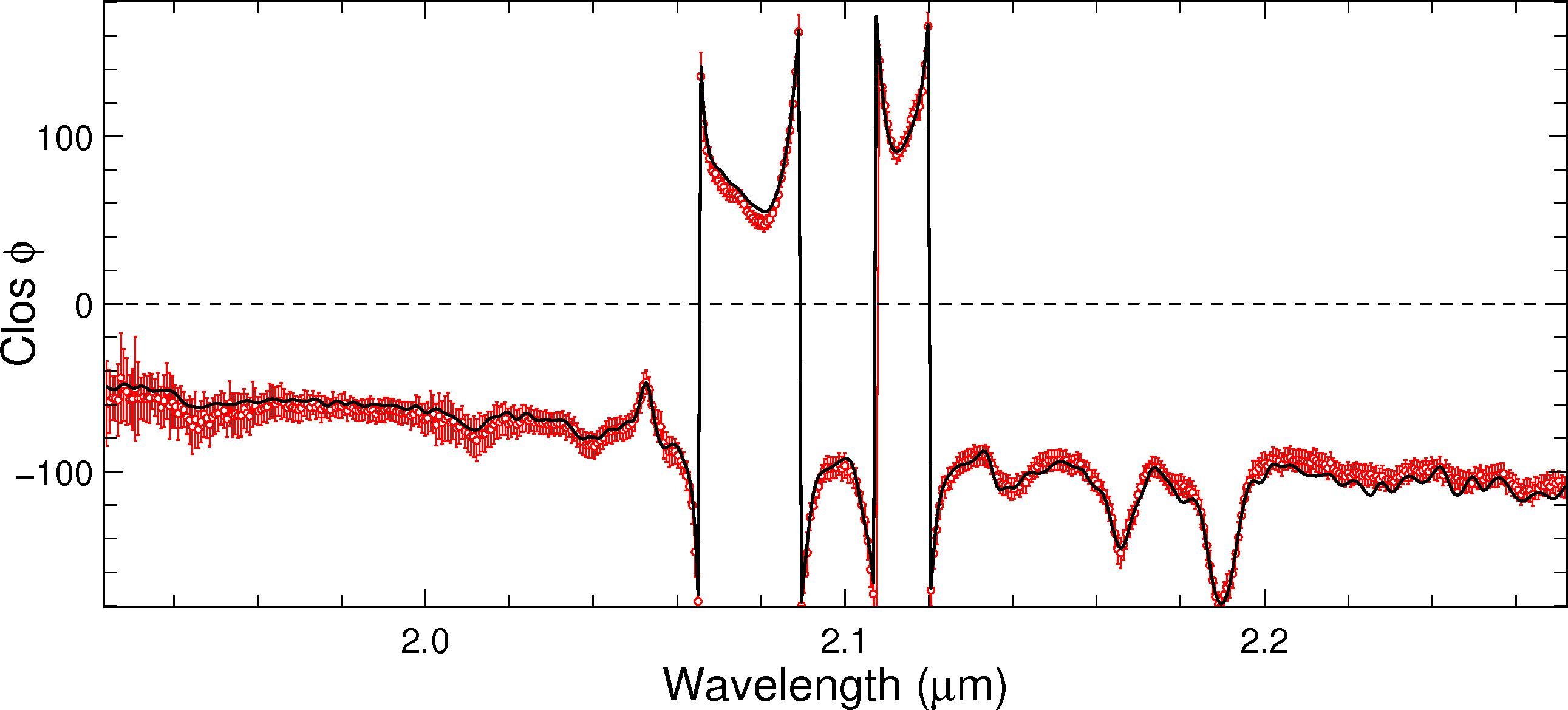}
 \includegraphics[width = .33\textwidth]{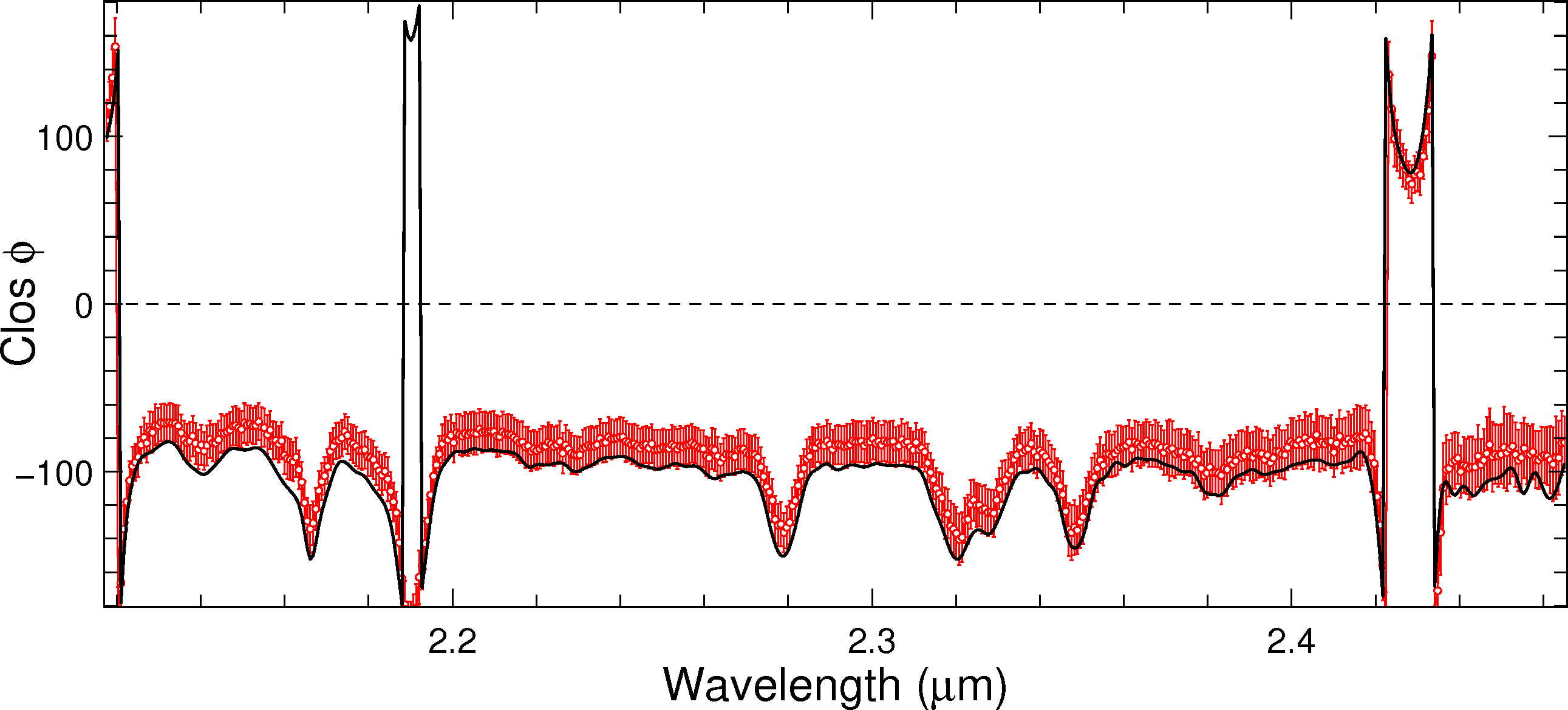}\\
 \vspace{0.2cm}
 \includegraphics[width = .33\textwidth]{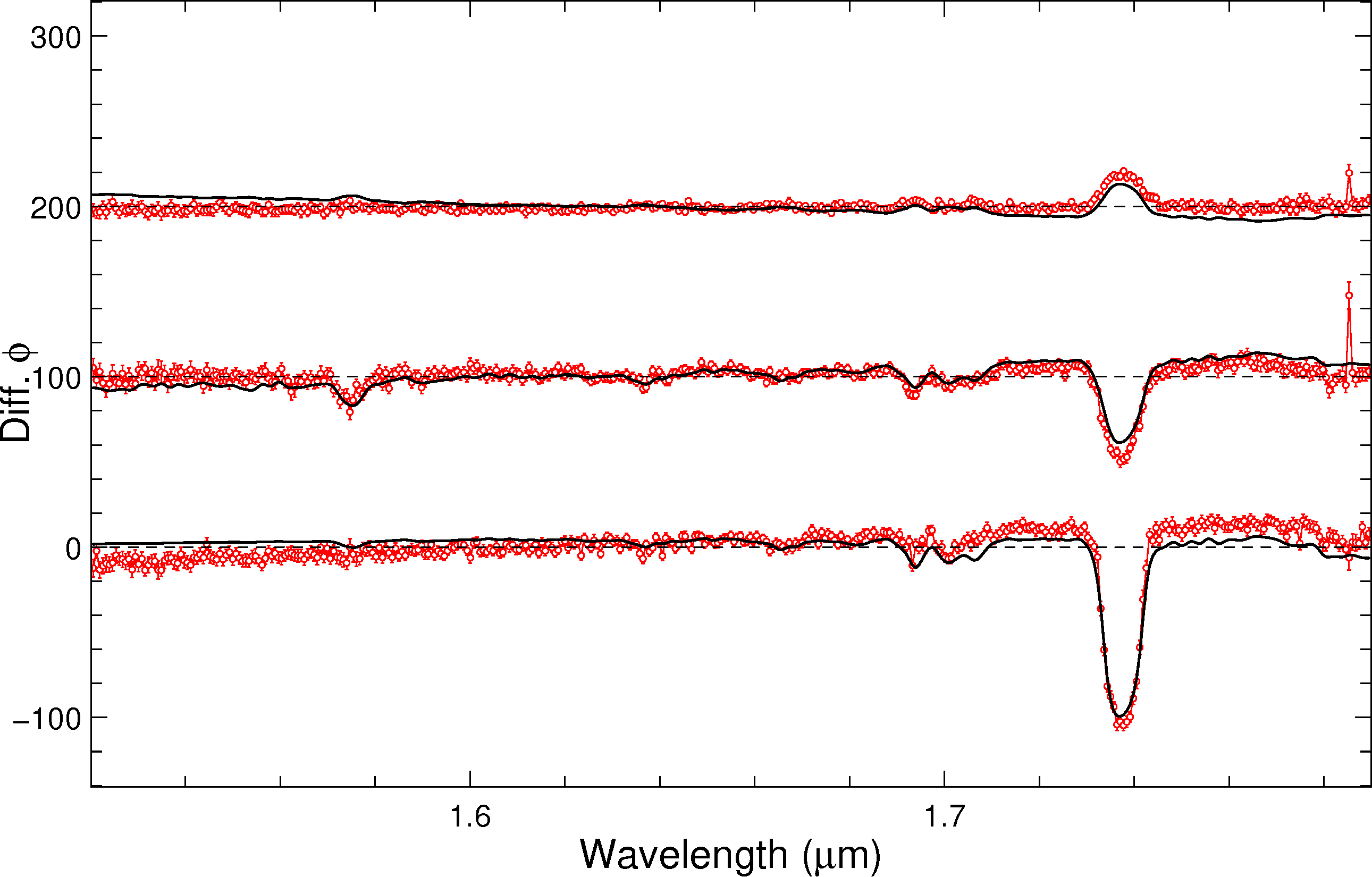}
 \includegraphics[width = .33\textwidth]{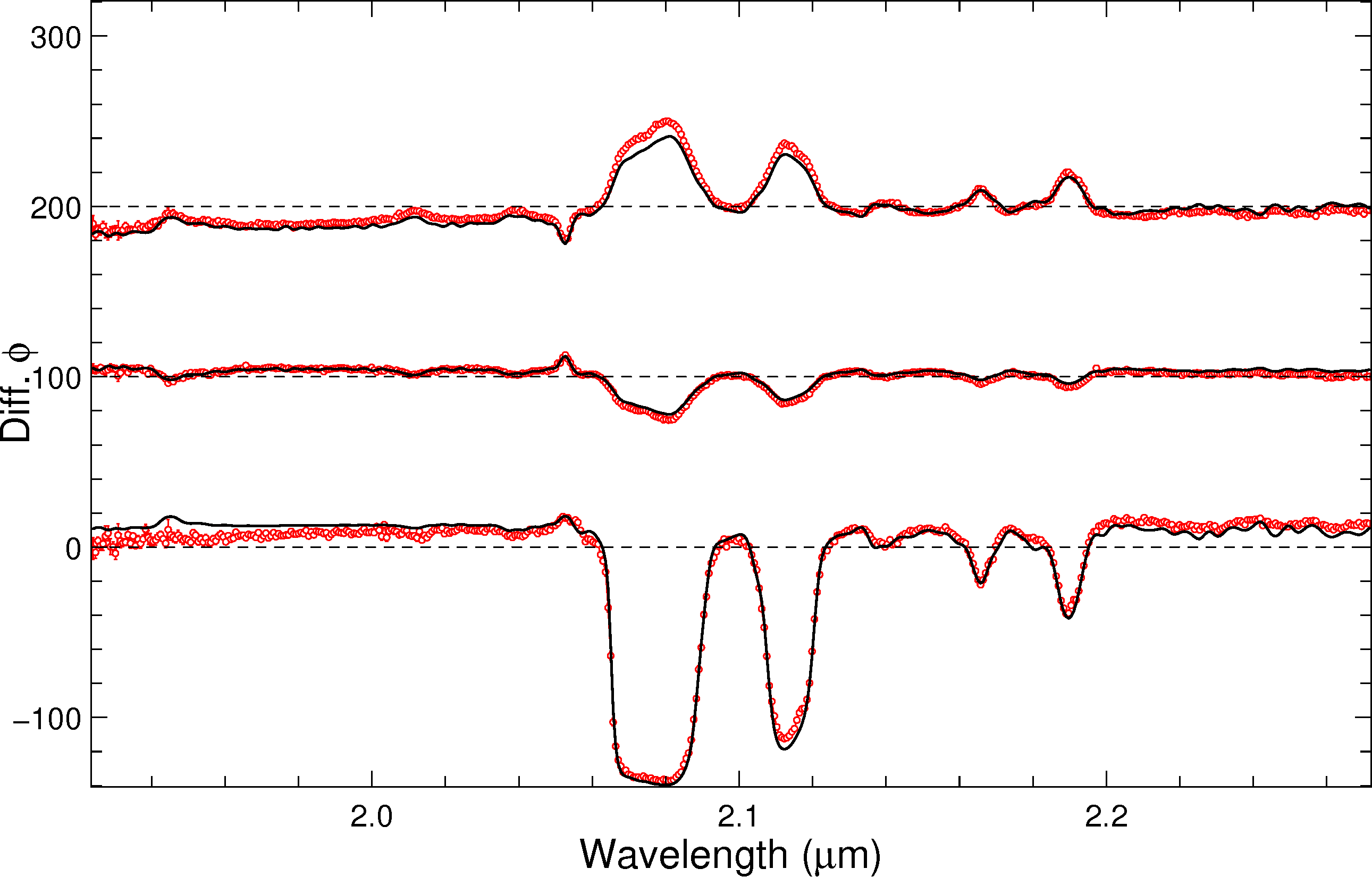}
 \includegraphics[width = .33\textwidth]{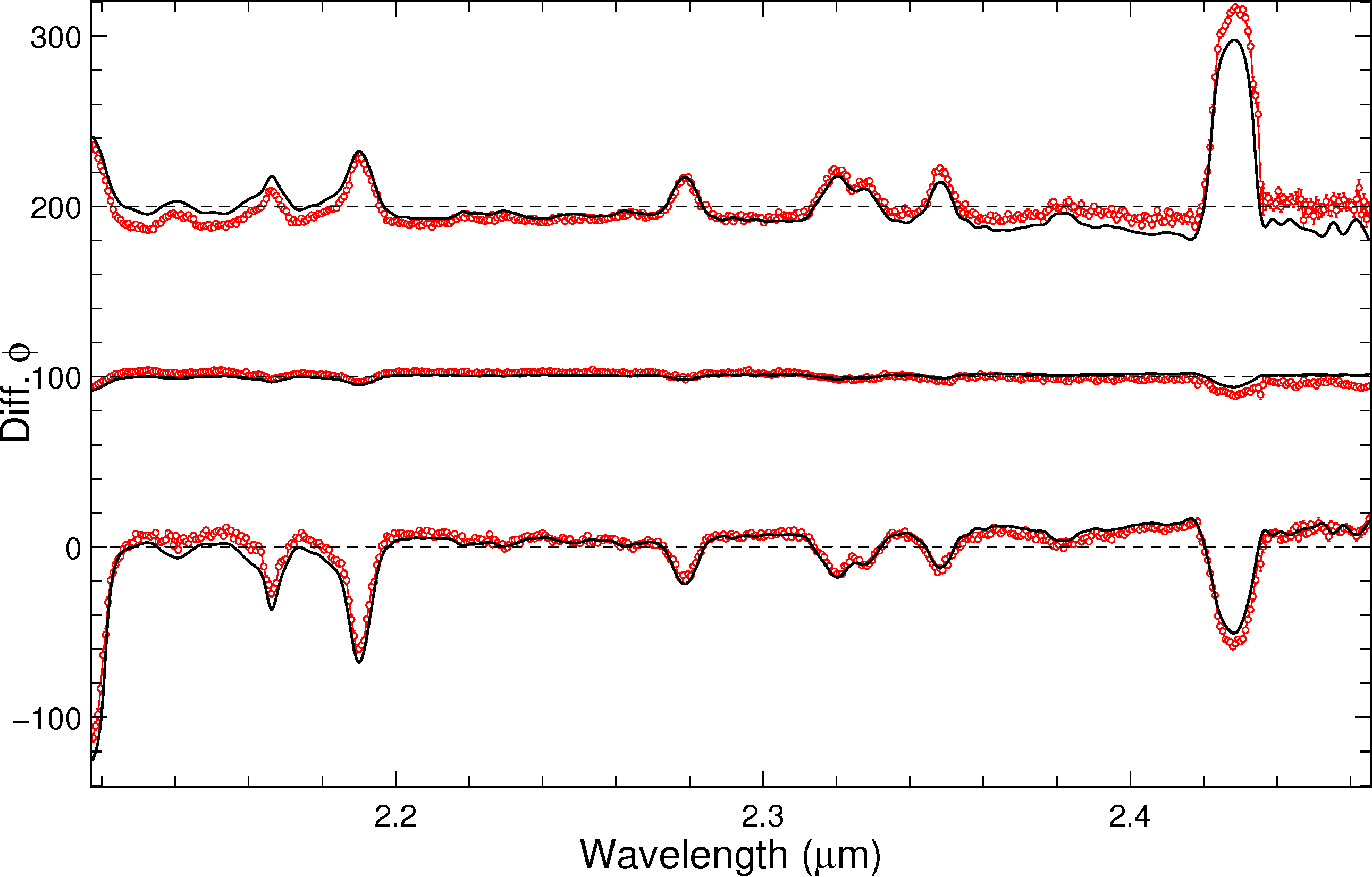}\\
 \vspace{0.2cm}
 \includegraphics[width = .33\textwidth]{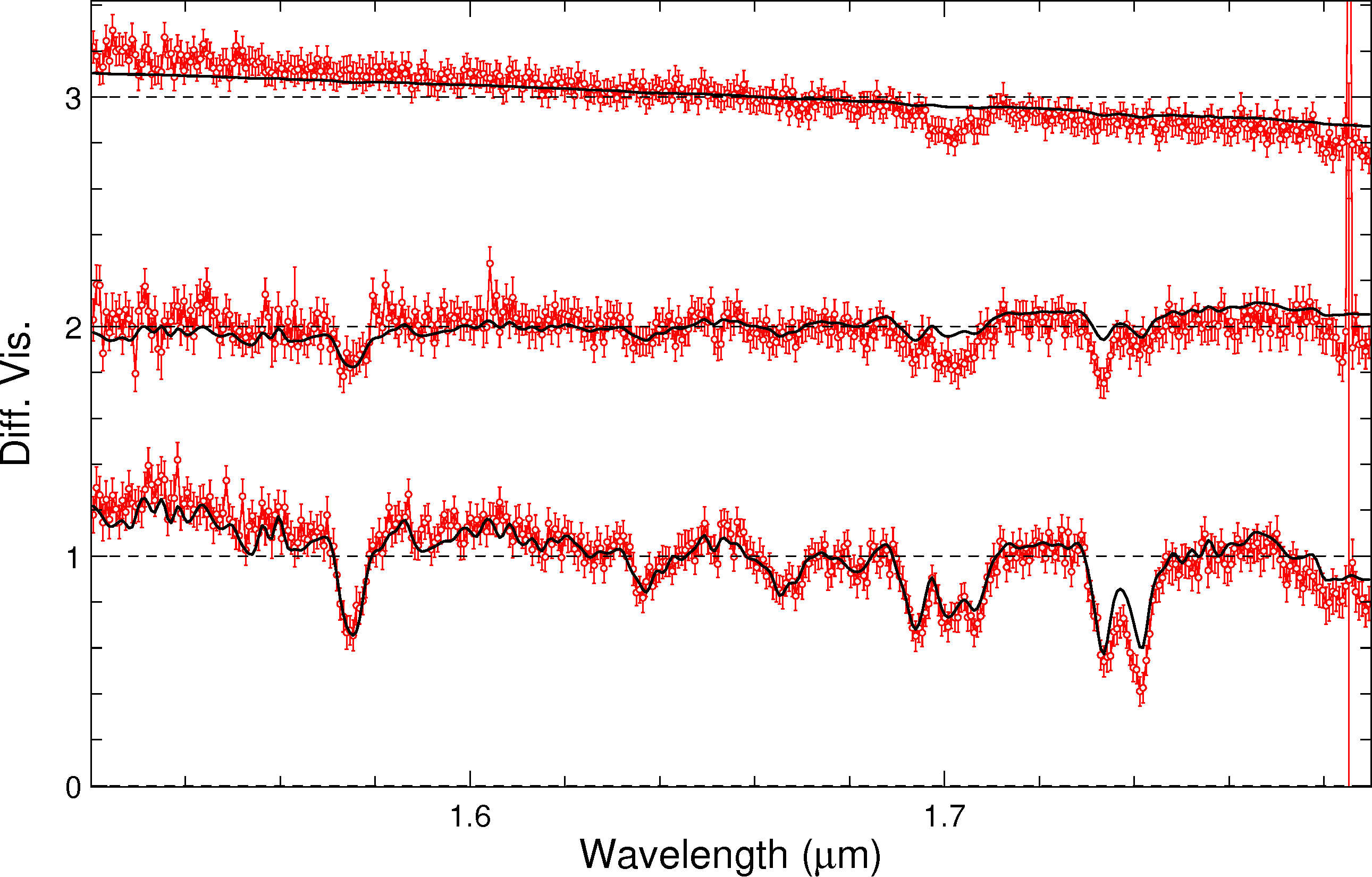}
 \includegraphics[width = .33\textwidth]{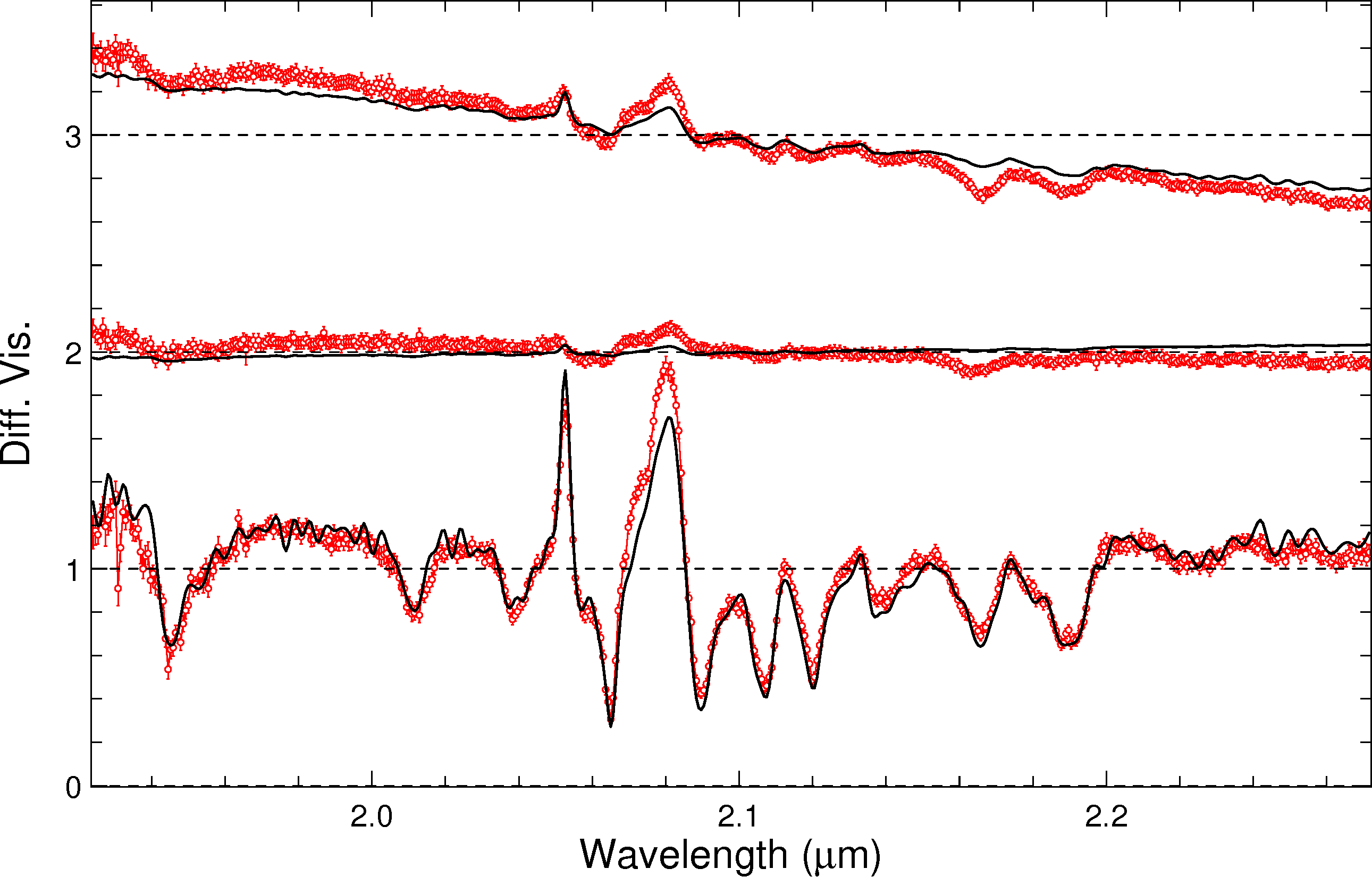}
 \includegraphics[width = .33\textwidth]{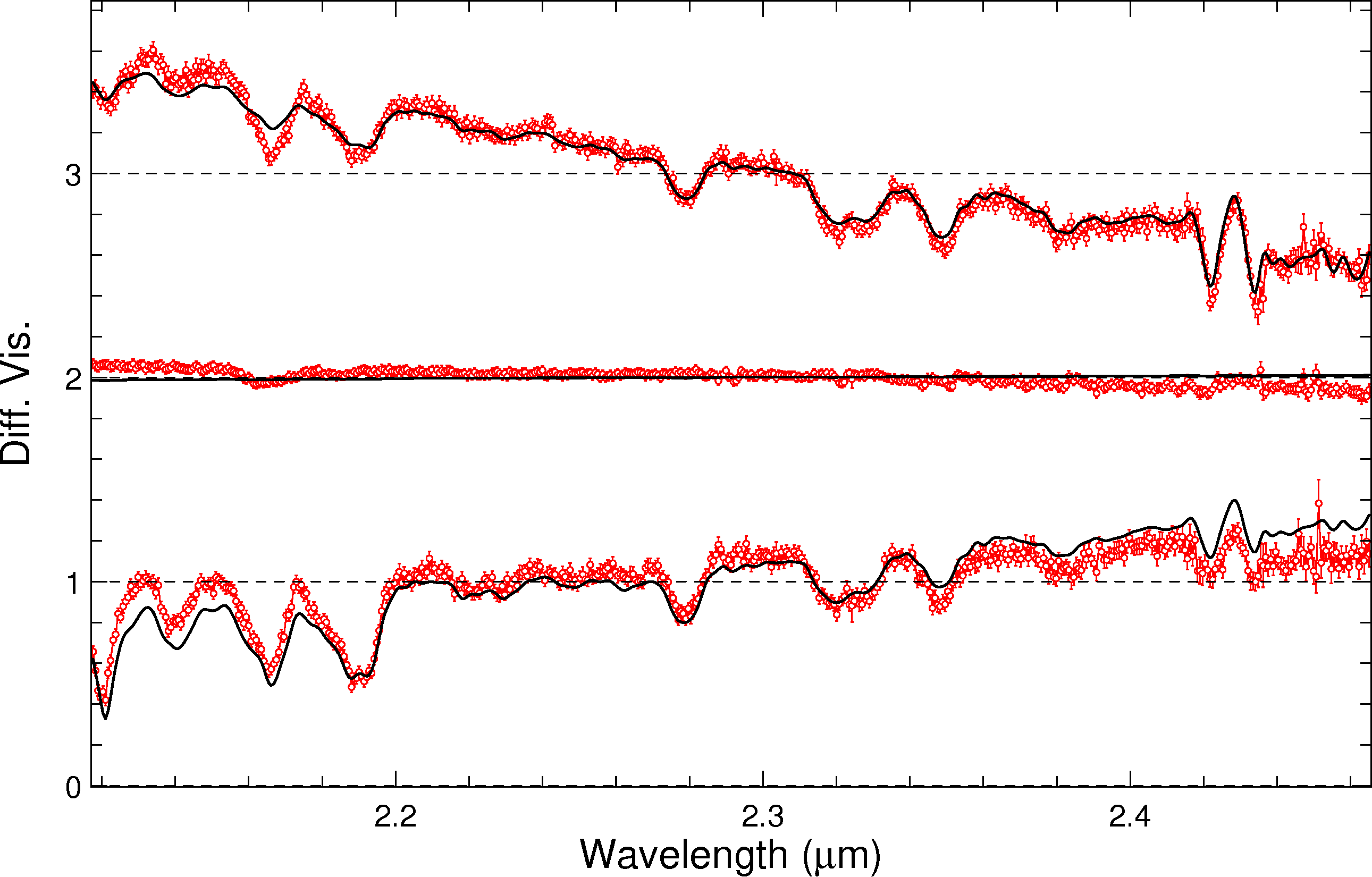}
  \caption{Illustration of the AMBER data as well as the model-fitting process used on the emission lines to extract the spectra of the WR star and the O star. The observed data is in red, while the best-fit two point-sources model is plot in black. {\bf From top to bottom:} Squared visibilities, closure phases, differential phases and differential visibilities. We selected similar baselines for the three nights. {\bf Left:} Night of 22/01/2010,  {\bf Middle:} Night of 22/12/2008, {\bf Right:} Night of 21/12/2008.} 
  \label{fig:modelfit_lines}
 \end{figure*}

In Fig.~\ref{fig:modelfit_lines} we show a sub-sample of our 2008 and 2010 observations (red lines), together with the best-fit 2 point sources model (black). As the $H$-band 2010 dataset was obtained at a similar orbital phase as the 2008 datasets, we selected similar baselines so that the visibilities, closure phases, differential phases and differential visibilities can be compared directly.

%__________________________________________________ One column table
\begin{table}
  \caption[]{Astrometric measurements of $\gamma^2$ Vel. $\Delta\alpha$ and $\Delta\delta$ represent the angular separation in right ascension and declination respectively, in mas.$\sigma_{\rm major\ axis}$, $\sigma_{\rm minor\ axis}$ are the uncertainties on $\Delta\alpha$ and $\Delta\delta$, respectively, represented by the sigma of a Gaussian tilted by the angle $\theta_{\sigma_{\rm major\ axis}}$.$^1$}
  \label{tab:astrometry}
    \begin{tabular}{lcccccccc}
    \hline
    \noalign{\smallskip}
    Date & $\Delta\alpha$ & $\Delta\delta$ & $\sigma_{\rm major\ axis}$ &  $\sigma_{\rm minor\ axis}$ & $\theta_{\sigma_{\rm major\ axis}}$ \\ 
    \noalign{\smallskip}
  \hline
%     \noalign{\smallskip}
%     25/12/2004 & -3.48 & -0.87 & 1.04 & 0.33 &   -8.5 \\
%     07/02/2006 & -3.93 &  0.39 & 4.48 & 0.50 &  126.5 \\ 
%     29/12/2006 &  0.14 &  1.92 & 0.34 & 0.09 &  -40.8 \\
%     07/03/2007 & -2.15 &  1.07 & 0.38 & 0.07 &  -193.4 \\
%     31/03/2007 &  2.24 &  1.73 & 0.42 & 0.13 &  176.3 \\
%     20/12/2008 &  2.52 &  1.44 & 0.31 & 0.18 &  179.2 \\
%     21/12/2008 &  2.54 &  1.39 & 0.19 & 0.13 &  -29.7 \\
%     22/12/2008 &  2.57 &  1.26 & 0.14 & 0.09 &  -16.7 \\
%     22/01/2010 &  2.37 &  0.67 & 0.21 & 0.17 &  -19.4 \\
%     06/01/2012 &  1.05 & -0.52 & 0.35 & 0.12 &  172.2 \\
    25/12/2004 & -3.48 & -0.87 & 0.31 & 0.10 &   -8.5 \\
    07/02/2006 & -3.93 &  0.39 & 3.11 & 0.35 &  126.5 \\ 
    29/12/2006 &  0.14 &  1.92 & 0.24 & 0.06 &  -40.8 \\
    07/03/2007 & -2.15 &  1.07 & 0.27 & 0.05 &  -193.4 \\
    31/03/2007 &  2.24 &  1.73 & 0.19 & 0.06 &  176.3 \\
    20/12/2008 &  2.52 &  1.44 & 0.07 & 0.04 &  179.2 \\
    21/12/2008 &  2.54 &  1.39 & 0.04 & 0.03 &  -29.7 \\
    22/12/2008 &  2.57 &  1.26 & 0.03 & 0.02 &  -16.7 \\
    22/01/2010 &  2.37 &  0.67 & 0.04 & 0.04 &  -19.4 \\
    06/01/2012 &  1.05 & -0.52 & 0.24 & 0.08 &  172.2 \\
   \noalign{\smallskip}
    \hline
  \end{tabular}
$^1$ $\sigma_{\rm major\ axis}$ and $\sigma_{\rm minor\ axis}$ have been normalized by the square root of the number of observations.
\end{table}

\subsection{Line identification}
\label{sec:lineident}

Table ~\ref{tab:linelist} provides the lines identified in the WR and O-star spectrum.

\begin{table}
      \caption[]{Line list for $H$- and $K$-band spectra. Line identification made use of the NIST database~\citep{NIST_ASD} and \citet{Varricatt+2004} ."?" denote tentative identifications.}
         \label{tab:linelist}
\centering                          % used for centering table
\begin{tabular}{l l | l l }        % centered columns (4 columns)
\hline\hline                 % inserts double horizontal lines
\multicolumn{2}{c|}{WR star} & \multicolumn{2}{c}{O star}\\
$\lambda$ [$\mu$m] & line &$\lambda$ [$\mu$m] & line\\    % table heading
\hline                        % inserts single horizontal line
\noalign{\smallskip}
 1.4764 & He\,II 9-6 & 1.4764  & He\,II 9-6\\
 1.4885 & He\,II 14-7& 1.5022  & N\,V\\
 1.491  & C\,IV      & 1.5188  & N\,V / He\,I\\
 1.5088 & He\,I      & 1.5310  & He\,I\\
 1.5723 & He\,II 13-7& 1.5399  & N\,V\\
 1.588  & He\,I      & 1.5723  & He\,II 13-7 \\
 1.641  & He\,I      & 1.5885  & H\,I 14-4\\
 1.664  & C\,IV      & 1.6114  & H\,I 13-4\\
 1.6922 & He\,II 12-7& 1.619   & N\,II?\\
 1.7007 & He\,I      & 1.6411  & H\,I 12-4\\
 1.7359 & He\,II 20-8& 1.6811  & H\,I 11-4 \\
 1.736/737 & C\,IV   & 1.6922  & He\,II 12-7\\
 1.801  & C\,IV      & 1.7007  & He\,I\\
        &            & 1.7367  & H\,I 10-4\\
        &            & 1.7521  & N\,III?\\
        &            & 1.7838/858/860 & C\,II\\
\hline                                   %inserts single line*
\noalign{\smallskip}
 1.9442 & He\,II 16-8    & 1.9442  & He\,I\\
 1.9548 & He\,I          & 1.9451     & H\,I 8-4\\
 2.012  & C\,IV          & 2.0587     & He\,I\\
 2.0378 & He\,II 15-8    & 2.0705/796/842 & C\,IV\\
 2.0587 & He\,I          &   2.1126     & He\,I\\
 2.0705/796/842  & C\,IV & 2.1138     & He\,I\\
 2.1061 & C\,IV          & 2.1614     & He\,I\\
 2.1081 & C\,III         & 2.1651     & He\,II 14-8\\
 2.1126 & He\,I          & 2.1661     & H\,I 7-4\\
 2.139  & C\,III         & 2.1890     & He\,II 10-7\\
 2.1651 & He\,II 14-8    & 2.2929     & O\,III?\\
 2.1890 & He\,II 10-7    & 2.3234/47 & C\,III?\\
 2.278  & C\,IV          & 2.3469     & He\,II 13-8\\
 2.3234/247/265 & C\,III & 2.4219-4318 &  C\,IV\\
 2.3270 & C\,IV          & 2.4460     & He\,I\\
 2.3469 & He\,II 13-8\\
 2.4219-4318 & C\,IV \\
\hline                                   %inserts single line
\end{tabular}
\end{table}

\end{document}